\documentclass[reprint,superscriptaddress,amsmath,amssymb,aps,prb,longbibliography]{revtex4-2}
\usepackage{ulem}
\usepackage{multirow}
\usepackage{siunitx} 
\usepackage{braket}
\usepackage{mathtools}
\usepackage{graphicx}
\usepackage{color}
\usepackage[bookmarks=true,colorlinks=true,urlcolor=blue,linkcolor=blue,citecolor=blue,breaklinks]{hyperref}

\usepackage{ulem}
\usepackage{xspace}
\usepackage{amsmath}
\usepackage{amssymb}
\usepackage{amsfonts}
\usepackage{bm}
\usepackage{tikz-feynman}
\tikzfeynmanset{compat=1.1.0}
\usepackage{minitoc}
\usepackage{tocloft}

\definecolor{pu}{RGB}{200,50,200}
\definecolor{gr}{RGB}{0,187,0}
\definecolor{bl}{RGB}{68,34,200}
\definecolor{re}{RGB}{200,34,68}
\definecolor{ye}{RGB}{255,165,0}
\definecolor{oran}{RGB}{255,170,0}

\newcommand{\moire}[0]{moir\'{e}\xspace}

\newcommand{\RNum}[1]{\uppercase\expandafter{\romannumeral #1\relax}}

\begin{document}

\title{Many-body perturbation theory for \moire systems}

\author{Liangtao Peng}
\affiliation{Department of Physics, Washington University in St. Louis, St. Louis, Missouri 63130, United States}
\author{Giovanni Vignale}
\affiliation{The Institute for Functional Intelligent Materials, National University of Singapore, 4 Science Drive 2, Singapore 117544}
\author{Shaffique Adam}
\affiliation{Department of Physics, Washington University in St. Louis, St. Louis, Missouri 63130, United States}
\affiliation{Department of Materials Science and Engineering, 
National University of Singapore, 9 Engineering Drive 1, 
Singapore 117575}

\begin{abstract}
Moir\'{e} systems such as magic-angle twisted bilayer graphene have attracted significant attention due to their ability to host correlated phenomena including superconductivity and strongly correlated insulating states. By defining the single-particle Green's function in the band basis, we systematically develop a many-body perturbation theory framework to address correlations beyond the usual mean-field Hartree-Fock approaches.  As a specific example, we first analyze twisted bilayer graphene within the Hartree-Fock approximation.  We derive analytical solutions for symmetry-breaking states at integer fillings and the finite-temperature metal-insulator transition that closely match previously known numerical results in the literature. Moving beyond Hartree-Fock, we incorporate self-consistent GW corrections demonstrating that first-order diagrams significantly overestimate the filling-dependent fluctuations in the electronic compressibility.  This framework provides a comprehensive pathway for exploring strong electronic correlations in \moire systems beyond mean-field, giving new insights into the interplay of symmetry breaking and electron correlations.
\end{abstract}

\maketitle

\section{Introduction}

Moir\'{e} systems have garnered significant interest in part due to their ability to host strongly correlated phenomena. A prime example is twisted bilayer graphene (TBG), which exhibits remarkable superconductivity and correlated insulating behavior near the magic angle~\cite{Cao_Nature2018_correlated, Cao_Nature2018_unconventional}. These discoveries have spurred exploration of other twisted materials, such as twisted trilayer graphene (TTG)~\cite{Park_Nature2021_TTG, Kim_2022Nature_TTG} and twisted transition metal dichalcogenides (TMDs)~\cite{Wang_2020Naturematerials_TMDs, Zhang_2020NaturePhysics_TMDs, Devakul_2021NatureComm_TMDs}. The high tunability of these systems makes them ideal platforms for investigating long-standing questions in strongly correlated physics.

Understanding the strongly correlated phenomena in \moire systems involves tackling the many-body problem in two steps: first, determining the band structure at the single-particle level, and second, incorporating electron-electron interactions. A pivotal breakthrough in the first step was achieved by Refs.~\cite{CastroNeto_PRL2007_TBG, MacDonald_PNAS2011_Moireband}, who developed a non-interacting low-energy model that predicts 
flat bands near the so-called magic angles. This model captures the essential physics of low-twist-angle graphene and aligns well with experimental observations in the twist-angle window before strong correlations emerge \cite{Michael_PRB2015_Localspectroscopy}. 

The next crucial step is to systematically incorporate Coulomb interactions.  This is challenging due to the system's large unit cell, extremely low bandwidths, and numerous degrees of freedom e.g.~spin, valley, layer, and sublattices (that inherent to honeycomb lattices).  Numerous studies focused on Hartree and Fock mean-field approaches  (see e.g.~ Refs.~\cite{Ezzi_arxiv2024_Hartree, Guinea_PNAS2018_Hartree, Cea_RPB2019_Hartree, Po_PRX2018_MIT, Bultinck_PRX2020_HiddenSymmetry, Kwan_PRX2021_Kekul, Xie_PRL2020_HartreeFock, Wagner_PRL2022_HartreeFock, Rademaker_PRB2019_Hartree, Goodwin_Electronic2020_Hartree, Zhang_PRB2020_HartreeFock, Bernevig_PRB2021_HartreeFock, Bernevig_PRB2021_HartreeFock2, Liu_PRB2021_HartreeFock, Liu_PRR2021_HartreeFock, Xie_PRB2023_HartreeFock}).  These mean-field approaches reveal several interesting findings including a Kramers intervalley-coherent (K-IVC) ground state \cite{Bultinck_PRX2020_HiddenSymmetry}, Kekul\'{e} spiral order \cite{Kwan_PRX2021_Kekul}, and a ``heavy fermion" pocket and an ultra-flat band near the magic angle \cite{Bernevig_PRB2021_HartreeFock, Cea_RPB2019_Hartree, Ezzi_arxiv2024_Hartree}.  These advances although based on mean-field theory and therefore, strictly speaking, ignoring correlations, have nevertheless greatly expanded our intuition for possible strongly correlated states in TBG.

However, the current Hartree-Fock framework has several limitations. First, the observation of superconductivity necessitates theoretical descriptions that include higher-order contributions~\cite{Peng_PRB2024_Plasmon, Guinea_PNAS2021_Coulomb, Lewandowski_PRB2021_umklapp}. Any microscopic mechanism of superconductivity requires a framework beyond Hartree-Fock~\cite{Christos_NaturmComm2023_NodalSC}, and the common approach of solving the gap equation based on bands obtained from the Hartree-Fock theory is not a controlled approximation. Moreover, Hartree-Fock theory relies on order parameters derived from the equal-time density matrix, making it challenging to extend to higher-order diagrams -- especially time-dependent ring diagrams or GW diagrams beyond mean-field. Very often, the Hartree-Fock theory is often unable to correctly assess the relative stability of states characterized by different order parameters, or none at all. This problem becomes particularly acute when, as in the case of \moire systems, the mean field theory predicts  multiple solutions with ground state energy differences on the order of meV per electron. It remains a real possibility that the energy ordering of these ground states can be changed by correlation effects captured through the higher-order diagrams.

Thus, a new framework is needed to address these challenges. Many-body perturbation theory based on the Green's function method offers a promising solution. The Green's function method for many-body theory emerged in the 1960s, inspired by the success of Feynman diagrams in high-energy physics. Early successes include the calculation of the correlation energy of the high-density electron gas by Gell-Mann and Brueckner~\cite{PhysRev.106.364}. The approach was codified in classic textbooks such as those by Kadanoff and Baym~\cite{kadanoff2018quantum} and Abrikosov, Gorkov, and Dzyaloshinskii~\cite{Abrikosov1964MethodsOQ}, among others~\cite{PhysRev.139.A796, PhysRev.118.1417, fetter2012quantum}.

Green's functions are particularly useful for going beyond mean-field approximations and provide a solid microscopic foundation for the concept of quasiparticles, including calculations of renormalized energies and lifetimes. The method is also a cornerstone of the theory of disordered systems—for example, the Coherent Potential Approximation~\cite{PhysRev.156.809}—and is regularly used in the description of transport phenomena via the Kubo formalism. Although initially introduced in the context of perturbation theory it was soon realized that the exact formulation leads to integral equations—such as the Dyson equation~\cite{PhysRev.75.1736} and the Bethe-Salpeter equation~\cite{PhysRev.84.1232}—which are intrinsically non-perturbative and offer a natural way to introduce mean-field-like self-consistency in the treatment of strongly correlated systems. Moreover, Green's function methods have played a central role in establishing the strong coupling theory of superconductivity due to electron-phonon interactions (Eliashberg theory)~\cite{eliashberg1960interactions}. Therefore, applying the Green's function approach to \moire materials could enable significant strides in understanding these systems. To our knowledge, such an approach has not yet been developed for \moire systems.

In this paper, we aim to fill this gap by systematically developing a framework of many-body perturbation theory for \moire systems.  Working in the basis that diagonalizes the single-particle Hamiltonian, we construct the  Green's function and introduce the diagrammatic representation of the Coulomb interaction perturbation series.  As a case study, we analyze first-order and ring diagrams for TBG. Using the Green's function method, we derive exact analytical solutions for the symmetry-breaking ground states at integer fillings in the chiral flat limit. We prove that different symmetry-breaking states at integer filling factors are degenerate in energy.  Away from the chiral flat limit, our framework analytically demonstrates that the intervalley coherent (IVC) state is the ground state, aligning with prior numerical studies.

Our work goes further by incorporating finite-temperature effects, allowing us to calculate the closing of the mean-field gap at a critical temperature, in close analogy to the Bardeen–Cooper–Schrieffer (BCS) superconducting-metal transition.  The transition temperature obtained analytically with this approach is in good agreement with existing numerical results. Another key advancement of this study is the investigation of ring diagrams within the GW approximation. Our comprehensive numerical calculations reveal that first-order diagrams tend to overestimate the electronic compressibility. Unlike previous studies, our approach treats all diagrams, including ring diagrams, in a fully self-consistent manner. This methodological innovation not only improves the accuracy of the compressibility predictions but also lays the groundwork for future extensions to study superconducting phenomena.

The rest of the paper is organized as follows: In Sec.~\ref{sec:model}, we introduce the full Hamiltonian and establish the band basis. In Sec.~\ref{sec:manybodyperturbationtheory}, we define the Green's function in the band basis and apply many-body perturbation theory. Section~\ref{sec:TBG} provides an example focusing on TBG in the chiral flat limit; here, we discuss the symmetry-breaking states and the metal-insulator transition at integer fillings. In Sec.~\ref{sec:GW}, we derive the ring diagrams and apply the self-consistent GW calculation. Finally, we present our conclusions and suggest future directions in Sec.~\ref{sec:Conclusion}. All derivations and technical details are provided in Appendices~\ref{appendix:full_coulomb}–\ref{appendix:MIT}.

\section{Model and Hamiltonian}
\label{sec:model}

We begin by defining the full Hamiltonian of the system as  $\mathcal{H} = \mathcal{H}_{0} + \mathcal{H}_{\mathrm{I}}$, where $\mathcal{H}_{0}$ represents the single-particle (non-interacting) Hamiltonian
\begin{equation}\label{eq:ham_h0}
\mathcal{H}_{0}
=\sum_{\tilde{\mathbf{k}},\tilde{\mathbf{k}}^{\prime}}\sum_{\alpha,\alpha^{\prime},\sigma}  \left[\mathcal{H}^{\sigma}(\tilde{\mathbf{k}},\tilde{\mathbf{k}}^{\prime})\right]_{\alpha\alpha^{\prime}} \hat{c}^{\dagger}_{\tilde{\mathbf{k}},\alpha,\sigma} \hat{c}_{\tilde{\mathbf{k}}^{\prime},\alpha^{\prime},\sigma},
\end{equation}
where $\hat{c}^{\dagger}_{\tilde{\mathbf{k}},\alpha,\sigma}$ and $\hat{c}_{\tilde{\mathbf{k}},\alpha,\sigma}$ are the creation and annihilation operators for electrons with momentum  $\tilde{\mathbf{k}}$, internal degree of freedom $\alpha$ encompassing valley and sublattice, and conserved quantum number $\sigma$ -- the electron spin. The matrix elements  $\left[\mathcal{H}^{\sigma}(\tilde{\mathbf{k}},\tilde{\mathbf{k}}^{\prime})\right]_{\alpha\alpha^{\prime}}$ constitute the single-particle Hamiltonian in momentum space, which depends on $\sigma$ and acts on the internal degrees of freedom. 

The interacting part of the Hamiltonian, $\mathcal{H}_{\mathrm{I}}$, is expressed as
\begin{equation}\label{eq:ham_hi}
\begin{aligned}
\mathcal{H}_{\mathrm{I}}&=
\frac{1}{2}
\sum_{\substack{\tilde{\mathbf{k}},\tilde{\mathbf{k}}^{\prime},\tilde{\mathbf{q}}}}
\sum_{\substack{\alpha,\alpha^\prime}}
\sum_{\substack{\sigma,\sigma^{\prime}}}
V_{\tilde{\mathbf{q}}}
\hat{c}_{\tilde{\mathbf{k}}+\tilde{\mathbf{q}},\alpha,\sigma}^{\dagger}
\hat{c}_{\tilde{\mathbf{k}}^{\prime}-\tilde{\mathbf{q}},\alpha^{\prime},\sigma^{\prime}}^{\dagger}
\hat{c}_{\tilde{\mathbf{k}}^{\prime},\alpha^{\prime},\sigma^{\prime}}
\hat{c}_{\tilde{\mathbf{k}},\alpha,\sigma},\\
\end{aligned}
\end{equation}
where $\tilde{\mathbf{k}}$, $\tilde{\mathbf{k}}^{\prime}$, and $\tilde{\mathbf{q}}$ are constrained to the Brillouin zone (BZ), and $V_{\tilde{\mathbf{q}}}=2\pi e^2/\epsilon \tilde{q}$ represents the Fourier transform of the Coulomb interaction. By defining the \moire Brillouin zone (mBZ) with reciprocal lattice vector $\mathbf{G}$, we can expand $\tilde{\mathbf{k}}=\mathbf{k}+\mathbf{G}$ with $\mathbf{k}$ restricted to the mBZ. Due to the \moire potential, the single-particle Hamiltonian mixes different $\mathbf{G}$ components, leading to a momentum-space Hamiltonian that is periodic in the mBZ~\cite{MacDonald_PNAS2011_Moireband}.

While the full Hamiltonian $\mathcal{H}$ contains all the necessary information, working on this basis presents practical challenges. First, the single-particle Hamiltonian $\mathcal{H}_{0}$ includes both diagonal and off-diagonal terms in momentum space, complicating perturbative analyses. Second, achieving convergence requires a large number of bands (i.e., internal degrees of freedom $\alpha$ and reciprocal lattice vectors $\mathbf{G}$), significantly increasing computational complexity. A natural solution is to introduce the band basis by diagonalizing the single-particle Hamiltonian
\begin{equation}\label{eq:nibasis}
\hat{c}_{\mathbf{k},n,\sigma}=\sum_{\alpha,\mathbf{G}}u_{n}(\mathbf{k},\mathbf{G};\alpha,\sigma)\hat{c}_{\mathbf{k},\mathbf{G},\alpha,\sigma},
\end{equation}
where $n$ indexes the energy bands and $\hat{c}_{\mathbf{k}, \mathbf{G}, \alpha, \sigma}$ is shorthand for $\hat{c}_{\mathbf{k}+\mathbf{G}, \alpha, \sigma}$. In this new basis, the non-interacting Hamiltonian becomes diagonal
\begin{equation}\label{eq:ham_ni}
\begin{aligned}
\mathcal{H}_{0}
&=\sum_{\mathbf{k}} \sum_{n,\sigma}  E^{\sigma}_{n}(\mathbf{k})\hat{c}^{\dagger}_{\mathbf{k},n,\sigma} \hat{c}_{\mathbf{k},n,\sigma},\\
\end{aligned}
\end{equation}
where $E^{\sigma}_{n}(\mathbf{k})$ are the eigenvalues (band energy) of the single-particle Hamiltonian for each $\sigma$.

Applying this basis transformation to the interaction Hamiltonian, we obtain (see Appendix.~\ref{appendix:full_coulomb} for details):
\begin{widetext}
    \begin{equation}
\begin{aligned}
\mathcal{H}_{\mathrm{I}}
=\frac{1}{2}
\sum_{\substack{\mathbf{k},\mathbf{k}^{\prime},\mathbf{q}}}
\sum_{\{n_i\}}
\sum_{\sigma,\sigma^\prime}
&V^{\sigma\sigma^{\prime}}_{\mathbf{q},\{n_i\}}
\hat{c}_{\mathbf{k}+\mathbf{q},n_1,\sigma}^{\dagger}
\hat{c}_{\mathbf{k}^{\prime}-\mathbf{q},n_2,\sigma^{\prime}}^{\dagger}
\hat{c}_{\mathbf{k}^{\prime},n_3,\sigma^{\prime}}
\hat{c}_{\mathbf{k},n_4,\sigma},
\end{aligned}
\end{equation}
\begin{equation}
\begin{aligned}
V^{\sigma\sigma^{\prime}}_{\mathbf{q},\{n_i\}}=
\sum_{\mathbf{G}}
V_{\mathbf{q}+\mathbf{G}}
\left[\Lambda^{*}_{\mathbf{k},\mathbf{q}+\mathbf{G}}\right]^{\sigma}_{n_4n_1}
\left[\Lambda^{*}_{\mathbf{k}^{\prime},-\mathbf{q}-\mathbf{G}}\right]^{\sigma^\prime}_{n_3n_2},
\end{aligned}
\end{equation}
\begin{equation}\label{eq:form_factor_full}
\left[\Lambda_{\mathbf{k},\mathbf{q}+\mathbf{G}}\right]^{\sigma}_{m n}=\sum_{\alpha, \mathbf{G}^{\prime}}
u^*_{m} \left(\mathbf{k},\mathbf{G}^{\prime} ; \alpha,\sigma\right) 
u_{n} \left(\mathbf{k},\mathbf{G}^{\prime}+\mathbf{q}+\mathbf{G} ; \alpha,\sigma\right).
\end{equation}
\end{widetext}
where $V^{\sigma\sigma^{\prime}}_{\mathbf{q},\{n_i\}}$ is the interaction matrix element and 
$\left[\Lambda_{\mathbf{k},\mathbf{q}+\mathbf{G}}\right]^{\sigma}_{m n}$
is the form factor.  In the band basis, the interaction Hamiltonian resembles the standard Coulomb interaction but with modified interaction strengths that depend on the form factors and band indices. It is important to note that at magic angle, the single-particle bandwidth becomes much smaller than the Coulomb interaction strength, enhancing the relative importance of electron-electron interactions. The form factors play a critical role in this case. Although calculating the form factors analytically can be difficult, their symmetry properties and dependence on the \moire structure nonetheless offer valuable insights that allow us to make precise statements about the properties of the ground state and the emergence of correlated phenomena.

\section{Many-body perturbation theory}
\label{sec:manybodyperturbationtheory}

To include the Coulomb interaction in a systematic way, we employ many-body perturbation theory by treating the interaction Hamiltonian $\mathcal{H}_{\mathrm{I}}$ as a perturbation to the single-particle Hamiltonian $\mathcal{H}_{0}$. The key idea of this method is to define the single-particle imaginary-time Green's function in the band basis~\cite{mahan2013many}
\begin{equation}
    \left[G(\mathbf{k},\tau)\right]_{\eta\eta^{\prime}}=-\langle \mathcal{T}_{\tau} \hat{c}_{\mathbf{k},\eta}(\tau)\hat{c}^{\dagger}_{\mathbf{k},\eta^{\prime}}(0) \rangle,
\end{equation}
where $\eta = (n, \sigma)$ combines the band index $n$ and the internal quantum number $\sigma$. Here, $\hat{c}_{\mathbf{k},\eta}(\tau)$ and $\hat{c}^{\dagger}_{\mathbf{k},\eta^{\prime}}(0)$ are the annihilation and creation operators in the Heisenberg picture, $\mathcal{T}_{\tau}$ denotes the time-ordering operator, and $\langle \cdots \rangle$ represents the thermal average over the interacting system. By performing a Fourier transform with respect to imaginary time $\tau$, we obtain the Green's function in frequency space
\begin{equation}
\left[G(\mathbf{k},i\omega_n)\right]_{\eta\eta^{\prime}}=\int_{0}^{\beta}e^{i\omega_n\tau}\left[\mathcal{G}(\mathbf{k},\tau)\right]_{\eta\eta^{\prime}}d\tau,
\end{equation}
where $\beta = 1/(k_{\mathrm{B}} T)$ is the inverse temperature, $k_{\mathrm{B}}$ is Boltzmann's constant, and $i\omega_n = (2n + 1)\pi/\beta$ are the fermionic Matsubara frequencies.

In the non-interacting limit, the Green's function is diagonal in $\eta$ and simplifies to
\begin{equation}
\left[G_{0}(\mathbf{k},i\omega_n)\right]_{\eta\eta^{\prime}} = \frac{\delta_{\eta,\eta^{\prime}}}{i\omega_n - E_{n}^{\sigma}(\mathbf{k}) +\mu}, 
\end{equation} 
where $E_{n}^{\sigma}(\mathbf{k})$ are the eigenvalues of the single-particle Hamiltonian, and $\mu$ is the global chemical potential.

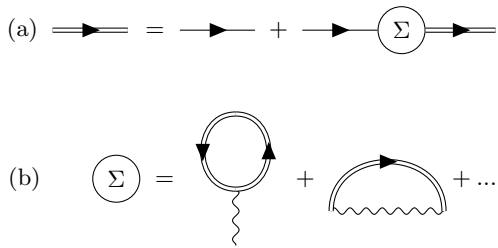
\begin{figure}[!t]
\centering
\begin{minipage}{\linewidth}
\[ 
\vcenter{\hbox{\begin{tikzpicture}
  \begin{feynman}
    \vertex (a) at (0,0){(a)};
    \vertex (i0);
    \vertex [right=0.4cm of i0] (i);
    \vertex [right=1cm of i] (o);
    \diagram*{
      (i) --[double,double distance=0.3ex,with arrow=0.5] (o)      
    };
  \end{feynman}
\end{tikzpicture}}}
~=~
\vcenter{\hbox{\begin{tikzpicture}
  \begin{feynman}
    \vertex (i);
    \vertex [right=1cm of i] (o);
    \diagram*{
      (i) --[fermion] (o)     
    };
  \end{feynman}
\end{tikzpicture}}}
~+~
\vcenter{\hbox{\begin{tikzpicture}
  \begin{feynman}
    \vertex (i);
    \node[right=1cm of i,draw,fill=white,circle] (v){$\Sigma$};
    \vertex [right=1.25cm of v] (o);
    \diagram*{
      (i) --[fermion] (v),        
      (v) --[double,double distance=0.3ex,with arrow=0.5]  (o)
    };
  \end{feynman}
\end{tikzpicture}}}
\]

\[
\vcenter{\hbox{\begin{tikzpicture}
  \begin{feynman}
    \vertex (a) at (-0.5,0){(b)};
    \vertex (i);
    \node[right=0.4cm of i,draw,fill=white,circle] (v){$\Sigma$};
    \diagram*{  (v) };
  \end{feynman}
\end{tikzpicture}}}
~=~
\vcenter{\hbox{\begin{tikzpicture}
    \begin{feynman}
        \vertex (a1);
        \vertex[right=of a1] (a2);
        \vertex[right=of a2] (a3);
        \vertex[above=0.75cm of a2] (a4);
        \vertex[above=1cm of a4] (a5);
        \diagram* {
        (a2)--[photon](a4),
        (a4)--[double,double distance=0.3ex,with arrow=0.5, half right](a5)--[double,double distance=0.3ex,with arrow=0.5, half right](a4)
        };
    \end{feynman}
\end{tikzpicture}}}
~+~
\vcenter{\hbox{\begin{tikzpicture}
\begin{feynman}
        \vertex (a1);
        \vertex[right=1cm of a1] (a2);
        \vertex[right=1.5cm of a2] (a3);
        \vertex[right=1cm of a3] (a4);
        \diagram* {
        (a3)--[photon](a2),
        (a2)--[double,double distance=0.3ex,with arrow=0.5,half left](a3),
        };
    \end{feynman}
\end{tikzpicture}}}
+...
\]
\end{minipage}
\caption{(a) Diagrammatic representation of Dyson's equation in terms of the self-energy. Single and double solid lines represent the bare Green's function and dressed Green's function, respectively. By including the proper electron self-energy, the dressed Green's function can be fully determined. (b) Expansion of self-energy. Here we draw the first-order diagrams known as the Hartree and Fock diagrams.}
\label{fig:Dyson_SelfEnergy}
\end{figure}

The interacting Green's function, $\hat{G}(\mathbf{k},i\omega_n)$, is determined by the Dyson equation
\begin{equation}
\hat{G}^{-1}(\mathbf{k},i\omega_n) = \hat{G}^{-1}_0(\mathbf{k},i\omega_n) - \hat{\Sigma}(\mathbf{k},i\omega_n),
\end{equation}
where $\hat{\Sigma}(\mathbf{k},i\omega_n)$ represents the electron self-energy, and $\hat{G}_0(\mathbf{k},i\omega_n)$ denotes the non-interacting Green's function. As illustrated in Fig.~(\ref{fig:Dyson_SelfEnergy}a), this equation is typically solved self-consistently, since the self-energy, $\hat{\Sigma}(\mathbf{k},i\omega_n)$, is a functional of the interacting Green's function, $\hat{G}(\mathbf{k},i\omega_n)$. During this iterative process, the global chemical potential is also adjusted self-consistently to ensure that the  total number of electrons $N_e$ is conserved, i.e. 
\begin{equation}\label{eq:electron_conservation}
N_e=\sum_{\mathbf{k}}\mathrm{Tr}[\hat{\rho}(\mathbf{k})]\,,  
\end{equation}
where the single-particle density matrix $\hat{\rho}(\mathbf{k})$ is related to the Green's function by
\begin{equation}\label{eq:density_matrix}
\hat{\rho}(\mathbf{k})=\hat{G}(\mathbf{k},\tau=0^-)=\frac{1}{\beta}\sum_{n}e^{-i\omega_n0^+}\hat{G}(\mathbf{k},i\omega_n),
\end{equation}
with $0^{\pm}$ ensuring proper analytic continuation.

Using Feynman diagram techniques~\cite{mahan2013many, Jishi_2013}, the self-energy $\hat{\Sigma}(\mathbf{k},i\omega_n)$ can be expanded order by order. As illustrative examples, we show here the first-order diagrams known as the Hartree and Fock diagrams. The Hartree self-energy is given by
\begin{equation}\label{eq:diagram_hartree}
\hat{\Sigma}_{\mathrm{H}}(\mathbf{k},i\omega_n) = \sum_{\mathbf{G}}
V_{\mathbf{G}}
\hat{\Lambda}^{\mathrm{T}}_{\mathbf{k},\mathbf{G}}
\frac{1}{\beta}\sum_{\mathbf{k}^{\prime},m}
\mathrm{Tr}\left[\hat{\Lambda}^{*}_{\mathbf{k}^{\prime},\mathbf{G}}\hat{G}(\mathbf{k}^{\prime},i\omega_m)\right],
\end{equation}
and the Fock self-energy is expressed as
\begin{equation}\label{eq:diagram_fock}
\begin{aligned}
\hat{\Sigma}_{\mathrm{F}}(\mathbf{k},i\omega_n)=&-\frac{1}{\beta}\sum_{m}
\sum_{\substack{\mathbf{q},\mathbf{G}}}
V_{\mathbf{q}+\mathbf{G}}
\hat{\Lambda}^{*}_{\mathbf{k},\mathbf{q}+\mathbf{G}}\times\\
&\hat{G}(\mathbf{k}+\mathbf{q},i\omega_n-i\omega_m)
\hat{\Lambda}^{\mathrm{T}}_{\mathbf{k},\mathbf{q}+\mathbf{G}}.
\end{aligned}
\end{equation}
Detailed derivations for these diagrams can be found in Appendix.~\ref{appendix:diagrams}.

Once the interacting Green's function is determined, 
the total energy of the system can be calculated using the Galitskii-Migdal formula~\cite{Galitskii_JETP1958_Etot, Holm_PRB2000_Etot}
\begin{equation}
E_{\mathrm{tot}}
=\frac{1}{\beta}\sum_{\mathbf{k},n}e^{-i\omega_n0^+}\mathrm{Tr}\left[\left(\hat{\mathcal{H}}_0(\mathbf{k})
+\frac{1}{2}\hat{\Sigma}(\mathbf{k},i\omega_n)\right)\hat{G}(\mathbf{k},i\omega_n)\right].
\end{equation}
If the self-energy is frequency-independent, this expression simplifies to
\begin{equation}\label{eq:total_energy}
E_{\mathrm{tot}}
=\sum_{\mathbf{k}}\mathrm{Tr}\left[\left(\hat{\mathcal{H}}_0(\mathbf{k})
+\frac{1}{2}\hat{\Sigma}(\mathbf{k})\right)\hat{\rho}(\mathbf{k})\right].
\end{equation}

These formulas presented in this section provide a framework for systematically incorporating the Coulomb interaction in \moire systems. We can systematically account for higher-order diagrams to compute both ground-state properties and quasiparticle spectra beyond the limitations of mean-field approximations.  A crucial aspect of this approach is the Green's function, which, although defined in the sublattice-layer basis, is formally equivalent to that in the band basis through a unitary transformation, as outlined in Eq.~\eqref{eq:nibasis}. One key advantage of working on the band basis is that it allows a targeted focus on a limited number of relevant bands near the Fermi level. This focus not only simplifies the problem but also significantly reduces the degrees of freedom in the system, making the calculations more computationally efficient without compromising accuracy in the low-energy regime.

\begin{table*}[!ht]
    \renewcommand{\arraystretch}{1.25} 
    \centering
    \begin{tabular}{| c | c | c | c |} 
    \hline
    $\nu$ & Phase & Order parameter [Eq.~(\ref{eq:order_parameter})] & Self-energy [Eq.~(\ref{eq:self_energy_HF})] \\ [0.5ex] 
    \hline
    \multirow{4}{*}{0} & Polarized & $\pm\hat{\sigma}_z, \pm\hat{\tau}_z, \pm\hat{\tau}_z\hat{\sigma}_z$ & $\mp\Sigma_\mathrm{F}\hat{\sigma}_z, \mp\Sigma_\mathrm{F}\hat{\tau}_z, \mp\Sigma_\mathrm{F}\hat{\tau}_z\hat{\sigma}_z$ \\ 
    \cline{2-4}
     & \multirow{2}{*}{Hall} & $\pm\hat{\gamma}_y, \pm\hat{\tau}_z\hat{\gamma}_y$ & $\mp\Sigma_\mathrm{F}\hat{\gamma}_y, \mp\Sigma_\mathrm{F}\hat{\tau}_z\hat{\gamma}_y$ \\ 
    \cline{3-4}
    & & $\pm\hat{\sigma}_z\hat{\gamma}_y, \pm\hat{\tau}_z\hat{\sigma}_z\hat{\gamma}_y$ & $\mp\Sigma_\mathrm{F}\hat{\sigma}_z\hat{\gamma}_y, \mp\Sigma_\mathrm{F}\hat{\tau}_z\hat{\sigma}_z\hat{\gamma}_y$ \\ 
    \cline{2-4}
    & IVC & $\pm\hat{\gamma}_{0,y}\hat{\tau}_{\mathrm{IVC}}$ & $\mp\Sigma_\mathrm{F}\hat{\gamma}_{0,y}\hat{\tau}_{\mathrm{IVC}}$ \\ 
    \cline{1-4}
    \multirow{2}{*}{-1} & Hall & ${\color{green}\pm}\left(1-\hat{\tau}_{{\color{blue}\pm}}\hat{\sigma}_{{\color{red}\pm}}\right)\hat{\gamma}_y-\hat{\tau}_{{\color{blue}\pm}}\hat{\sigma}_{{\color{red}\pm}}$ & $\Sigma_\mathrm{F}\left[{\color{green}\mp}\left(1-\hat{\tau}_{{\color{blue}\pm}}\hat{\sigma}_{{\color{red}\pm}}\right)\hat{\gamma}_y+\hat{\tau}_{{\color{blue}\pm}}\hat{\sigma}_{{\color{red}\pm}}\right]-\Sigma_\mathrm{H}$ \\ 
    \cline{2-4}
    & IVC+Hall & ${\color{yellow}\pm}\hat{\tau}_{\mathrm{IVC}} \hat{\sigma}_{{\color{red}\pm}} \hat{\gamma}_{0,y} {\color{green}\pm} \hat{\tau}_{{\color{blue}\pm}}\hat{\sigma}_{{\color{red}\mp}}\hat{\gamma}_y - \hat{\tau}_{{\color{blue}\mp}}\hat{\sigma}_{{\color{red}\mp}}$ & $\Sigma_\mathrm{F}\left({\color{yellow}\mp}\hat{\tau}_{\mathrm{IVC}}\hat{\sigma}_{{\color{red}\pm}}\hat{\gamma}_{0,y}{\color{green}\mp}\hat{\tau}_{{\color{blue}\pm}}\hat{\sigma}_{{\color{red}\mp}}\hat{\gamma}_y+\hat{\tau}_{{\color{blue}\mp}}\hat{\sigma}_{{\color{red}\mp}}\right)-\Sigma_\mathrm{H}$ \\ 
    \cline{1-4}
    \multirow{3}{*}{-2} & Polarized & ${\color{red}\pm}\hat{\tau}_{{\color{blue}\pm}}\hat{\sigma}_z-\hat{\tau}_{{\color{blue}\mp}}$ & $\Sigma_\mathrm{F}\left({\color{red}\mp}\hat{\tau}_{{\color{blue}\pm}}\hat{\sigma}_z+\hat{\tau}_{{\color{blue}\mp}}\right)-2\Sigma_\mathrm{H}$ \\ 
    \cline{2-4}
    & Hall & ${\color{red}\pm}\hat{\tau}_{{\color{blue}\pm}}\hat{\sigma}_z\hat{\gamma}_y-\hat{\tau}_{{\color{blue}\mp}}$ & $\Sigma_\mathrm{F}\left({\color{red}\mp}\hat{\tau}_{{\color{blue}\pm}}\hat{\sigma}_z\hat{\gamma}_y+\hat{\tau}_{{\color{blue}\mp}}\right)-2\Sigma_\mathrm{H}$ \\ 
    \cline{2-4}
    & IVC & ${\color{yellow}\pm}\hat{\sigma}_{{\color{red}\pm}}\hat{\gamma}_{0,y} \hat{\tau}_{\mathrm{IVC}}-\hat{\sigma}_{{\color{red}\mp}}$ & $\Sigma_\mathrm{F}\left({\color{yellow}\mp}\hat{\tau}_{\mathrm{IVC}}\hat{\sigma}_{{\color{red}\pm}}\hat{\gamma}_{0,y}+\hat{\sigma}_{{\color{red}\mp}}\right)-2\Sigma_\mathrm{H}$ \\ 
    \cline{1-4}
    -3 & Hall & ${\color{green}\pm}\hat{\tau}_{{\color{blue}\pm}}\hat{\sigma}_{{\color{red}\pm}}\hat{\gamma}_y-\hat{\tau}_{{\color{blue}\mp}}-\hat{\tau}_{{\color{blue}\pm}}\hat{\sigma}_{{\color{red}\mp}}$ & $\Sigma_\mathrm{F}\left({\color{green}\mp}\hat{\tau}_{{\color{blue}\pm}}\hat{\sigma}_{{\color{red}\pm}}\hat{\gamma}_y+\hat{\tau}_{{\color{blue}\mp}}+\hat{\tau}_{{\color{blue}\pm}}\hat{\sigma}_{{\color{red}\mp}}\right)-3\Sigma_\mathrm{H}$ \\ 
    \hline
\end{tabular}
\caption{Analytical expressions for the order parameter and corresponding self-energy for different symmetry-breaking states at each integer filling factor $\nu$. Here, $\hat{\tau}$, $\hat{\sigma}$, and $\hat{\gamma}$ represent the Pauli matrices in the valley, spin, and band bases, respectively. We also introduce $\hat{\tau}_{\pm} = \frac{1}{2}(1 \pm \hat{\tau}_z)$ and $\hat{\sigma}_{\pm} = \frac{1}{2}(1 \pm \hat{\sigma}_z)$ for notational convenience. The color-coded terms are used to distinguish between different components of the order parameters and the corresponding self-energies for clarity. For each filling factor, there are mainly three types of symmetry-breaking states: (i) Polarized states, (ii) Hall states, and (iii) Intervalley coherent states. All the symmetry-breaking degenerate in energy at each filling factor.}
\label{tb:ground_state_chiralflat}
\end{table*}

\section{Exact Hartree-Fock ground state for twisted bilayer graphene}
\label{sec:TBG}

So far, our discussion has been general and applicable to all \moire systems. To demonstrate the practical application of our theory, we now focus on TBG and analyze the first-order diagrams within our many-body perturbation theory.  An important outstanding question for TBG systems is determining the exact ground state after including electron-electron interactions. This is crucial not only for a better understanding of flat-band systems but also for providing a starting point for investigating many-body phenomena, such as superconductivity. In earlier studies, Bultinck \textit{et al.}~\cite{Bultinck_PRX2020_HiddenSymmetry} numerically demonstrated that the Hartree-Fock ground state is an inter-valley coherent state (IVC). This state involves a coherent superposition of states in the two valleys, but breaks both time-reversal symmetry $\mathcal{T}$ and the valley charge conservation symmetry $\mathrm{U}(1)$. More recently, Becker \textit{et al.}~\cite{becker2023exactgroundstateinteracting} proved that for a two-band TBG model at charge neutrality without spin and valley degrees of freedom, there are only two possible ground states in the chiral-flat limit namely, the Slater determinant states. Following that, Stubbs \textit{et al.}~\cite{stubbs2024hartreefockgroundstatemanifold} incorporated both spin and valley degrees of freedom and showed that the manifold of Hartree-Fock ground states at half-filling is characterized by a $\mathrm{U}(4) \times \mathrm{U}(4)$ symmetry (see e.g. Fig.~5 in Ref.~\cite{Bultinck_PRX2020_HiddenSymmetry}) applied to five unique Slater determinant states. To our knowledge, an analytical proof for the exact Hartree-Fock ground state beyond this chiral-flat limit and beyond the special case of charge neutrality has not been previously provided in the literature.

In this section, we aim to analytically understand the ground state of TBG close to magic angle within the first-order Hartree-Fock theory. Our analysis is based on the Bistritzer-MacDonald (BM) model~\cite{MacDonald_PNAS2011_Moireband}. Unless stated otherwise, we adopt the parameters $\hbar v_F/a_0=2.365$ eV and $\epsilon=10$ throughout this work, where $v_F$ denotes the Fermi velocity of monolayer graphene, and $a_0=0.246$ nm is the lattice constant. The intersublattice interlayer coupling is set to $w_1 = 0.11$ eV, which corresponds to the magic angle $\theta=1.086$ degree. We assume that the interaction-induced self-energy is much smaller than the band gap between the central flat bands and the higher-energy bands, allowing us to restrict our focus to the lowest eight bands and thereby reduce the system's degrees of freedom. This assumption is roughly satisfied near magic angle~\cite{Po_PRX2018_MIT, Bultinck_PRX2020_HiddenSymmetry}. Under this approximation, the complete Green's function for TBG can be represented as an $8 \times 8$ matrix in the $\hat{\tau} \otimes \hat{\sigma} \otimes \hat{\gamma}$ space, where $\hat{\tau}$, $\hat{\sigma}$, and $\hat{\gamma}$ denote the valley, spin, and band pseudospin degrees of freedom, respectively. For simplicity, we omit identity matrices in our notation; for example, $\hat{\tau}_z \hat{\gamma}_x$ represents $\hat{\tau}_z \otimes \hat{\sigma}_0 \otimes \hat{\gamma}_x$.

We note that when the parameters of the BM model are obtained by fitting to DFT results or by comparing with experimental data, the effects of interactions are already partially incorporated.  As a result, incorporating interactions in this way risks double-counting these contributions (see e.g. discussion in Refs.~\cite{Bultinck_PRX2020_HiddenSymmetry, Xie_PRL2020_HartreeFock}). For example, if the interlayer coupling is turned off, effectively decoupling the two layers, the Fock contribution leads to a logarithmically divergent renormalization of the Dirac velocity~\cite{HoKin_Science2018, DasSarma_PRB2014}. However, this renormalization is unnecessary if the tight-binding parameters were specifically chosen to reproduce the experimentally measured Dirac velocity.  

To avoid such double counting, we first regularize our Hartree-Fock diagrams by subtracting the density of the decoupled graphene bilayer up to the charge-neutral point. This procedure is equivalent to the ``subtraction scheme'' introduced by Ref.~\cite{Xie_PRL2020_HartreeFock} where they define the density matrix relative to isolated rotated graphene layers filled to their charge neutrality point. Specifically, we replace the density matrix defined in Eq.~\eqref{eq:density_matrix} by $\hat{\rho} \rightarrow \hat{\rho} - \frac{1}{2} \hat{\mathbb{I}}$. Recognizing that $\sum_{n}\frac{e^{-i\omega_n0^+}}{i\omega_n} = \frac{1}{2}$, this is equivalent to regularizing our diagrams by replacing $\hat{G}$ in Eq.~(\ref{eq:diagram_hartree}) and Eq.~(\ref{eq:diagram_fock}) with $\hat{G} \rightarrow \hat{G} - \frac{1}{i\omega_n} \hat{\mathbb{I}}$.  If all the diagrams we consider are time-independent, then the sum over Matsubara frequencies can be performed explicitly, and many-body perturbation theory formalism developed here exactly reduces to the previously developed mean-field theory method.  However, as we discuss below, the method still offers advantages such as analytical tractability.  To account for the subtraction discussed above, we define our mean-field order parameter as the shifted single-particle density matrix
\begin{equation}\label{eq:order_parameter}
\hat{Q}(\mathbf{k}) = 2\left(\hat{\rho}(\mathbf{k})-\frac{1}{2}\hat{\mathbb{I}}\right).
\end{equation}
Given that the density matrix $\hat{\rho}(\mathbf{k})$ satisfies the following properties
\begin{equation}
\sum_{\mathbf{k}}\mathrm{Tr}[\hat{\rho}(\mathbf{k})]=4+\nu; \quad \hat{\rho}^2(\mathbf{k})=\hat{\rho}(\mathbf{k}),
\end{equation}
and it can be shown that the order parameter satisfies
\begin{equation}\label{eq:constrain_Q}
\sum_{\mathbf{k}}\mathrm{Tr}[\hat{Q}(\mathbf{k})]=2\nu; \quad \hat{Q}^2(\mathbf{k})=\hat{\mathbb{I}}.
\end{equation}
The Hartree and Fock self-energies can then be rewritten in terms of $\hat{Q}(\mathbf{k})$ as
\begin{equation}
\hat{\Sigma}_{\mathrm{H}}(\mathbf{k})
= \frac{1}{2}\sum_{\mathbf{G}}
V_{\mathbf{G}}
\hat{\Lambda}^{\mathrm{T}}_{\mathbf{k},\mathbf{G}}
\sum_{\mathbf{k}^{\prime}}
\mathrm{Tr}\left[\hat{\Lambda}^{*}_{\mathbf{k}^{\prime},\mathbf{G}}\hat{Q}(\mathbf{k}^\prime)\right],
\end{equation}
and
\begin{equation}\label{eq:Fock_Q}
\hat{\Sigma}_{\mathrm{F}}(\mathbf{k})
=-\frac{1}{2}\sum_{\substack{\mathbf{q},\mathbf{G}}}
V_{\mathbf{q}+\mathbf{G}}
\hat{\Lambda}^{*}_{\mathbf{k},\mathbf{q}+\mathbf{G}}\hat{Q}(\mathbf{k}+\mathbf{q})
\hat{\Lambda}^{\mathrm{T}}_{\mathbf{k},\mathbf{q}+\mathbf{G}},
\end{equation}
with form factor given by Eq.~\eqref{eq:form_factor_full}. As discussed earlier, we have regularized our diagrams by subtracting the contribution of the decoupled graphene bilayer up to the charge-neutrality point. The self-energy at the Hartree-Fock level is then given by 
\begin{equation}\label{eq:self_energy_HF}
\hat{\Sigma}_{\mathrm{HF}}(\mathbf{k})=\hat{\Sigma}_{\mathrm{H}}(\mathbf{k})+\hat{\Sigma}_{\mathrm{F}}(\mathbf{k})\,,
\end{equation}

The Hartree and Fock energies can also be rewritten in terms of $\hat{Q}(\mathbf{k})$ by using Eq.~\eqref{eq:order_parameter} giving
\begin{equation}
\begin{aligned}
E_{\mathrm{H}}
&=\frac{1}{8}\sum_{\mathbf{G}}V_{\mathbf{G}}
\sum_{\mathbf{k}}\mathrm{Tr}\left[\hat{\Lambda}^{\mathrm{T}}_{\mathbf{k},\mathbf{G}}\hat{Q}(\mathbf{k})\right]
\sum_{\mathbf{k}^{\prime}}\mathrm{Tr}\left[\hat{\Lambda}^{*}_{\mathbf{k}^{\prime},\mathbf{G}}\hat{Q}(\mathbf{k}^\prime)\right],\\
\end{aligned}
\label{eq:hartree_energy}
\end{equation}
and 
\begin{equation}
\begin{aligned}
E_{\mathrm{F}}
&=-\frac{1}{8}\sum_{\mathbf{k},\mathbf{q},\mathbf{G}}
V_{\mathbf{q}+\mathbf{G}}
\mathrm{Tr}\left[
\hat{\Lambda}^{*}_{\mathbf{k},\mathbf{q}+\mathbf{G}}\hat{Q}(\mathbf{k}+\mathbf{q})
\hat{\Lambda}^{\mathrm{T}}_{\mathbf{k},\mathbf{q}+\mathbf{G}}\hat{Q}(\mathbf{k})\right].\\
\end{aligned}
\label{eq:fock_energy}
\end{equation}

\subsection{Gauge Fixing}

Before delving into detailed calculations, we briefly address the issue of gauge fixing. Near the magic angle, the bands become flat, making Coulomb interactions significant and the form factor introduced in Eq.~\eqref{eq:form_factor_full} crucial. The form factor can be generically parameterized as
\begin{equation}
\hat{\Lambda}_{\mathbf{k},\mathbf{q}+\mathbf{G}} = \sum_{a=0,z} \sum_{b=0,x,y,z} \Lambda^{ab}_{\mathbf{k},\mathbf{q}+\mathbf{G}} \hat{\tau}_{a} \hat{\gamma}_{b},
\end{equation}
where only $a = \{ 0, z \}$ are allowed because the non-interacting Hamiltonian is diagonal in the valley basis. However, because valley and Bloch momentum \(\mathbf{k}\) in the mBZ are good quantum numbers, the Bloch wave function retains a gauge degree of freedom under a unitary transformation, $\ket{\psi_{\mathbf{k},n,\eta}} \rightarrow  e^{i\varphi_{n,\eta}(\mathbf{k})}\,\ket{\psi_{\mathbf{k},n,\eta}}$,  where  $\varphi_{n,\eta}(\mathbf{k})$ is an arbitrary phase acting on the band and valley basis. The wave function is independent of spin due to the absence of spin-orbit coupling. Although physical observables are independent of the gauge choice, improper gauge fixing can result in redundant matrix elements $ \Lambda^{ab}_{\mathbf{k},\mathbf{q}+\mathbf{G}}$ in the form factor and random phases in symmetry-breaking order parameters.

Following Ref.~\cite{Bernevig_PRB2021_HartreeFock}, we adopt a specific gauge-fixing scheme in which the combined symmetry of rotational symmetry about the $z$-axis and time-reversal symmetry, $C_{2z}\mathcal{T}$, becomes a complex conjugation, while the combined symmetry of rotational symmetry about the $z$-axis and particle-hole symmetry, $C_{2z}\mathcal{P}$, is represented by $\hat{\tau}_y \hat{\gamma}_y$. Consequently, the form factor satisfies
\begin{equation}\label{eq:gauge_fixing}
\begin{aligned}
\hat{\Lambda}_{\mathbf{k},\mathbf{q}+\mathbf{G}} &= \hat{\Lambda}_{\mathbf{k},\mathbf{q}+\mathbf{G}}^{*}, \\
\hat{\Lambda}_{\mathbf{k},\mathbf{q}+\mathbf{G}} &= (\hat{\tau}_y \hat{\gamma}_y)^\dagger \hat{\Lambda}_{\mathbf{k},\mathbf{q}+\mathbf{G}} (\hat{\tau}_y \hat{\gamma}_y),
\end{aligned}
\end{equation}
as imposed by \(C_{2z}\mathcal{T}\) and \(C_{2z}\mathcal{P}\), respectively. These symmetries require the form factor \(\hat{\Lambda}_{\mathbf{k},\mathbf{q}+\mathbf{G}}\) to commute with \(\hat{\tau}_y \hat{\gamma}_y\) and ensure that \(\Lambda^{ab}_{\mathbf{k},\mathbf{q}+\mathbf{G}}\) is a real function. Therefore, the form factor matrix can be decomposed into four terms  
\begin{eqnarray}\label{eq:form_factor}
\hat{\Lambda}_{\mathbf{k},\mathbf{q}+\mathbf{G}} = && \hat{\gamma}_0 \Lambda^0_{\mathbf{k}, \mathbf{q}+\mathbf{G}} + \hat{\gamma}_x \hat{\tau}_z \Lambda^1_{\mathbf{k}, \mathbf{q}+\mathbf{G}} + i \hat{\gamma}_y \Lambda^2_{\mathbf{k}, \mathbf{q}+\mathbf{G}} \nonumber \\
&& + \hat{\gamma}_z \hat{\tau}_z \Lambda^3_{\mathbf{k}, \mathbf{q}+\mathbf{G}},
\end{eqnarray}
where \(\Lambda^i_{\mathbf{k}, \mathbf{q}+\mathbf{G}}\ (i=0,1,2,3)\) are real scalar functions.

In the chiral limit, where the hopping amplitude between AA regions, \(w_0 = 0\), the system acquires an additional chiral symmetry \(C\). Again, following Ref.~\cite{Bernevig_PRB2021_HartreeFock}, the chiral symmetry is represented by \(\hat{\tau}_z \hat{\gamma}_y\). This symmetry imposes an additional condition
\begin{equation}\label{eq:chiral_symmetry}
\hat{\Lambda}_{\mathbf{k},\mathbf{q}+\mathbf{G}} = (\hat{\tau}_z \hat{\gamma}_y)^\dagger \hat{\Lambda}_{\mathbf{k},\mathbf{q}+\mathbf{G}} (\hat{\tau}_z \hat{\gamma}_y),
\end{equation}
which requires the form factor to commute with \(\hat{\tau}_z \hat{\gamma}_y\). Combining Eqs.~\eqref{eq:form_factor} and \eqref{eq:chiral_symmetry}, the form factor in the chiral limit simplifies to
\begin{equation}\label{eq:form_factor_chiral}
\hat{\Lambda}_{\mathbf{k},\mathbf{q}+\mathbf{G}} = \Lambda^0_{\mathbf{k}, \mathbf{q}+\mathbf{G}} \hat{\gamma}_0 + \Lambda^2_{\mathbf{k}, \mathbf{q}+\mathbf{G}} i \hat{\gamma}_y,
\end{equation}
where only two components, \(\Lambda^0_{\mathbf{k}, \mathbf{q}+\mathbf{G}}\) and \(\Lambda^2_{\mathbf{k}, \mathbf{q}+\mathbf{G}}\), remain.

\subsection{Ground States at Chiral-flat Limit}

We first focus on the chiral-flat band limit~\cite{Tarnopolsky_PRL2019_Origin, Bernevig_PRB2021_HartreeFock}. In this limit, a well-defined magic angle with flat bands is achieved by setting the hopping amplitude between the AA regions, $w_0 = 0$ (the chiral condition). At this magic angle, the single-particle part of the Hamiltonian can be neglected by setting $E^{\sigma}_{n}(\mathbf{k}) = 0$ in Eq.~(\ref{eq:ham_ni}) (the flat-band condition). As discussed above, the form factor matrix can be decomposed into two components, as shown in Eq.~\eqref{eq:form_factor_chiral}.

The initial step is to identify all the possible ground states. While in principle the order parameters may depend on momentum, the simplest ground states have a uniform occupation in momentum space.  Consistent with previous studies, we assume that the order parameter $\hat{Q}(\mathbf{k}) = \hat{Q}$ is momentum-independent. Similar to the form factor, the order parameter can be written as
\begin{equation}
    \hat{Q} = \sum_{a,b,c=0,x,y,z} Q^{abc} \hat{\tau}_a \otimes \hat{\sigma}_b \otimes \hat{\gamma}_c,
\end{equation}
where the different components represent different symmetry-breaking phases subject to constraints given by Eq.~\eqref{eq:constrain_Q}.

To illustrate the method, we first analyze the spinless, valleyless model, which is entirely described by the spinor $\hat{\gamma}$. A self-consistent solution requires that the order parameter remains unchanged throughout the self-consistency loop. In other words, starting from an initial ansatz $\hat{Q}$, the order parameter obtained in the next iteration should remain the same. In the chiral-flat limit, a sufficient condition for this is that the commutator vanishes, i.e., $[\hat{Q},\hat{\Lambda}]=0$. It can be shown that both $\hat{\gamma}_0$ and $\hat{\gamma}_y$ satisfy this condition (see Appendix~\ref{appendix:conditions} for more detailed discussions). However, $\hat{\gamma}_0$ acts as a uniform shift across all bands, lacking any physical consequences. Therefore, the only possible order parameter is $\hat{\gamma}_y$, which breaks the $C_{2z}\mathcal{T}$ symmetry. The physical consequence of this state is that it gaps out the Dirac points \cite{Po_PRX2018_MIT} and carries a non-zero Chern number \cite{Bernevig_PRB2021_HartreeFock}, forming a Hall state.

For the full spinful and valleyful model, incorporating $\hat{\gamma}_0$ with spin and valley degrees of freedom gives rise to physically distinct symmetry-breaking phases. These include spin-polarized, valley-polarized, and spin-valley-locked states, characterized by the order parameters $\hat{\sigma}_z$, $\hat{\tau}_z$, and $\hat{\tau}_z \hat{\sigma}_z$, respectively. The interplay between spin and valley degrees of freedom also introduces additional Hall states, such as the quantum Hall state $\hat{\gamma}_y$ and the valley Hall state $\hat{\tau}_z \hat{\gamma}_y$. Moreover, it is possible to break the $U(1)$ symmetry while still satisfying the self-consistency condition. This scenario is characterized by the order parameter $\hat{\gamma}_{0} \hat{\tau}_{\mathrm{IVC}}$ or $\hat{\gamma}_{y} \hat{\tau}_{\mathrm{IVC}}$, where $\hat{\tau}_{\mathrm{IVC}} = \hat{\tau}_x \cos{\phi_{\mathrm{IVC}}} + \hat{\tau}_y \sin{\phi_{\mathrm{IVC}}}$.  

There are therefore two subtypes of intervalley coherent states:  the Kramers inter-valley coherent state (K-IVC), with the order parameter \(\hat{\gamma}_y \hat{\tau}_{\mathrm{IVC}}\), and the time-reversal inter-valley coherent state (\(\mathcal{T}\)-IVC), with order parameter \(\hat{\gamma}_0 \hat{\tau}_{\mathrm{IVC}}\). The former breaks time-reversal symmetry but preserves a modified time-reversal symmetry, \(\mathcal{T}^\prime\), which combines the regular (spinless) time-reversal symmetry \(\mathcal{T}\) with a \(\pi\)-phase shift in the IVC phase, while the latter preserves time-reversal symmetry. The K-IVC state can exhibit valley currents (circulating currents associated with valley polarization), which are absent in the T-IVC state.

To summarize, there are three families of symmetry-breaking states at integer fillings: 
\begin{enumerate}
\item \textbf{Polarized States}: These include spin-polarized, valley-polarized, and spin-valley-locked states. The corresponding order parameters are $\hat{\sigma}_{z}$, $\hat{\tau}_{z}$, and $\hat{\tau}_{z} \hat{\sigma}_{z}$, respectively. 
\item \textbf{Hall States}: These states have a non-zero Chern number and all involve the band-switching operator $\hat{\gamma}_y$, which creates a quasiparticle state that mixes the two lower bands. Examples are the quantum Hall state $\hat{\gamma}_y$, valley Hall state $\hat{\tau}_z\hat{\gamma}_{y}$, spin-Hall state $\hat{\sigma}_{z}\hat{\gamma}_{y}$, and valley-spin-Hall state $\hat{\tau}_{z}\hat{\sigma}_{z}\hat{\gamma}_{y}$. 
\item \textbf{Intervalley Coherent States}: These states break the \(U(1)\) symmetry, and involve a coherent superposition of states in the two valleys. The corresponding order parameter are Kramers inter-valley coherent state $\hat{\gamma}_{y} \hat{\tau}_{\mathrm{IVC}}$ and the time-reversal inter-valley coherent state $\hat{\gamma}_{0} \hat{\tau}_{\mathrm{IVC}}$, where $\hat{\tau}_{\mathrm{IVC}} = \hat{\tau}_x \cos{\phi_{\mathrm{IVC}}} + \hat{\tau}_y \sin{\phi_{\mathrm{IVC}}}$. The former breaks time-reversal symmetry but preserves the modified time-reversal symmetry, \(\mathcal{T}^\prime\), which combines the regular (spinless) time-reversal symmetry \(\mathcal{T}\) with a \(\pi\)-phase shift of the IVC phase, while the latter preserves time-reversal symmetry.
\end{enumerate}

Using the many-body perturbation theory, we derive exact analytical solutions for integer filling factors, which are summarized in Table~\ref{tb:ground_state_chiralflat} (details are provided in Appendix~\ref{appendix:diagrams}). We find that the self-energies share a universal structure
\begin{equation}\label{eq:OP_and_selfenergy}
\hat{\Sigma}_{\mathrm{HF}}(\mathbf{k})=\nu\Sigma_{\mathrm{H}}(\mathbf{k})-\Sigma_{\mathrm{F}}(\mathbf{k})\hat{Q},
\end{equation}
where $\Sigma_{\mathrm{H}}$ is the Hartree-induced correction
\begin{equation}
\Sigma_{\mathrm{H}}(\mathbf{k})=\sum_{\mathbf{k}^{\prime},\mathbf{G}} V_{\mathbf{G}}\Lambda^0_{\mathbf{k},\mathbf{G}}\Lambda^0_{\mathbf{k}^{\prime},\mathbf{G}},
\end{equation}
which is proportional to the filling factor $\nu$, and the Fock-induced correction
\begin{equation}
\Sigma_{\mathrm{F}}(\mathbf{k})=\frac{1}{2}\sum_{\mathbf{q},\mathbf{G}} V_{\mathbf{q}+\mathbf{G}}\left[\left(\Lambda^{0}_{\mathbf{k},\mathbf{q}+\mathbf{G}}\right)^2+\left(\Lambda^{2}_{\mathbf{k},\mathbf{q}+\mathbf{G}}\right)^2\right],
\end{equation}
that applies uniformly across different symmetry-breaking spinors, independent of $\nu$. Both the Hartree and Fock energies depend only on the bare Coulomb interaction and the form factors, which are determined by the specific twist angle. 

The universality of the Hartree and Fock corrections arises from the fact that the form factor in the chiral-flat limit [Eq.~(\ref{eq:form_factor_chiral})] is identical for all spin and valley components. This allows us to write the ground-state energy as
\begin{equation}
E_{\mathrm{tot}}=\frac{4+\nu}{2}\left[\nu \sum_{\mathbf{k}} \Sigma_{\mathrm{H}}(\mathbf{k})-\sum_{\mathbf{k}} \Sigma_{\mathrm{F}}(\mathbf{k})\right].
\end{equation}

For the chiral flat band limit (left panel of  Fig.~(\ref{fig:EnergySplit}a)), the $\mathrm{U}(4) \times \mathrm{U}(4)$ symmetry implies that in spite of different order parameters, all symmetry-breaking phases have the same total energy at each filling factor (see also Ref.~\cite{Bernevig_PRB2021_HartreeFock2}). The coefficient $4+\nu$ represents the number of filled bands, with each band contributing a fixed energy value. The term $\nu$ in the Hartree energy reflects the deviation from charge neutrality. We note that the vanishing Hartree energy at the charge-neutral point in TBG can also be understood through symmetry considerations~\cite{Ezzi_arxiv2024_Hartree}. An even stronger statement can be made that the $\mathbf{k}$-dependent self-energies of different ordered states are related to each other, at each $\mathbf{k}$, by a $\mathrm{U}(4) \times \mathrm{U}(4)$ transformations.  This property is inherited by the $\mathbf{k}$-dependent Green's function and implies that many-body corrections arising from additional diagrams (for example ring diagrams, as described in Section VI) will still be identical for different ordered phases. The degeneracy cannot be removed as long as exact $\mathrm{U}(4) \times \mathrm{U}(4)$ symmetry remains in force.

\subsection{Ground States at Integer Filling away from chiral-flat limit}

\begin{figure}
    \centering
    \begin{minipage}{.5\textwidth}
        \centering
        \includegraphics[trim={0cm 0 1.1cm 0},clip,width=0.9\columnwidth]{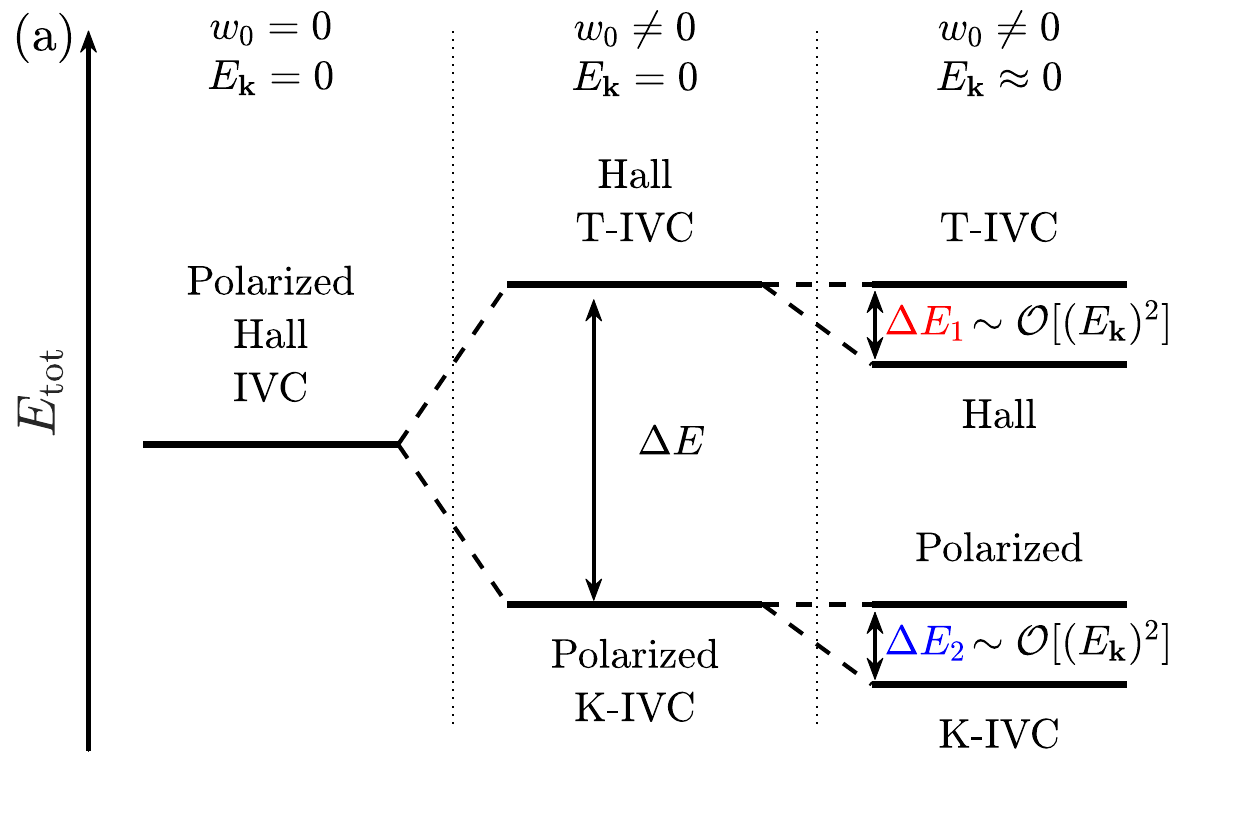}
    \end{minipage}
    \begin{minipage}{.5\textwidth}
        \centering
        \includegraphics[width=0.9\columnwidth]{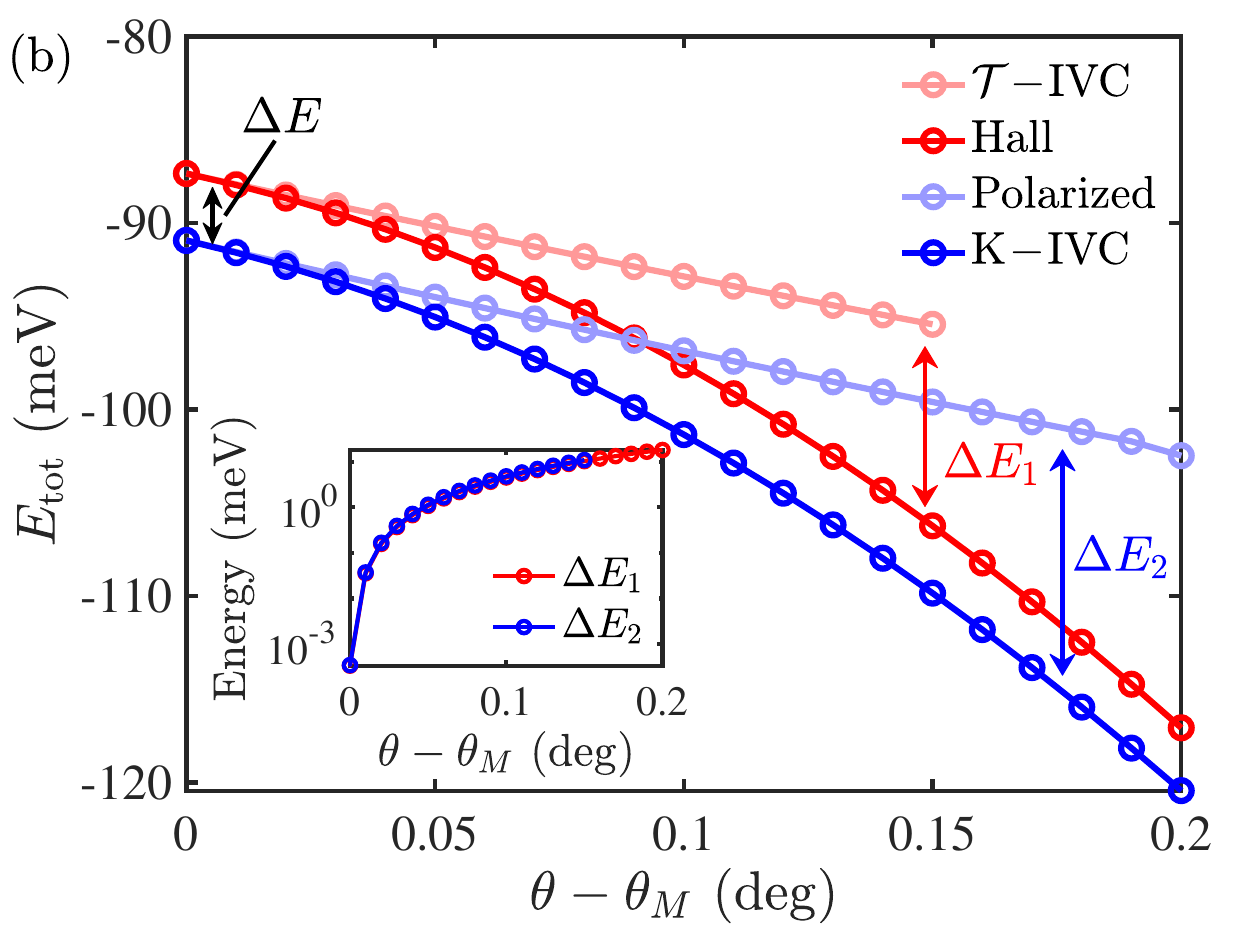}
    \end{minipage}
    \caption{(a) Schematic for the evolution of the ground-state energy for twisted bilayer graphene at charge neutrality based on our analytical results. Left: In the chiral flat limit, all symmetry-breaking phases are degenerate in energy.  Middle: Away from the chiral limit ($w_0 \ne 0$) but maintaining the flat-band condition ($E_k =0$) separates these phases states into two groups: (i) polarized states and K-IVC state; and (ii) Hall states and $\mathcal{T}$-IVC states.  Right: Finally, introducing finite single-particle dispersion ($\theta \neq \theta_M$) adds a second-order correction to the ground-state energy, resulting in the K-IVC state being the ground state. (b) Numerical solution for the ground-state energy as a function of twist angle at charge neutrality for the fully relaxed model.  The numerics agree with our analytical expectations.  For $\theta=\theta_M$ we observe the splitting into two groups similar to the middle panel above.  Away from magic angle, the expectations from the analytical model (right panel above) persist for small range in angles, but then breaks down due to single-particle effects. 
    The inset shows the numerically obtained ground-state energy differences between the $\mathcal{T}$-IVC/Hall state and the Polarized/K-IVC states that is consistent with analytical results.}
    \label{fig:EnergySplit}
\end{figure}

Next, we relax the condition $w_0 = 0$ while still assuming that the bands are flat. This situation is particularly relevant to a realistic model for twisted bilayer graphene close to magic angle where the hopping amplitude between the AA regions becomes nonzero.  The flat band assumption corresponds to being sufficiently close to the magic angle that Coulomb interaction is much larger than the bandwidth.  The absence of chiral symmetry 
in this situation introduces additional terms to the form factor (see Eq.~\eqref{eq:form_factor}).  These extra terms result in symmetry-breaking -- separating the T-IVC and Hall states from the polarized and K-IVC states (see middle panel of Fig.~(\ref{fig:EnergySplit}a)).

For concreteness, we focus on the case of charge neutrality $\nu=0$ and defer the discussion of scenarios away from charge neutrality to Appendix~\ref{appendix:solutions_non-chiralflat}. In the non-chiral flat limit, the commutator condition derived in the chiral-flat limit no longer holds, and self-consistent solutions emerge that violate the commutator condition, $[\hat{Q},\hat{\Lambda}]=0$. Furthermore, the degeneracy among all symmetry-breaking states observed in the chiral-flat limit is also lifted. To classify the ground states more precisely, we define the difference between the form factors in the chiral (Eq.~\eqref{eq:form_factor}) and non-chiral limits (Eq.~\eqref{eq:form_factor_chiral}) as 
\begin{equation}\label{eq:delta_lambda}
\delta\hat{\Lambda} = \hat{\gamma}_x \hat{\tau}_z \Lambda^1_{\mathbf{k}, \mathbf{q}+\mathbf{G}} + \hat{\gamma}_z \hat{\tau}_z \Lambda^3_{\mathbf{k}, \mathbf{q}+\mathbf{G}}.
\end{equation}
where $\delta\hat{\Lambda}$ is shorthand for $\delta\hat{\Lambda}_{\mathbf{k},\mathbf{q}+\mathbf{G}}$. Depending on the sign between $\delta\hat{\Lambda}^{*}\hat{Q}\delta\hat{\Lambda}^{\mathrm{T}}$ and $\hat{Q}$, we find that the symmetry-breaking phases considered in the previous case fall into two groups:

\begin{enumerate}
\item \textbf{Polarized States and K-IVC States}: These states satisfy the condition $\delta\hat{\Lambda}^{*}\hat{Q}\delta\hat{\Lambda}^{\mathrm{T}} \propto +\hat{Q}$. This group exhibit the same order parameter and self-energy as for the chiral-flat limit, as summarized in Table~\ref{tb:ground_state_chiralflat}, except that the Fock-induced self-energy is replaced as a matrix
\begin{equation}\label{eq:fock_group1}
\hat{\Sigma}^{\mathrm{\RNum{1}}}_{\mathrm{F}}(\mathbf{k})=
\Sigma^1_{\mathrm{F}}(\mathbf{k})
+\hat{\gamma}_x\hat{\tau}_z
\Sigma^2_{\mathrm{F}}(\mathbf{k})
+\hat{\gamma}_z\hat{\tau}_z\Sigma^3_{\mathrm{F}}(\mathbf{k}),
\end{equation}
\noindent where
\begin{equation}\label{eq:fock_group1_componants}
\begin{aligned}
\Sigma^1_{\mathrm{F}}(\mathbf{k})&=\frac{1}{2}\sum_{\mathbf{q},\mathbf{G}}
V_{\mathbf{q}+\mathbf{G}}\sum^4_{i=1} \left(\Lambda^i_{\mathbf{k}, \mathbf{q}+\mathbf{G}}\right)^2,\\
\Sigma^2_{\mathrm{F}}(\mathbf{k})&=\frac{1}{2}\sum_{\mathbf{q},\mathbf{G}}
V_{\mathbf{q}+\mathbf{G}}
\left[
\Lambda^0_{\mathbf{k}, \mathbf{q}+\mathbf{G}}
\Lambda^1_{\mathbf{k}, \mathbf{q}+\mathbf{G}}
-\Lambda^2_{\mathbf{k}, \mathbf{q}+\mathbf{G}}
\Lambda^3_{\mathbf{k}, \mathbf{q}+\mathbf{G}}
\right],\\
\Sigma^3_{\mathrm{F}}(\mathbf{k})&=\frac{1}{2}\sum_{\mathbf{q},\mathbf{G}}
V_{\mathbf{q}+\mathbf{G}}
\left[
\Lambda^0_{\mathbf{k}, \mathbf{q}+\mathbf{G}}
\Lambda^3_{\mathbf{k}, \mathbf{q}+\mathbf{G}}
+\Lambda^1_{\mathbf{k}, \mathbf{q}+\mathbf{G}}
\Lambda^2_{\mathbf{k}, \mathbf{q}+\mathbf{G}}
\right].\\
\end{aligned}
\end{equation}
The relationship between the order parameter and self-energy, as outlined in Eq.~\eqref{eq:OP_and_selfenergy}, remains valid, but the Fock-induced self-energy and order parameter product should now be interpreted as a matrix product.
\item \textbf{Hall States and $\mathcal{T}$-IVC States}: These states satisfy the condition $\delta\hat{\Lambda}^{*}\hat{Q}\delta\hat{\Lambda}^{\mathrm{T}} \propto -\hat{Q}$. Again, the corresponding order parameter and self-energy are also consistent with Table~\ref{tb:ground_state_chiralflat}, except the Fock-induced self-energy takes the form
\begin{equation}\label{eq:fock_group2}
\Sigma^{\mathrm{\RNum{2}}}_{\mathrm{F}}(\mathbf{k})=\frac{1}{2}\sum_{\mathbf{q},\mathbf{G}}
V_{\mathbf{q}+\mathbf{G}}
\left[
\sum_{i=0,2} \left(\Lambda^i_{\mathbf{k}, \mathbf{q}+\mathbf{G}}\right)^2
-\sum_{i=1,3} \left(\Lambda^i_{\mathbf{k}, \mathbf{q}+\mathbf{G}}\right)^2
\right].
\end{equation}
\end{enumerate}

\noindent The Hartree-induced correction remains the same across both groups of states. While the ground-state energy is degenerate within each group, energy splitting occurs between the two groups. By computing the ground state energy via Eq.~\eqref{eq:total_energy}, we find that the polarized states and K-IVC states are energetically favored, with the energy splitting given by
\begin{equation}\label{eq:energy_splitting}
\Delta E=(4+\nu)\sum_{\mathbf{k},\mathbf{q},\mathbf{G}} V_{\mathbf{q}+\mathbf{G}}\left[\left(\Lambda^{1}_{\mathbf{k},\mathbf{q}+\mathbf{G}}\right)^2+\left(\Lambda^{3}_{\mathbf{k},\mathbf{q}+\mathbf{G}}\right)^2\right].
\end{equation}

This lower energy of the polarized and K-IVC states can be understood through the expressions for the total Hartree and Fock energies, as given in Eq.~\eqref{eq:hartree_energy} and Eq.~\eqref{eq:fock_energy}. Since the Hartree-induced correction and the corresponding Hartree energy are identical for both symmetry groups, we only focus on the Fock energy. The Fock energy can be viewed as the trace of the product $\hat{\Lambda}^{*}\hat{Q}\hat{\Lambda}^{\mathrm{T}}$ and the order parameter $\hat{Q}$. 

As inferred from Eq.~\eqref{eq:delta_lambda}, the form factor in the non-chiral case can be written as $\hat{\Lambda}=\hat{\Lambda}^0+\delta\hat{\Lambda}$, where $\hat{\Lambda}^0$ represents the form factor in the chiral-flat limit, i.e., Eq.~\eqref{eq:form_factor}. The ground Fock energy can be therefore decomposed into four terms
\begin{equation}
\begin{aligned}
\mathrm{Tr}[\hat{\Lambda}^{*}\hat{Q}\hat{\Lambda}^{\mathrm{T}}\hat{Q}]&=\mathrm{Tr}[\hat{\Lambda}^{0*}\hat{Q}\hat{\Lambda}^{0\mathrm{T}}\hat{Q}]+\mathrm{Tr}[\delta\hat{\Lambda}^{*}\hat{Q}\delta\hat{\Lambda}^{\mathrm{T}}\hat{Q}]\\&+\mathrm{Tr}[\hat{\Lambda}^{0*}\hat{Q}\delta\hat{\Lambda}^{\mathrm{T}}\hat{Q}]+\mathrm{Tr}[\delta\hat{\Lambda}^{*}\hat{Q}\hat{\Lambda}^{0\mathrm{T}}\hat{Q}],
\end{aligned}
\end{equation}
where the first term recovers the ground-state energy of the chiral-flat band limit. It can be easily verified that the last two terms vanish. Therefore, the energy difference relative to the chiral-flat limit across all symmetry-breaking phases is solely determined by $\mathrm{Tr}[\delta\hat{\Lambda}^{*}\hat{Q}\delta\hat{\Lambda}^{\mathrm{T}}\hat{Q}]$. 

This trace of the product of Pauli matrices yields a non-zero value only when the matrices are in the same channel. Consequently, any order parameter satisfying $\delta\hat{\Lambda}^{*}\hat{Q}\delta\hat{\Lambda}^{\mathrm{T}} \propto +\hat{Q}$ (e.g.,  polarized states and K-IVC states) lowers the ground-state energy, while those satisfying $\delta\hat{\Lambda}^{*}\hat{Q}\delta\hat{\Lambda}^{\mathrm{T}} \propto -\hat{Q}$ (e.g., Hall states and $\mathcal{T}$-IVC states) increase the total energy, resulting in the energy splitting between the two symmetry groups.  This is illustrated in the rightmost panel of Fig.~(\ref{fig:EnergySplit}a).

Finally, we consider finite single-particle dispersion around the magic angle by adding a perturbation $E_n^\sigma(\mathbf{k}) \hat{\gamma}_z$ to the Hamiltonian. It has been demonstrated~\cite{Christos_PRX2022_HartreeTTG} that any perturbation commuting with $\hat{Q}$  does not contribute to the ground-state energy within second-order perturbation theory. Conversely, perturbations anticommuting with $\hat{Q}$ lower the energy to order $\mathcal{O} \left[ \left( E_n^\sigma(\mathbf{k}) \right)^2 \right]$. The anticommutation relation $\{ \hat{\gamma}_y \hat{\tau}_{\mathrm{IVC}}, \hat{\gamma}_z \} = 0$ for K-IVC states and the commutation relation $[ \hat{\sigma}_z, \hat{\gamma}_z ] = 0$ for polarized states indicate that the degeneracy between these states is lifted. Ultimately, the K-IVC state emerges as the ground state, consistent with previous studies~\cite{Bultinck_PRX2020_HiddenSymmetry} (see right panel of Fig.~(\ref{fig:EnergySplit}a)).

In this section, we discussed the evolution of the ground-state energy for TBG, summarized in Fig.~(\ref{fig:EnergySplit}a). We have shown that the chiral-flat limit serves as a clear starting point for understanding symmetry-breaking phases in TBG, with the realistic model viewed as a perturbation to the chiral-flat limit near the magic angle. By introducing a $w_0$ and finite single-particle dispersion, we enable a perturbative analysis where the energy differences between various symmetry-breaking states remain small. Additionally, in Fig.~(\ref{fig:EnergySplit}b), we extend our calculations to the fully relaxed model (see Ref.~\cite{Ezzi_arxiv2024_Hartree} for details on the model). As predicted by theory, at the magic angle, these four states separate into two distinct subgroups: (i) Polarized and K-IVC states, and (ii) Hall and $\mathcal{T}$-IVC states, with the energy difference within each subgroup approaching zero, as shown in the inset. Away from the magic angle, the theoretically predicted order persists near the magic angle but eventually breaks down due to the competition between single-particle effects and interactions. Our analytical predictions align closely with the full numerical solution highlighting the success of this approach.

\subsection{Finite Temperature Metal-to-Insulator Transition}

\begin{figure}
    \centering
    \includegraphics[width=\columnwidth]{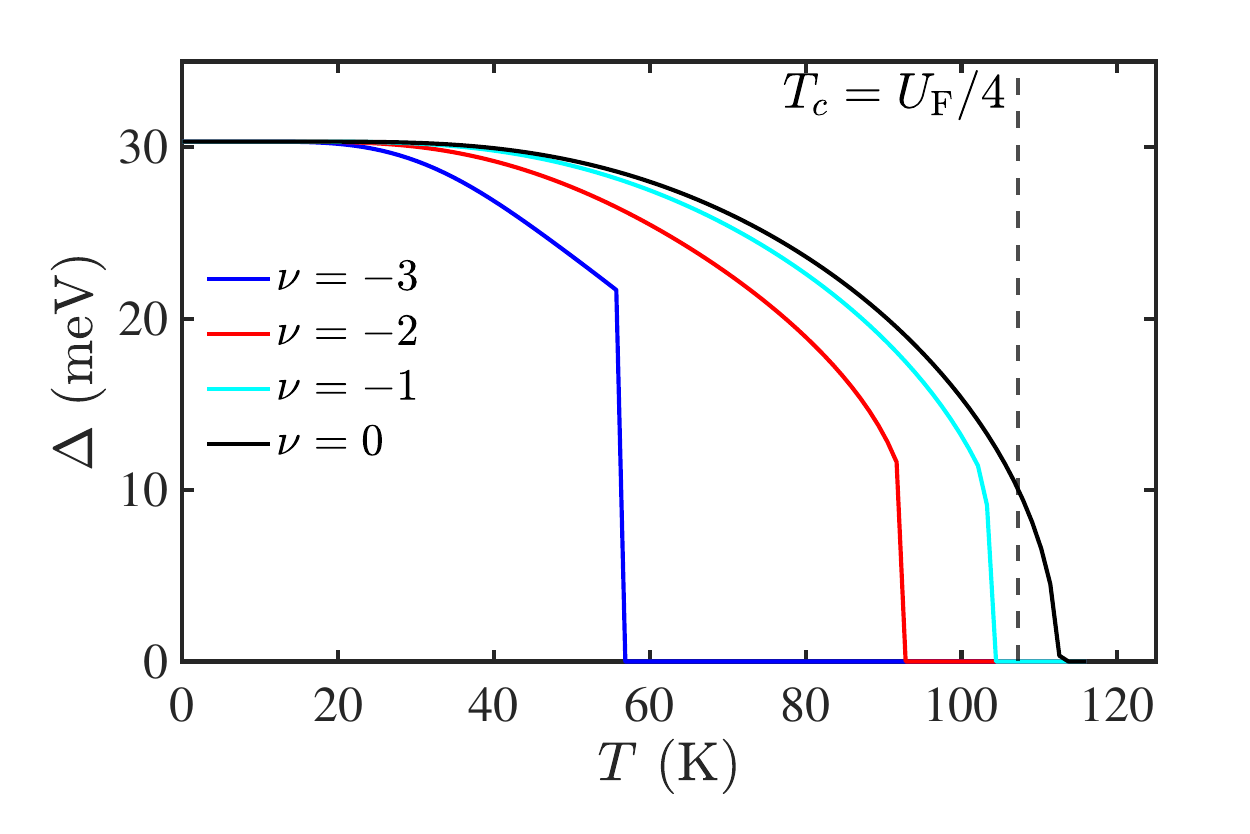}
    \caption{Numerical calculation of the Hartree-Fock gap versus temperature for different filling factors. A metal-insulator transition is observed, closely resembling the superconducting-metal transition. At the charge-neutral point, the absence of Hartree diagrams allows us to derive the BCS-like gap equation given in Eq.~(\ref{eq:BCS}). The corresponding solution for the transition temperature agrees well with numerical results. Away from charge neutrality, the interplay between Hartree and Fock effects becomes more complicated, making it challenging to derive an analytical solution. Our numerical results indicate that the transition temperature decreases as the filling factor moves away from charge neutrality.}
    \label{fig:MIT}
\end{figure}

We now extend our discussion to integer fillings at finite temperatures. Within the framework of many-body perturbation theory, incorporating finite temperatures is straightforward, as the Fermi-Dirac distribution is naturally embedded in the Matsubara summation. The mean-field theory predicts the appearance of a self-consistent gap in the single-particle spectrum, even though the wave function remains uncorrelated, i.e. a single Slater determinant. Figure~(\ref{fig:MIT}) illustrates the Hartree-Fock gap as a function of temperature for various integer filling factors in the chiral-flat limit. Notably, we have verified that, for each filling, different symmetry-breaking phases yield identical results, highlighting the robustness of the insulating behavior against specific symmetry-breaking configurations.

At zero temperature, all filling factors exhibit the same Hartree-Fock gap. This can be understood by initially neglecting the Hartree effect, which results in identical gaps across the fillings. The gap remains unchanged because the Hartree effect primarily shifts the bands upward without significantly altering the gap within the range of fillings considered. As temperature increases, the gap gradually closes, leading to a transition from the insulating to the metallic phase—a phenomenon known as the metal-to-insulator transition. This transition occurs sharply around a specific transition temperature, which generally decreases as the filling moves away from the charge-neutral point. This behavior arises because, at high temperatures, the Fermi-Dirac distribution approaches a constant value. This uniformity causes contributions from all degrees of freedom to cancel out, leading to the vanishing of the Hartree and Fock diagrams. As a result, thermal fluctuations effectively restore symmetry, driving the system into a metallic state.

Interestingly, the temperature dependence of the Hartree-Fock gap resembles that of the superconducting-to-metal transition, suggesting a similar underlying mechanism. At charge neutrality, we can simplify the equations by noting that the Hartree self-energy is zero, i.e., $\Sigma_{\mathrm{H}}(\mathbf{k}) = 0$. The gap equation reduces to a form reminiscent of the BCS gap equation. Assuming the gap is constant within the Brillouin zone, $\Delta = \Sigma_{\mathrm{F}}(\mathbf{k})$, we obtain
\begin{equation}\label{eq:BCS}
1=\frac{U_{\mathrm{F}}}{2}\frac{\tanh{(\Delta/2T)}}{\Delta},
\end{equation}
where $U_{\mathrm{F}}$ is the pairing strength given by
\begin{equation}
U_{\mathrm{F}}=\sum_{\mathbf{k},\mathbf{q},\mathbf{G}}
V_{\mathbf{q}+\mathbf{G}}\left[\left(\Lambda^{0}_{\mathbf{k},\mathbf{q}+\mathbf{G}}\right)^2+\left(\Lambda^2_{\mathbf{k},\mathbf{q}+\mathbf{G}}\right)^2\right],
\end{equation}
Taking the limit $\Delta \rightarrow 0$ in Eq.~(\ref{eq:BCS}), we find the {\it analytic} transition temperature
\begin{equation}
T=\frac{U_{\mathrm{F}}}{4}.
\end{equation}

Figure (\ref{fig:MIT}) compares our analytical predictions with numerical results, showing good agreement. Importantly, we find that the transition temperature at charge neutrality is solely determined by the Fock energy, as the Hartree term vanishes at this point. Away from charge neutrality, the interplay between Hartree and Fock effects becomes more complex, making an analytical solution challenging. Our numerical results indicate that the transition temperature decreases as the filling factor deviates from charge neutrality. This trend arises because the Hartree effect pushes the bands closer together, an effect that becomes more pronounced with increased deviation. This finding aligns with experimental observations that the correlated insulating gap at charge neutrality is stronger than at half-filling \cite{Lu_Nature2019_CI, Stepanov_Nature2020_Untying}.

\section{Self-consistent GW calculations}\label{sec:GW}

In this last section, we apply our formalism to a calculation that has not been done previously in the \moire literature by including terms beyond the Hartree and Fock diagrams. Specifically, we consider the combination of Hartree-Fock diagrams with ring diagrams within the random phase approximation (RPA), collectively known as the GW approximation.

The dynamic properties of screened Coulomb interactions in two-dimensional materials have long been recognized as crucial for understanding elementary excitation spectra and determining transport properties~\cite{DasSarma_RevModPhys2011, CastroNeto_RevModPhys2009}. In the context of \moire systems, the intrinsically undamped plasmon modes in TBG near the magic angle have been thoroughly investigated theoretically~\cite{Levitov_PNAS2019_Plasmon, Kuang_PRB2021_Plasmon, Cavicchi_PRB2024_Plasmon} and observed experimentally~\cite{Hu_PRL2017_Plasmon, Hesp_NaturePhysics2021_Plasmon, Huang_Nature2022_Plasmon}. Moreover, several theoretical studies suggest that these dynamic properties are pivotal for understanding the superconducting phase in \moire systems~\cite{Peng_PRB2024_Plasmon, Guinea_PNAS2021_Coulomb, Lewandowski_PRB2021_umklapp}. These findings underscore the importance of including dynamic screening effects in the GW approximation to gain a comprehensive understanding of both normal and superconducting phases in TBG.

However, the current RPA framework is incompatible with symmetry-breaking scenarios within the Hartree-Fock framework. Previous RPA formulations typically assume no symmetry breaking, whereas mean-field calculations have demonstrated that symmetry breaking is essential for understanding the cascade of phase transitions experimentally observed in TBG. Moreover, standard RPA approaches neglect Hartree diagrams, despite the fact that Hartree interactions have been shown to be crucial for band renormalization (see e.g. Refs.\cite{Ezzi_arxiv2024_Hartree, Cea_RPB2019_Hartree}). We note that these issues are peculiar to \moire materials.  For the conventional electron gas the Hartree diagrams at $\mathbf{q}=0$ exactly cancel the positive background contributions to the total energy. Similarly, they are not problematic for monolayer graphene, where the Hartree effect near charge neutrality is negligible and the system does not exhibit symmetry-breaking phases.

\begin{figure}
\centering
\begin{minipage}{\linewidth}
\[
\vcenter{\hbox{\begin{tikzpicture}
  \begin{feynman}
    \vertex (i);
    \vertex (a) at (-0.2,0){(a)};
    \node[right=0.4cm of i,draw,fill=white,circle] (v){$\Pi$};
    \diagram*{  (v) };
  \end{feynman}
\end{tikzpicture}}}
~=~
\vcenter{\hbox{\begin{tikzpicture}\begin{feynman}
    \vertex (a1);
    \vertex[right=1cm of a1] (a2);
    \diagram* {
    (a1)--[double,double distance=0.3ex,with arrow=0.5, half left, looseness=0.7](a2)--[double,double distance=0.3ex,with arrow=0.5, half left, looseness=0.7](a1)
    };
\end{feynman}\end{tikzpicture}}}
~+~
\vcenter{\hbox{\begin{tikzpicture}\begin{feynman}
    \vertex (a1);
    \vertex[right=1cm of a1] (a2);
    \vertex[above=0.5cm of a2] (a3);
    \vertex[right=1cm of a3] (a4);
    \diagram* {
    (a1)--[double,double distance=0.3ex,with arrow=0.5, half left, looseness=0.7](a2)--[double,double distance=0.3ex,with arrow=0.5, half left, looseness=0.7](a1),
    (a2)--[photon](a3),
    (a3)--[double,double distance=0.3ex,with arrow=0.5, half left, looseness=0.7](a4)--[double,double distance=0.3ex,with arrow=0.5, half left, looseness=0.7](a3)
    };
\end{feynman}\end{tikzpicture}}}
+...
~=~
\frac{
\vcenter{\hbox{\begin{tikzpicture}\begin{feynman}
    \vertex (a1);
    \vertex[right=1cm of a1] (a2);
    \diagram* {
    (a1)--[double,double distance=0.3ex,with arrow=0.5, half left, looseness=0.7](a2)--[double,double distance=0.3ex,with arrow=0.5, half left, looseness=0.7](a1)
    };
\end{feynman}\end{tikzpicture}}}
}{\hat{\mathbb{I}}-
\vcenter{\hbox{\begin{tikzpicture}\begin{feynman}
    \vertex (a1);
    \vertex[right=1cm of a1] (a2);
    \vertex[above=0.5cm of a2] (a3);
    \diagram* {
    (a1)--[double,double distance=0.3ex,with arrow=0.5, half left, looseness=0.7](a2)--[double,double distance=0.3ex,with arrow=0.5, half left, looseness=0.7](a1),
    (a2)--[photon](a3),
    };
\end{feynman}\end{tikzpicture}}}
}
\]

\[
\vcenter{\hbox{\begin{tikzpicture}\begin{feynman}
    \vertex (a) at (-1.7,0){(b)};
    \vertex (i);
    \node[right=0.4cm of i,draw,fill=white,circle] (v){$\Sigma$};
    \diagram*{  (v) };
\end{feynman}\end{tikzpicture}}}
~=~
\vcenter{\hbox{\begin{tikzpicture}\begin{feynman}
    \vertex (a1);
    \vertex[right=of a1] (a2);
    \vertex[right=of a2] (a3);
    \vertex[above=0.75cm of a2] (a4);
    \vertex[above=1cm of a4] (a5);
    \diagram* {
    (a2)--[photon](a4),
    (a4)--[double,double distance=0.3ex,with arrow=0.5, half right](a5)--[double,double distance=0.3ex,with arrow=0.5, half right](a4)
    };
\end{feynman}\end{tikzpicture}}}
~+~
\vcenter{\hbox{\begin{tikzpicture}\begin{feynman}
    \vertex (a1);
    \vertex (a) at (5,0){};
    \vertex[right=1cm of a1] (a2);
    \vertex[right=1.5cm of a2] (a3);
    \vertex[right=1cm of a3] (a4);
    \diagram* {
    (a2)--[double,double distance=0.3ex,with arrow=0.5,half left](a3),
    };
    \draw[double distance=0.3ex,decorate,decoration={snake,amplitude=1pt,segment length=5pt}] (a2) -- (a3);
\end{feynman}\end{tikzpicture}}}
\]

\[
\vcenter{\hbox{\begin{tikzpicture}\begin{feynman}
    \vertex (a) at (-1.9,0){(c)};
    \vertex (a1) at (0, 0);
    \vertex (a2) at (1, 0);
    \draw[double distance=0.3ex,decorate,decoration={snake,amplitude=1pt,segment length=5pt}] (a1) -- (a2);
\end{feynman}\end{tikzpicture}}}
~=~
\vcenter{\hbox{\begin{tikzpicture}\begin{feynman}
    \vertex (a1) at (0, 0);
    \vertex (a2) at (1, 0);
    \diagram* {
    (a1)--[photon](a2),
    };
\end{feynman}\end{tikzpicture}}}
~+~
\vcenter{\hbox{\begin{tikzpicture}\begin{feynman}
    \vertex (a1) at (0,0);
    \vertex (a) at (5,0){};
    \node[right=0.75cm of a1,draw,fill=white,circle] (a2){$\Pi$};
    \vertex[right=1cm of a2] (a3);
    \diagram* {
    (a1)--[photon](a2),
    (a2)--[photon](a3),
    };
\end{feynman}\end{tikzpicture}}}
\]
\end{minipage}
\caption{(a) The dressed bubble diagram in RPA, resulting from summing over ring diagrams. (b) Diagrammatic representation of the self-energy within the GW approximation. The diagrams considered include the Hartree diagram with the bare Coulomb interaction and the Fock diagram with the screened Coulomb interaction. It is important to noted that for Hartree diagrams one need to use bare Coulomb interaction to avoid double counting. (c) Ring contribution to the self-energy. The double-dashed line represents the screened Coulomb interaction $V(\mathbf{q},i\omega_n)$, and the single-dashed line represents the bare Coulomb interaction $V(\mathbf{q})$.}
\label{fig:Ring_SelfEnergy}
\end{figure}
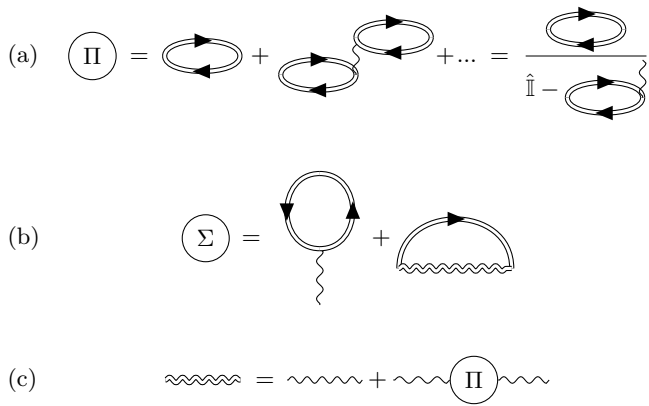

The GW approximation considers both the Hartree diagram and the RPA diagrams
\begin{equation}
\hat{\Sigma}_{\mathrm{GW}}(\mathbf{k},i\omega_n)=\hat{\Sigma}_{\mathrm{H}}(\mathbf{k},i\omega_n)+\hat{\Sigma}_{\mathrm{RPA}}(\mathbf{k},i\omega_n).
\end{equation}
\noindent In this context, we note a recent innovative approach presented in two recent works~\cite{Jihang_PRB2024_GW, Cavicchi_PRB2024_Plasmon}, which calculated one-shot GW diagrams and RPA corrections based on self-consistent Hartree bands.  These authors circumvented the need for directly computing dynamic ring diagrams.  However, while these studies demonstrated the importance of beyond mean-field corrections, their formalism was unable to compute dynamic ring diagrams self-consistently, a limitation that we overcome in the present work.  Following our many-body perturbation theory approach a schematic representation of the included diagrams is shown in Fig.~\eqref{fig:Ring_SelfEnergy}.   We introduce the ring diagrams in terms of the screened Coulomb interaction
\begin{widetext}
\begin{equation}
\hat{\Sigma}_{\mathrm{RPA}}(\mathbf{k},i\omega_n)
=-\frac{1}{\beta}\sum_{m}
\sum_{\mathbf{q},\mathbf{G}_1,\mathbf{G}_2}
\hat{V}(-\mathbf{q},i\omega_m)
\left[
\hat{\Lambda}^*_{\mathbf{k},\mathbf{q}+\mathbf{G}_1}
\hat{G}(\mathbf{k}+\mathbf{q},i\omega_n-i\omega_m)
\hat{\Lambda}^{\mathrm{T}}_{\mathbf{k},\mathbf{q}+\mathbf{G}_2}
\right],
\end{equation}
where $\hat{V}(\mathbf{q},i\omega_m)$ denotes the dynamically screened Coulomb interaction given by
\begin{equation}
\hat{V}(\mathbf{q},i\omega_m)
=\hat{V}_{\mathbf{q}}\left[\hat{\mathbb{I}}-\hat{V}_{\mathbf{q}}\hat{\Pi}(\mathbf{q},i\omega_m)\right]^{-1},
\end{equation}
and $\hat{\Pi}(\mathbf{q},i\omega_m)$ is the polarizability
\begin{equation}
\hat{\Pi}(\mathbf{q},i\omega_m)
=\frac{1}{\beta}\sum_{o}
\sum_{\mathbf{k}}
\mathrm{Tr}\left[
\hat{G}(\mathbf{k},i\omega_o)
\hat{\Lambda}^*_{\mathbf{k},\mathbf{q}+\mathbf{G}_1}
\hat{G}(\mathbf{k}+\mathbf{q},i\omega_o+i\omega_m)
\hat{\Lambda}^{\mathrm{T}}_{\mathbf{k},\mathbf{q}+\mathbf{G}_2}\right].
\end{equation}
\end{widetext}

\begin{figure}
    \centering
    \includegraphics[width=0.9\columnwidth]{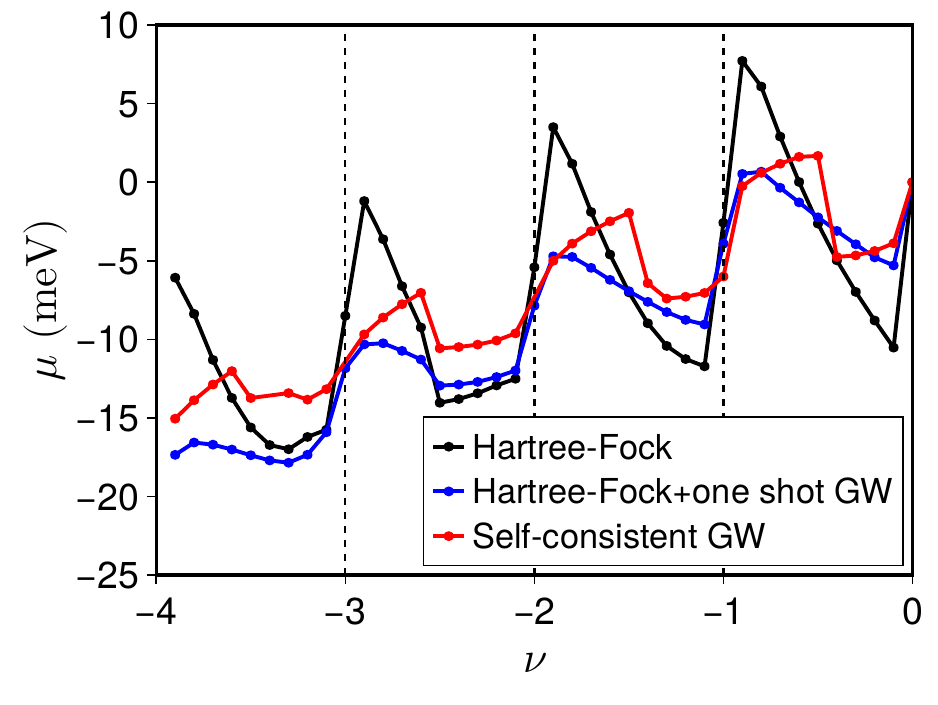}
    \caption{Chemical potential as a function of the filling factor is shown for different sets of included diagrams. Here, we present half of the filling factor range; the other half can be obtained via particle-hole symmetry. Our results show that first-order diagrams overestimate the oscillation of the chemical potential, which is directly related to the electronic compressibility, defined as \(d\mu/dn\), where \(n\) is the carrier density.
    Only the GW approaches show a weak oscillation superimposed over an overall increasing function consistent with experimental observations~\cite{Yazdani_Nature2020_Cascade, Zondiner_2020Nature_Cascade}.}
    \label{fig:ChemicalPotential}
\end{figure}

\noindent The detailed derivation can be found in Appendix~\ref{appendix:ring_diagrams}.  Figure~\eqref{fig:ChemicalPotential} compares the chemical potential obtained by including only the Hartree-Fock diagrams to the one calculated with the full self-consistent GW diagrams. As in the Hartree-Fock case, the chemical potential for GW calculations was determined by enforcing the conservation of the total number of electrons, as described by Eq.~\eqref{eq:electron_conservation}, for a given filling factor $\nu$, which relates to the total number of electrons $N_e$ via $N_e = 4 + \nu$ in the context of TBG with two bands. The only difference in the GW approach is the inclusion of additional diagrams. All approaches show oscillatory features  known as ``cascades" in the recent literature~\cite{Yazdani_Nature2020_Cascade, Zondiner_2020Nature_Cascade}.  However, we note that the Hartree-Fock approximation gives oscillations that are much larger than what is seen in experiment. These oscillations --  superimposed over a roughly linear Hartree increase -- can be easily understood by referring to the Hartree-Fock band structure plotted in Fig.~\eqref{fig:hf_bands}.  The chemical potential, interpreted as the energy of the highest occupied level, shows an overall decrease as a band goes from being empty to being fully occupied with increasing $\nu$: this is caused by the exchange energy which becomes more negative as the density increases and strongly dominates the effect of the small upward shift of the Fermi level within the band. However, when a new band first becomes occupied, which happens at integer filling factors, one observes a sharp positive jump in $\mu$, corresponding to the band gap.

The RPA ring diagrams included in the GW approach provide extra screening to the Fock contribution but not the Hartree contribution in the low wave-length limit thereby reducing the magnitude of the oscillations while not changing the overall increase of chemical potential with filling.  We can explore this idea further by first considering the Hartree and Fock diagrams and then adding the ring diagrams sequentially, order by order. To first order, the expression for the polarizability can be derived analytically
\begin{equation}
\left[\Pi(\mathbf{q}, i \omega)\right]_{\mathbf{G}_1=\mathbf{G}_2=0}=\sum_{\mathbf{k}} \sum_{\zeta, \zeta^{\prime}} \frac{\left(f_{\mathbf{k}+\mathbf{q}}^\zeta-f_{\mathbf{k}}^{\zeta^{\prime}}\right) F_{\mathbf{k}, \mathbf{k}+\mathbf{q}}^{\zeta \zeta^{\prime}}}{E_{\mathbf{k}+\mathbf{q}}^\zeta-E_{\mathbf{k}}^{\zeta^{\prime}}-i \omega},
\end{equation}
where the composite indices $\zeta, \zeta^{\prime}$ run over electron bands, valleys, and spins. Here, $f_{\mathbf{k}}^\zeta$ represents the Fermi-Dirac distribution for a state with energy $E_{\mathbf{k}}^\zeta$, and $F_{\mathbf{k}, \mathbf{k}+\mathbf{q}}^{\zeta \zeta^{\prime}}=\left|\psi_{\zeta, \mathbf{k}+\mathbf{q}}^{\dagger} \psi_{\zeta^{\prime}, \mathbf{k}}\right|^2$ is the form factor associated with different Bloch states. The energy $E_{\mathbf{k}}^\zeta$ and wave function $\psi_{\zeta^{\prime}, \mathbf{k}}$ should be understood as the quasi-particle state including Hartree-Fock diagram. Since the main contribution arises from the long-wavelength limit, we have
\begin{equation}
\left[\Pi(\mathbf{q}, i \omega=0)\right]_{\mathbf{G}_1=\mathbf{G}_2=0}\sim \frac{1}{\Delta},
\end{equation}
where $\Delta$ is the energy gap around the Fermi level. At integer filling factors, the Hartree-Fock approximation provides a gapped state with a gap around 30 meV, which effectively suppresses the contributions from the ring diagrams. This makes the Hartree-Fock approximation accurate for integer fillings.  However, if the Fock gap is diminished for any reason e.g. from a high-$\kappa$ dielectric substrate, then this would also reduce the accuracy of the Hartree-Fock approach since the contributions of the ring diagrams become (relatively) more important.  For non-integer fillings, the Fermi level is inside the band, making the RPA ring diagrams significant strongly suppressing the Fock contributions that are responsible for the oscillations in the chemical potential.  

Finally, we note that throughout our GW calculation, we find that adding ring diagrams does not remove the degeneracy of the ground state energy. This can be understood in terms of the $\mathrm{U}(4) \times \mathrm{U}(4)$ symmetry of the chiral-flat limit discussed earlier.  This symmetry ensures that any rotation within this group preserves the ground state degeneracy. The quasiparticle states, obtained by diagonalizing the self-energy $\Sigma_{\mathrm{HF}}(\mathbf{k})$, remain invariant under these rotations, as they correspond to a similarity transformation of the self-energy. In addition, the energy corrections from the ring diagrams are found to be too small to change the ordering of the broken-symmetry states in the non-chiral non-flat situation,as discussed in Fig.~(\ref{fig:EnergySplit}b). 

\section{Conclusion}
\label{sec:Conclusion}

In this work we have developed a framework for many-body perturbation theory based on the continuum model of \moire systems. Compared to previous mean-field theories, our approach offers several advantages. Notably, Green's function representation is highly versatile and can be systematically extended to higher-order diagrams and dynamic interactions. This framework can also incorporate anomalous diagrams to investigate the superconducting phase, potentially enabling a comprehensive understanding of interactions in twisted systems within a single consistent framework.

As a first application, we applied our theory to twisted bilayer graphene, which offers significant insights into correlated physics in a minimal setting. By including first-order diagrams—specifically, the Hartree and Fock diagrams—we analytically solved for the ground state. Remarkably, we found that different symmetry-breaking states within each filling number are degenerate in energy.  We extended our analysis to finite temperatures at integer filling factors, where our calculations revealed a metal-to-insulator transition. Inspired by the similarity between this transition and the superconducting-to-metal transition, we derived a BCS-like gap equation and connected the critical temperature to the Fock energy. The decrease in transition temperature away from the charge-neutral point is attributed to the Hartree effect, aligning with experimental observations. Finally, our GW calculations show that ring diagrams are irrelevant at integer filling factors but provide significant corrections at non-integer filling factors. These results suggest that ring diagrams might contribute greatly to the superconducting phase appearing at non-integer fillings.  Moreover, as we have shown here, the theory also allows for analytical insights into problems that were previously only addressed numerically. In future work we plan to extend this framework to non-chiral and non-flat cases examining the competition between the single-particle energy and the interaction-induced renormalization.  We also plan to incorporate on equal footing the effects of the Coulomb interaction, superconductivity, and external screening~\cite{Peng_PRB2024_Plasmon}.  With this, we anticipate to accurately address recent experiments~\cite{Stepanov_Nature2020_Untying, Xiaoxue_Science2021_BLG, Alexey_arxiv2024_Screening, JeanieLau_arxiv2024_Screening, Tutuc_arxiv2024_Screening} that explored how both the correlated insulators and superconductivity was modified when the Coulomb interaction was screened.  \\

\section{Acknowledgments}

It is a pleasure to thank Mohammed Al Ezzi, Christophe De Beule, Darryl Foo, and Shangjian Jin for their helpful comments and for collaboration on related projects.  We acknowledge the financial support from the Singapore National Research Foundation Investigator Award (NRF-NRFI06-2020-0003).

\bibliography{refs}

\newpage
\appendix
\setcounter{equation}{0}
\setcounter{figure}{0}
\setcounter{table}{0}
\setcounter{page}{1}
\makeatletter
\renewcommand{\theequation}{S\arabic{equation}}
\renewcommand{\thefigure}{S\arabic{figure}}
\renewcommand{\bibnumfmt}[1]{[S#1]}
\renewcommand*{\thepage}{A\arabic{page}}
\renewcommand{\thesection}{\Alph{section}}
\renewcommand{\thesubsection}{\arabic{subsection}}

\onecolumngrid

\section{Interacting Hamiltonian}
\label{appendix:full_coulomb}

The Hamiltonian of the \moire system can be written down as
\begin{equation}
    \mathcal{H}=\mathcal{H}_{0}+\mathcal{H}_{\mathrm{I}}.
\end{equation}
Here, $\mathcal{H}_{0}$ represents the single-particle Hamiltonian
\begin{equation}
\begin{aligned}
\mathcal{H}_{0}
&=\sum_{\tilde{\mathbf{k}},\tilde{\mathbf{k}}^{\prime}}\sum_{\alpha,\alpha^{\prime},\sigma}  \left[\mathcal{H}^{\sigma}(\tilde{\mathbf{k}},\tilde{\mathbf{k}}^{\prime})\right]_{\alpha\alpha^{\prime}} \hat{c}^{\dagger}_{\tilde{\mathbf{k}},\alpha,\sigma} \hat{c}_{\tilde{\mathbf{k}}^{\prime},\alpha^{\prime},\sigma}\\
&=\sum_{\mathbf{k}}\sum_{\mathbf{G},\mathbf{G}^{\prime}}\sum_{\alpha,\alpha^{\prime},\sigma}  \left[\mathcal{H}_{\mathbf{G}\mathbf{G}^{\prime}}^{\sigma}(\mathbf{k})\right]_{\alpha\alpha^{\prime}} \hat{c}^{\dagger}_{\mathbf{k},\mathbf{G},\alpha,\sigma} \hat{c}_{\mathbf{k},\mathbf{G}^{\prime},\alpha^{\prime},\sigma},\\
\end{aligned}
\end{equation}
where $\hat{c}^{\dagger}_{\mathbf{k},\alpha,\sigma}$ ($\hat{c}_{\mathbf{k},\alpha,\sigma}$) are the creation (annihilation) operators of momentum $\tilde{\mathbf{k}}$ in the Brillouin zone (BZ), with $\alpha$ representing the internal degree of freedom like sublattice and $\sigma$ denoting the quantum number like spin. $\mathcal{H}^{\sigma}(\tilde{\mathbf{k}},\tilde{\mathbf{k}}^{\prime})$ is the first-quantized momentum space Hamiltonian at quantum number $\sigma$ on the basis of the internal degree of freedom. In the second line, we use $\tilde{\mathbf{k}}=\mathbf{k}+\mathbf{G}$, assuming a \moire Brillouin zone (mBZ) with reciprocal lattice vector $\mathbf{G}$ and momentum $\mathbf{k}$ restricted to the mBZ. The corresponding first-quantized momentum space Hamiltonian $\mathcal{H}_{\mathbf{G}\mathbf{G}^{\prime}}^{\sigma}(\mathbf{k})$ is now periodic in the mBZ.

The interacting part of Hamiltonian $\mathcal{H}_{\mathrm{I}}$ takes the form
\begin{equation}
\begin{aligned}
\mathcal{H}_{\mathrm{I}}&=
\frac{1}{2}
\sum_{\tilde{\mathbf{k}},\tilde{\mathbf{k}}^{\prime},\tilde{\mathbf{q}}}
\sum_{\alpha,\alpha^\prime,\sigma,\sigma^{\prime}}
V_{\tilde{\mathbf{q}}}
\hat{c}_{\tilde{\mathbf{k}}+\tilde{\mathbf{q}},\alpha,\sigma}^{\dagger}
\hat{c}_{\tilde{\mathbf{k}}^{\prime}-\tilde{\mathbf{q}},\alpha^{\prime},\sigma^{\prime}}^{\dagger}
\hat{c}_{\tilde{\mathbf{k}}^{\prime},\alpha^{\prime},\alpha}
\hat{c}_{\tilde{\mathbf{k}},\alpha,\sigma},\\
\end{aligned}
\end{equation}
where $\tilde{\mathbf{k}},\tilde{\mathbf{k}}^{\prime},\tilde{\mathbf{q}}$ are restricted to the BZ, and $V_{\tilde{\mathbf{q}}}$ is the Fourier transform of the Coulomb interaction. By expanding the momentum as $\tilde{\mathbf{k}}=\mathbf{k}+\mathbf{G}$ and introducing the mBZ, the interacting Hamiltonian is reformulated as
\begin{equation}
\mathcal{H}_{\mathrm{I}}=\frac{1}{2}
\sum_{\mathbf{k},\mathbf{k}^{\prime},\mathbf{q}}
\sum_{\mathbf{G},\mathbf{G}^{\prime},\mathbf{G}^{\prime\prime}}
\sum_{\alpha,\alpha^\prime,\sigma,\sigma^{\prime}}
V_{\mathbf{q}+\mathbf{G}^{\prime\prime}}
\hat{c}_{\mathbf{k}+\mathbf{G}+\mathbf{q}+\mathbf{G}^{\prime\prime},\alpha,\sigma}^{\dagger}
\hat{c}_{\mathbf{k}^{\prime}+\mathbf{G}^{\prime}-\mathbf{q}-\mathbf{G}^{\prime\prime},\alpha^{\prime},\sigma^{\prime}}^{\dagger}
\hat{c}_{\mathbf{k}^{\prime}+\mathbf{G}^{\prime},\alpha^{\prime},\sigma^{\prime}}
\hat{c}_{\mathbf{k}+\mathbf{G},\alpha,\sigma}.
\end{equation}

Apply the basis transformation defined in Eq.~(\ref{eq:nibasis}), the interacting part of Hamiltonian can be recast as
\begin{equation}\label{eq:full_coulomb}
\begin{aligned}
\mathcal{H}_{\mathrm{I}}&=\frac{1}{2}
\sum_{\{n_i\}}
\sum_{\mathbf{k},\mathbf{k}^{\prime},\mathbf{q}}
\sum_{\mathbf{G},\mathbf{G}^{\prime},\mathbf{G}^{\prime\prime}}
\sum_{\alpha,\alpha^\prime,\sigma,\sigma^{\prime}}
V_{\mathbf{q}+\mathbf{G}^{\prime\prime}}
u^{*}_{n_1}(\mathbf{k}+\mathbf{q},\mathbf{G}+\mathbf{G}^{\prime\prime};\alpha,\sigma)
u^{*}_{n_2}(\mathbf{k}^{\prime}-\mathbf{q},\mathbf{G}^{\prime}-\mathbf{G}^{\prime\prime};\alpha^\prime,\sigma^\prime)\times\\
&\quad\quad
u_{n_3}(\mathbf{k}^{\prime},\mathbf{G}^{\prime};\alpha^\prime,\sigma^\prime)
u_{n_4}(\mathbf{k},\mathbf{G};\alpha,\sigma)
\hat{c}_{\mathbf{k}+\mathbf{q},n_1,\sigma}^{\dagger}
\hat{c}_{\mathbf{k}^{\prime}-\mathbf{q},n_2,\sigma^{\prime}}^{\dagger}
\hat{c}_{\mathbf{k}^{\prime},n_3,\sigma^{\prime}}
\hat{c}_{\mathbf{k},n_4,\sigma}\\
&=\frac{1}{2}
\sum_{\{n_i\}}
\sum_{\mathbf{k},\mathbf{k}^{\prime},\mathbf{q}}
\sum_{\mathbf{G}^{\prime\prime}}
\sum_{\sigma,\sigma^{\prime}}
V_{\mathbf{q}+\mathbf{G}^{\prime\prime}}
\left[\sum_{\alpha,\mathbf{G}}
u^{*}_{n_1}(\mathbf{k}+\mathbf{q},\mathbf{G}+\mathbf{G}^{\prime\prime};\alpha,\sigma)
u_{n_4}(\mathbf{k},\mathbf{G};\alpha,\sigma)
\right]\times\\
&\quad\quad
\left[\sum_{\alpha^{\prime},\mathbf{G}^{\prime}}
u^{*}_{n_2}(\mathbf{k}^{\prime}-\mathbf{q},\mathbf{G}^{\prime}-\mathbf{G}^{\prime\prime};\alpha^\prime,\sigma^\prime)
u_{n_3}(\mathbf{k}^{\prime},\mathbf{G}^{\prime};\alpha^\prime,\sigma^\prime)
\right]
\hat{c}_{\mathbf{k}+\mathbf{q},n_1,\sigma}^{\dagger}
\hat{c}_{\mathbf{k}^{\prime}-\mathbf{q},n_2,\sigma^{\prime}}^{\dagger}
\hat{c}_{\mathbf{k}^{\prime},n_3,\sigma^{\prime}}
\hat{c}_{\mathbf{k},n_4,\sigma}\\
&=\frac{1}{2}
\sum_{\{n_i\}}
\sum_{\mathbf{k},\mathbf{k}^{\prime},\mathbf{q}}
\sum_{\mathbf{G}^{\prime\prime}}
\sum_{\sigma,\sigma^{\prime}}
V_{\mathbf{q}+\mathbf{G}^{\prime\prime}}
\left[\Lambda^{*}_{\mathbf{k},\mathbf{q}+\mathbf{G}^{\prime\prime}}\right]^{\sigma}_{n_4n_1}
\left[\Lambda^{*}_{\mathbf{k}^{\prime},-\mathbf{q}-\mathbf{G}^{\prime\prime}}\right]^{\sigma^\prime}_{n_3n_2}
\hat{c}_{\mathbf{k}+\mathbf{q},n_1,\sigma}^{\dagger}
\hat{c}_{\mathbf{k}^{\prime}-\mathbf{q},n_2,\sigma^{\prime}}^{\dagger}
\hat{c}_{\mathbf{k}^{\prime},n_3,\sigma^{\prime}}
\hat{c}_{\mathbf{k},n_4,\sigma},\\
\end{aligned}
\end{equation}
where we have introduced the form factor
\begin{equation}
\left[\Lambda_{\mathbf{k},\mathbf{q}+\mathbf{G}}\right]^{\sigma}_{m n}=\sum_{\alpha,\mathbf{G}^{\prime}}
u^*_{m} \left(\mathbf{k},\mathbf{G}^{\prime} ; \alpha,\sigma\right) 
u_{n} \left(\mathbf{k},\mathbf{G}^{\prime}+\mathbf{q}+\mathbf{G} ; \alpha,\sigma\right).
\end{equation}
The form factor has the properties
\begin{equation}\label{eq:FF1}
\hat{\Lambda}_{\mathbf{k}+\mathbf{q},-\mathbf{q}-\mathbf{G}}=\hat{\Lambda}^{\dagger}_{\mathbf{k},\mathbf{q}+\mathbf{G}},
\end{equation}
from the definition and in particular, in the case of $\mathbf{q}=0$, we have
\begin{equation}\label{eq:FF2}
\hat{\Lambda}_{\mathbf{k},-\mathbf{G}}=\hat{\Lambda}^{\dagger}_{\mathbf{k},\mathbf{G}}.
\end{equation}
Equation.~(\ref{eq:full_coulomb}) can be rewrite as
\begin{equation}
\begin{aligned}
\mathcal{H}_{\mathrm{I}}&=\frac{1}{2}
\sum_{\{n_i\}}
\sum_{\mathbf{k},\mathbf{k}^{\prime},\mathbf{q}}
\sum_{\sigma,\sigma^{\prime}}
\left[
\sum_{\mathbf{G}}
V_{\mathbf{q}+\mathbf{G}}
\left[\Lambda^{*}_{\mathbf{k},\mathbf{q}+\mathbf{G}}\right]^{\sigma}_{n_4n_1}
\left[\Lambda^{*}_{\mathbf{k}^{\prime},-\mathbf{q}-\mathbf{G}}\right]^{\sigma^\prime}_{n_3n_2}
\right]
\hat{c}_{\mathbf{k}+\mathbf{q},n_1,\sigma}^{\dagger}
\hat{c}_{\mathbf{k}^{\prime}-\mathbf{q},n_2,\sigma^{\prime}}^{\dagger}
\hat{c}_{\mathbf{k}^{\prime},n_3,\sigma^{\prime}}
\hat{c}_{\mathbf{k},n_4,\sigma}\\
&=\frac{1}{2}
\sum_{\{n_i\}}
\sum_{\mathbf{k},\mathbf{k}^{\prime},\mathbf{q}}
\sum_{\sigma,\sigma^{\prime}}
V^{\sigma\sigma^{\prime}}_{\mathbf{q},\{n_i\}}
\hat{c}_{\mathbf{k}+\mathbf{q},n_1,\sigma}^{\dagger}
\hat{c}_{\mathbf{k}^{\prime}-\mathbf{q},n_2,\sigma^{\prime}}^{\dagger}
\hat{c}_{\mathbf{k}^{\prime},n_3,\sigma^{\prime}}
\hat{c}_{\mathbf{k},n_4,\sigma},\\
\end{aligned}
\end{equation}
where we have redefined the Coulomb interaction as
\begin{equation}
V^{\sigma\sigma^{\prime}}_{\mathbf{q},\{n_i\}}=
\sum_{\mathbf{G}}
V_{\mathbf{q}+\mathbf{G}}
\left[\Lambda^{*}_{\mathbf{k},\mathbf{q}+\mathbf{G}}\right]^{\sigma}_{n_4n_1}
\left[\Lambda^{*}_{\mathbf{k}^{\prime},-\mathbf{q}-\mathbf{G}}\right]^{\sigma^\prime}_{n_3n_2}.
\end{equation}

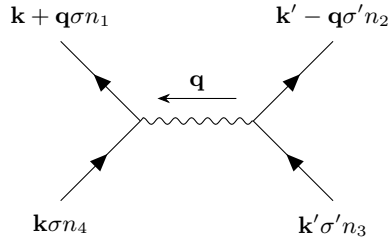
\begin{figure}
    \centering
    \begin{tikzpicture}
        \begin{feynman}
            \vertex (a1);
            \vertex[right=1.5cm of a1] (a2);
            \vertex[above left=1.5cm of a1] (a3);
            \vertex[below left=1.5cm of a1] (a4);
            \vertex[above right=1.5cm of a2] (a5);
            \vertex[below right=1.5cm of a2] (a6);
            \diagram* {
            (a4) -- [fermion] (a1) -- [fermion] (a3),
            (a2) -- [photon, momentum'={$\mathbf q$}] (a1),
            (a6) -- [fermion] (a2) -- [fermion] (a5),
            };
            \vertex [below=0.2em of a4] {$\mathbf k\sigma n_{4}$};
            \vertex [above=0.2em of a3] {$\mathbf k+\mathbf q \sigma n_{1}$};
            \vertex [above=0.2em of a5] {$\mathbf k^{\prime}-\mathbf{q} \sigma^{\prime} n_{2}$};
            \vertex [below=0.2em of a6] {$\mathbf k^{\prime} \sigma^{\prime} n_{3}$};
        \end{feynman}
    \end{tikzpicture}
    \caption{Schematic representation of the Coulomb interaction. The two particles initially have two vectors $\mathbf{k}$ and $\mathbf{k}^{\prime}$ in bands $n_4$ and $n_3$. The interaction is viewed as a collision in which one particle transfers momentum $\mathbf{q}$ to the other. After the collision, we obtain the vector $\mathbf{k}+\mathbf{q}$ and $\mathbf{k}^{\prime}-\mathbf{q}$ in bands $n_1$ and $n_2$.}
    \label{fig:diagrams_coulomb}
\end{figure}

In the new basis, the interacting Hamiltonian can be seen as the usual Coulomb interaction but with modified strength in different band indices, determined by the form factor as shown in Fig.~(\ref{fig:diagrams_coulomb}). It is important to note that the form factor plays a significant role in this system, especially near the magic angle where the single-particle energy is much smaller than the Coulomb interaction. Although the exact form of the form factor is unknown, its symmetry-constrained structure provides substantial information about the ground state.

\section{Diagram Expansion}

Based on the Hamiltonian, we now apply the Feynman technique to obtain the important diagrams. We will mainly focus on three types of diagrams: (i) Hartree diagram, (ii) Fock diagram, and (iii) Ring diagrams. Those diagrams has been shown their importance in conventional electron gas systems. 

\subsection{Hartree diagram}

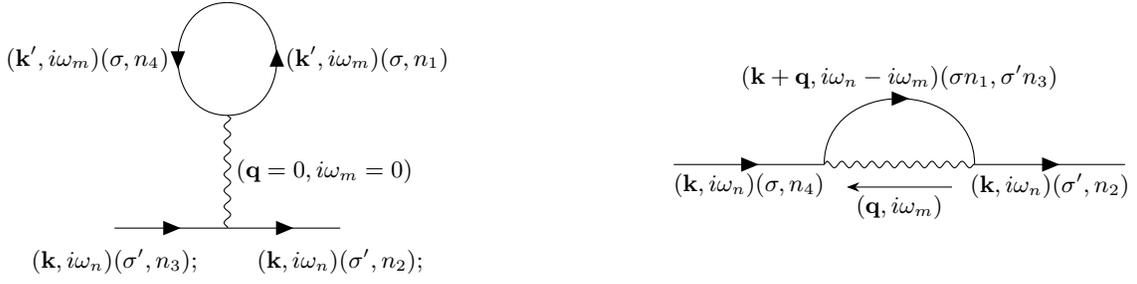
\begin{figure}[t]
\centering
\begin{minipage}{0.5\linewidth}
    \centering
    \begin{tikzpicture}
    \begin{feynman}
        \vertex (a1);
        \vertex[right=of a1] (a2);
        \vertex[right=of a2] (a3);
        \vertex[above=of a2] (a4);
        \vertex[above=of a4] (a5);
        \diagram* {
        (a1)--[fermion](a2)--[fermion](a3),
        (a2)--[photon, edge label'={$(\mathbf{q}=0,i\omega_m=0)$}](a4),
        (a4)--[fermion, edge label'={$(\mathbf{k}^{\prime},i\omega_{m})(\sigma,n_{1})$}, half right](a5)--[fermion, edge label'={$(\mathbf{k}^{\prime},i\omega_{m})(\sigma,n_{4})$}, half right](a4)
        };
        \vertex [below=0.5em of a1] {$(\mathbf{k},i\omega_{n})(\sigma^{\prime},n_{3});$};
        \vertex [below=0.5em of a3] {$(\mathbf{k},i\omega_{n})(\sigma^{\prime},n_{2});$};
    \end{feynman}
    \end{tikzpicture}
\end{minipage}%
\begin{minipage}{0.5\linewidth}
    \centering
    \begin{tikzpicture}
    \begin{feynman}
        \vertex (a1);
        \vertex[right=2cm of a1] (a2);
        \vertex[right=2cm of a2] (a3);
        \vertex[right=2cm of a3] (a4);
        \diagram* {
        (a1)--[fermion, edge label'={$(\mathbf{k},i\omega_{n}) (\sigma ,n_{4})$}](a2),
        (a3)--[photon, momentum={$(\mathbf q,i\omega_m$)}](a2),
        (a3)--[fermion, edge label'={$(\mathbf{k},i\omega_{n}) (\sigma^{\prime},n_{2})$}](a4),
        (a2)--[fermion, edge label={$(\mathbf{k}+\mathbf{q},i\omega_{n}-i\omega_{m}) (\sigma n_{1},\sigma^{\prime}n_{3})$}, half left](a3),
        };
    \end{feynman}
    \end{tikzpicture}
\end{minipage}
\caption{Diagrammatic representation of Hartree (left) and Fock (right) diagrams.}
\label{fig:hartreefock}
\end{figure}

The Hartree diagrams is shown in  Fig.~(\ref{fig:hartreefock}). According to the Feynman rule \cite{Jishi_2013}, Hartree self-energy could be written down as
\begin{equation}\label{eq:hartree_full}
\begin{aligned}
\Sigma_{n_3n_2}^{\sigma^{\prime}\sigma^{\prime}}(\mathbf{k},i\omega_n)
&= \frac{1}{\beta}\sum_{m}
\sum_{n_1,n_4,\sigma}
\sum_{\mathbf{k}^{\prime},\mathbf{G}}
V_{\mathbf{G}}
\left[\Lambda^{*}_{\mathbf{k}^{\prime},\mathbf{G}}\right]^{\sigma}_{n_4n_1}
\left[\Lambda^{*}_{\mathbf{k},-\mathbf{G}}\right]^{\sigma^\prime}_{n_3n_2}
\left[G(\mathbf{k}^{\prime},i\omega_m)\right]^{\sigma\sigma}_{n_1n_4}\\
&= \sum_{\mathbf{G}}
V_{\mathbf{G}}
\left[\Lambda_{\mathbf{k},\mathbf{G}}\right]^{\sigma^\prime}_{n_2n_3}
\sum_{\mathbf{k}^{\prime}}
\frac{1}{\beta}\sum_{m}
\sum_{n_1,n_4,\sigma}
\left[\Lambda^{*}_{\mathbf{k}^{\prime},\mathbf{G}}\right]^{\sigma}_{n_4n_1}
\left[G(\mathbf{k}^{\prime},i\omega_m)\right]^{\sigma\sigma}_{n_1n_4},\\
\end{aligned}
\end{equation}
where we have used the relationship Eq.~(\ref{eq:FF2}). Equation.~(\ref{eq:hartree_full}) can be written down using the matrix representation
\begin{equation}
\hat{\Sigma}_{\mathrm{H}}(\mathbf{k},i\omega_n) = \sum_{\mathbf{G}}
V_{\mathbf{G}}
\hat{\Lambda}^{\mathrm{T}}_{\mathbf{k},\mathbf{G}}
\sum_{\mathbf{k}^{\prime}}
\frac{1}{\beta}\sum_{m}
\mathrm{Tr}\left[\hat{\Lambda}^{*}_{\mathbf{k}^{\prime},\mathbf{G}}\hat{G}(\mathbf{k}^{\prime},i\omega_m)\right].
\end{equation}

\begin{figure}
    \centering
    \begin{tikzpicture}
        \begin{feynman}
            \vertex (a1);
            \vertex[right=2cm of a1] (a2);
            \vertex[right=6cm of a2] (a3);
            \vertex[right=2cm of a3] (a4);
            \vertex[above=of a2] (a5);
            \vertex[above=of a3] (a6);
            \diagram* {
            (a1)--[fermion, edge label'={$(\mathbf{k},i\omega_n)(\sigma_{2} n_{3})$}](a2)--[fermion, edge label'={$(\mathbf{k}-\mathbf{q},i\omega_n-i\omega_m)(\sigma_{2} n_{2},\sigma_{3} n_{8})$}](a3)--[fermion, edge label'={($\mathbf{k},i\omega_n) (\sigma_3 n_{5})$}](a4),
            (a2)--[photon, momentum={$(\mathbf q,i\omega_m)$}](a5),
            (a6)--[photon, momentum={$(\mathbf q,i\omega_m)$}](a3),
            (a5)--[fermion, edge label={$(\mathbf{k}^{\prime}+\mathbf{q},i\omega_o+i\omega_m)(\sigma_1 n_{1},\sigma_4 n_{7})$}, half left, looseness=0.3](a6)--[fermion, edge label={$(\mathbf{k}^{\prime},i\omega_o) (\sigma_1 n_{4}, \sigma_4 n_{6})$}, half left, looseness=0.3](a5)
            };
        \end{feynman}
        \end{tikzpicture}
    \caption{Lowest order ring diagram.}
    \label{fig:ring_diagrams}
\end{figure}
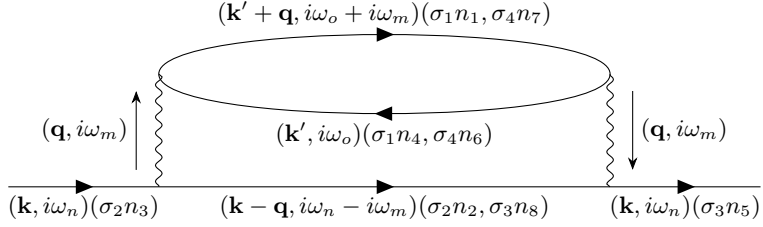

\subsection{Fock diagram}

Similarly, the Fock diagram can be expanded as
\begin{equation}\label{eq:fock_full}
\begin{aligned}
\Sigma^{\sigma\sigma^{\prime}}_{n_4n_2}(\mathbf{k},i\omega_n)
&=-\frac{1}{\beta}\sum_{m}
\sum_{n_1,n_3}
\sum_{\mathbf{q},\mathbf{G}}
V_{\mathbf{q}+\mathbf{G}}
\left[\Lambda^{*}_{\mathbf{k},\mathbf{q}+\mathbf{G}}\right]^{\sigma}_{n_4n_1}
\left[\Lambda^{*}_{\mathbf{k}^{\prime},-\mathbf{q}-\mathbf{G}}\right]^{\sigma^\prime}_{n_3n_2}
G^{\sigma\sigma^{\prime}}_{n_1n_3}(\mathbf{k}+\mathbf{q},i\omega_n-i\omega_m)\delta_{\mathbf{k}^{\prime},\mathbf{k}+\mathbf{q}}
\\
&=-\frac{1}{\beta}\sum_{m}
\sum_{n_1,n_3}
\sum_{\mathbf{q},\mathbf{G}}
V_{\mathbf{q}+\mathbf{G}}
\left[\Lambda^{*}_{\mathbf{k},\mathbf{q}+\mathbf{G}}\right]^{\sigma}_{n_4n_1}
\left[\Lambda^{*}_{\mathbf{k}+\mathbf{q},-\mathbf{q}-\mathbf{G}}\right]^{\sigma^\prime}_{n_3n_2}
G^{\sigma\sigma^{\prime}}_{n_1n_3}(\mathbf{k}+\mathbf{q},i\omega_n-i\omega_m)
\\
&=-\frac{1}{\beta}\sum_{m}
\sum_{n_1,n_3}
\sum_{\mathbf{q},\mathbf{G}}
V_{\mathbf{q}+\mathbf{G}}
\left[\Lambda^{*}_{\mathbf{k},\mathbf{q}+\mathbf{G}}\right]^{\sigma}_{n_4n_1}
G^{\sigma\sigma^{\prime}}_{n_1n_3}(\mathbf{k}+\mathbf{q},i\omega_n-i\omega_m)
\left[\Lambda_{\mathbf{k},\mathbf{q}+\mathbf{G}}\right]^{\sigma^\prime}_{n_2n_3},
\end{aligned}
\end{equation}
where we have used the relationship Eq.~(\ref{eq:FF1}), and the corresponding matrix representation is
\begin{equation}
\hat{\Sigma}_{\mathrm{F}}(\mathbf{k},i\omega_n)=-\frac{1}{\beta}\sum_{m}
\sum_{\mathbf{q},\mathbf{G}}
V_{\mathbf{q}+\mathbf{G}}
\hat{\Lambda}^{*}_{\mathbf{k},\mathbf{q}+\mathbf{G}}
\hat{G}(\mathbf{k}+\mathbf{q},i\omega_n-i\omega_m)
\hat{\Lambda}^{\mathrm{T}}_{\mathbf{k},\mathbf{q}+\mathbf{G}}.
\end{equation}

\subsection{Ring diagrams}
\label{appendix:ring_diagrams}

Next, we will derive the expression for ring diagrams within random phase approximation. We start with the first order ring diagrams, as shown in Fig.~\eqref{fig:ring_diagrams}. The corresponding self-energy can be evaluated as
\begin{equation}
\begin{aligned}
&\Sigma_{n_3n_5}^{\sigma_{2}\sigma_{3}}(\mathbf{k},i\omega_n)
= -\left(-\frac{1}{\beta}\right)^2\sum_{o,m}
\sum_{\{n_i\}} \sum_{\{\sigma_i\}}
\sum_{\mathbf{k}^{\prime},\mathbf{q},\mathbf{G}_1,\mathbf{G}_2}
V_{\mathbf{q}+\mathbf{G}_1}V_{\mathbf{q}+\mathbf{G}_2}
\left[\Lambda^{*}_{\mathbf{k}^{\prime},\mathbf{q}+\mathbf{G}_1}\right]^{\sigma_1}_{n_4n_1}
\left[\Lambda^{*}_{\mathbf{k},-\mathbf{q}-\mathbf{G}_1}\right]^{\sigma_2}_{n_3n_2}\times\\
&\left[\Lambda^{*}_{\mathbf{k}-\mathbf{q},\mathbf{q}+\mathbf{G}_2}\right]^{\sigma_3}_{n_8n_5}
\left[\Lambda^{*}_{\mathbf{k}^{\prime}+\mathbf{q},-\mathbf{q}-\mathbf{G}_2}\right]^{\sigma_4}_{n_7n_6}
\left[G(\mathbf{k}-\mathbf{q},i\omega_n-i\omega_m)\right]^{\sigma_2\sigma_3}_{n_2n_8}
\left[G(\mathbf{k}^{\prime},i\omega_o)\right]^{\sigma_4\sigma_1}_{n_6n_4}
\left[G(\mathbf{k}^{\prime}+\mathbf{q},i\omega_o+i\omega_m)\right]^{\sigma_1\sigma_4}_{n_1n_7}\\
&= -\frac{1}{\beta^2}\sum_{o,m}
\sum_{\{n_i\}} \sum_{\{\sigma_i\}}
\sum_{\mathbf{k}^{\prime},\mathbf{q},\mathbf{G}_1,\mathbf{G}_2}
V_{\mathbf{q}+\mathbf{G}_1}V_{\mathbf{q}+\mathbf{G}_2}
\left[\Lambda^{*}_{\mathbf{k}^{\prime},\mathbf{q}+\mathbf{G}_1}\right]^{\sigma_1}_{n_4n_1}
\left[\Lambda^{*}_{\mathbf{k},-\mathbf{q}-\mathbf{G}_1}\right]^{\sigma_2}_{n_3n_2}\times\\
&\left[\Lambda_{\mathbf{k},-\mathbf{q}-\mathbf{G}_2}\right]^{\sigma_3}_{n_5n_8}
\left[\Lambda_{\mathbf{k}^{\prime},\mathbf{q}+\mathbf{G}_2}\right]^{\sigma_4}_{n_6n_7}
\left[G(\mathbf{k}-\mathbf{q},i\omega_n-i\omega_m)\right]^{\sigma_2\sigma_3}_{n_2n_8}
\left[G(\mathbf{k}^{\prime},i\omega_o)\right]^{\sigma_4\sigma_1}_{n_6n_4}
\left[G(\mathbf{k}^{\prime}+\mathbf{q},i\omega_o+i\omega_m)\right]^{\sigma_1\sigma_4}_{n_1n_7}\\
&= -\frac{1}{\beta}\sum_{m}
\sum_{\substack{n_2,n_3\\n_5,n_8\\\sigma_2\sigma_3}}
\sum_{\mathbf{q},\mathbf{G}_1,\mathbf{G}_2}
V_{\mathbf{q}+\mathbf{G}_1}V_{\mathbf{q}+\mathbf{G}_2}
\left[\Lambda^{*}_{\mathbf{k},-\mathbf{q}-\mathbf{G}_1}\right]^{\sigma_2}_{n_3n_2}
\left[\Lambda_{\mathbf{k},-\mathbf{q}-\mathbf{G}_2}\right]^{\sigma_3}_{n_5n_8}
\left[G(\mathbf{k}-\mathbf{q},i\omega_n-i\omega_m)\right]^{\sigma_2\sigma_3}_{n_2n_8}\times\\
&\left[\frac{1}{\beta}\sum_{o}\sum_{\substack{n_1,n_4\\n_6,n_7\\\sigma_1\sigma_4}}
\sum_{\mathbf{k}^{\prime}}
\left[\Lambda^{*}_{\mathbf{k}^{\prime},\mathbf{q}+\mathbf{G}_1}\right]^{\sigma_1}_{n_4n_1}
\left[\Lambda_{\mathbf{k}^{\prime},\mathbf{q}+\mathbf{G}_2}\right]^{\sigma_4}_{n_6n_7}
\left[G(\mathbf{k}^{\prime},i\omega_o)\right]^{\sigma_4\sigma_1}_{n_6n_4}
\left[G(\mathbf{k}^{\prime}+\mathbf{q},i\omega_o+i\omega_m)\right]^{\sigma_1\sigma_4}_{n_1n_7}
\right]\\
&= -\frac{1}{\beta}\sum_{m}
\sum_{\substack{n_2,n_3\\n_5,n_8\\\sigma_2\sigma_3}}
\sum_{\mathbf{q},\mathbf{G}_1,\mathbf{G}_2}
V_{\mathbf{q}+\mathbf{G}_1}V_{\mathbf{q}+\mathbf{G}_2}
\left[\Lambda^{*}_{\mathbf{k},-\mathbf{q}-\mathbf{G}_1}\right]^{\sigma_2}_{n_3n_2}
\left[\Lambda_{\mathbf{k},-\mathbf{q}-\mathbf{G}_2}\right]^{\sigma_3}_{n_5n_8}
\left[G(\mathbf{k}-\mathbf{q},i\omega_n-i\omega_m)\right]^{\sigma_2\sigma_3}_{n_2n_8}
\left[\Pi(\mathbf{q},i\omega_m)\right]_{\mathbf{G}_1\mathbf{G}_2}\\
&=
-\frac{1}{\beta}\sum_{m}
\sum_{\mathbf{q},\mathbf{G}_1,\mathbf{G}_2}
V_{\mathbf{q}+\mathbf{G}_1}
\left[\Pi(\mathbf{q},i\omega_m)\right]_{\mathbf{G}_1\mathbf{G}_2}
V_{\mathbf{q}+\mathbf{G}_2}
\sum_{\substack{n_2,n_3\\n_5,n_8\\\sigma_2\sigma_3}}
\left[\Lambda^{*}_{\mathbf{k},-\mathbf{q}-\mathbf{G}_1}\right]^{\sigma_2}_{n_3n_2}
\left[G(\mathbf{k}-\mathbf{q},i\omega_n-i\omega_m)\right]^{\sigma_2\sigma_3}_{n_2n_8}
\left[\Lambda_{\mathbf{k},-\mathbf{q}-\mathbf{G}_2}\right]^{\sigma_3}_{n_5n_8}\\
&=-\frac{1}{\beta}\sum_{m}
\sum_{\mathbf{q},\mathbf{G}_1,\mathbf{G}_2}
V_{\mathbf{q}+\mathbf{G}_1}
\left[\Pi(\mathbf{q},i\omega_m)\right]_{\mathbf{G}_1\mathbf{G}_2}
V_{\mathbf{q}+\mathbf{G}_2}
\left[
\hat{\Lambda}^*_{\mathbf{k},-\mathbf{q}-\mathbf{G}_1}
\hat{G}(\mathbf{k}-\mathbf{q},i\omega_n-i\omega_m)
\hat{\Lambda}^{\mathrm{T}}_{\mathbf{k},-\mathbf{q}-\mathbf{G}_2}
\right],\\
\end{aligned}
\end{equation}
where we have introduced the polarizability defined as
\begin{equation}
\begin{aligned}
\left[\Pi(\mathbf{q},i\omega_m)\right]_{\mathbf{G}_1\mathbf{G}_2}
&=\frac{1}{\beta}\sum_{o}\sum_{\substack{n_1,n_4\\n_6,n_7\\\sigma_1\sigma_4}}
\sum_{\mathbf{k}^{\prime}}
\left[\Lambda^{*}_{\mathbf{k}^{\prime},\mathbf{q}+\mathbf{G}_1}\right]^{\sigma_1}_{n_4n_1}
\left[\Lambda_{\mathbf{k}^{\prime},\mathbf{q}+\mathbf{G}_2}\right]^{\sigma_4}_{n_6n_7}
\left[G(\mathbf{k}^{\prime},i\omega_o)\right]^{\sigma_4\sigma_1}_{n_6n_4}
\left[G(\mathbf{k}^{\prime}+\mathbf{q},i\omega_o+i\omega_m)\right]^{\sigma_1\sigma_4}_{n_1n_7}\\
&=\frac{1}{\beta}\sum_{o}\sum_{\substack{n_1,n_4\\n_6,n_7\\\sigma_1\sigma_4}}
\sum_{\mathbf{k}^{\prime}}
\left[G(\mathbf{k}^{\prime},i\omega_o)\right]^{\sigma_4\sigma_1}_{n_6n_4}
\left[\Lambda^{*}_{\mathbf{k}^{\prime},\mathbf{q}+\mathbf{G}_1}\right]^{\sigma_1}_{n_4n_1}
\left[G(\mathbf{k}^{\prime}+\mathbf{q},i\omega_o+i\omega_m)\right]^{\sigma_1\sigma_4}_{n_1n_7}
\left[\Lambda_{\mathbf{k}^{\prime},\mathbf{q}+\mathbf{G}_2}\right]^{\sigma_4}_{n_6n_7}\\
&=\frac{1}{\beta}\sum_{o}
\sum_{\mathbf{k}^{\prime}}
\mathrm{Tr}\left[
\hat{G}(\mathbf{k}^{\prime},i\omega_o)
\hat{\Lambda}^*_{\mathbf{k}^{\prime},\mathbf{q}+\mathbf{G}_1}
\hat{G}(\mathbf{k}^{\prime}+\mathbf{q},i\omega_o+i\omega_m)
\hat{\Lambda}^{\mathrm{T}}_{\mathbf{k}^{\prime},\mathbf{q}+\mathbf{G}_2}\right].\\
\end{aligned}
\end{equation}

Similarly, for the next-order diagram we have
\begin{equation}
\begin{aligned}
\Sigma_{n_3n_5}^{\sigma_{2}\sigma_{3}}(\mathbf{k},i\omega_n)
&=
-\frac{1}{\beta}\sum_{m}
\sum_{\mathbf{q},\mathbf{G}_1,\mathbf{G}_2,\mathbf{G}_3}
V_{\mathbf{q}+\mathbf{G}_1}
\left[\Pi(\mathbf{q},i\omega_m)\right]_{\mathbf{G}_1\mathbf{G}_3}
V_{\mathbf{q}+\mathbf{G}_3}
\left[\Pi(\mathbf{q},i\omega_m)\right]_{\mathbf{G}_3\mathbf{G}_2}
V_{\mathbf{q}+\mathbf{G}_2}\times\\
&\left[
\hat{\Lambda}^*_{\mathbf{k},-\mathbf{q}-\mathbf{G}_1}
\hat{G}(\mathbf{k}-\mathbf{q},i\omega_n-i\omega_m)
\hat{\Lambda}^{\mathrm{T}}_{\mathbf{k},-\mathbf{q}-\mathbf{G}_2}
\right].\\
\end{aligned}
\end{equation}

Therefore, up to infinite order,
\begin{equation}
\begin{aligned}
\Sigma_{n_3n_5}^{\sigma_{2}\sigma_{3}}(\mathbf{k},i\omega_n)
&=
-\frac{1}{\beta}\sum_{m}
\sum_{\mathbf{q},\mathbf{G}_1,\mathbf{G}_2,\mathbf{G}_3}
V_{\mathbf{q}+\mathbf{G}_1}
\left[\Pi_{\mathrm{RPA}}(\mathbf{q},i\omega_m)\right]_{\mathbf{G}_1\mathbf{G}_2}
V_{\mathbf{q}+\mathbf{G}_2}\left[
\hat{\Lambda}^*_{\mathbf{k},-\mathbf{q}-\mathbf{G}_1}
\hat{G}(\mathbf{k}-\mathbf{q},i\omega_n-i\omega_m)
\hat{\Lambda}^{\mathrm{T}}_{\mathbf{k},-\mathbf{q}-\mathbf{G}_2}
\right].\\
\end{aligned}
\end{equation}
where
\begin{equation}
\begin{aligned}
\left[\Pi_{\mathrm{RPA}}(\mathbf{q},i\omega_m)\right]_{\mathbf{G}_1\mathbf{G}_2}
&=
\left[\Pi(\mathbf{q},i\omega_m)\right]_{\mathbf{G}_1\mathbf{G}_2}
+\sum_{\mathbf{G}_3}
\left[\Pi(\mathbf{q},i\omega_m)\right]_{\mathbf{G}_1\mathbf{G}_3}
V_{\mathbf{q}+\mathbf{G}_3}
\left[\Pi(\mathbf{q},i\omega_m)\right]_{\mathbf{G}_3\mathbf{G}_2}+\cdots.
\\
\end{aligned}
\end{equation}

Using the matrix representation we have
\begin{equation}
\begin{aligned}
\hat{\Pi}_{\mathrm{RPA}}(\mathbf{q},i\omega_m)
&=\hat{\Pi}(\mathbf{q},i\omega_m)
+\hat{\Pi}(\mathbf{q},i\omega_m)\hat{V}_{\mathbf{q}}\hat{\Pi}(\mathbf{q},i\omega_m)+\cdots\\
&=\hat{\Pi}(\mathbf{q},i\omega_m)+\hat{\Pi}(\mathbf{q},i\omega_m)\hat{V}_{\mathbf{q}}\hat{\Pi}_{\mathrm{RPA}}(\mathbf{q},i\omega_m),\\
\\
\end{aligned}
\end{equation}
with the solution
\begin{equation}
\begin{aligned}
\hat{\Pi}_{\mathrm{RPA}}(\mathbf{q},i\omega_m)
&=\left[\hat{\mathbb{I}}-\hat{V}_{\mathbf{q}}\hat{\Pi}(\mathbf{q},i\omega_m)\right]^{-1}\hat{\Pi}(\mathbf{q},i\omega_m).\\
\end{aligned}
\end{equation}

The screened Coulomb interaction is given by
\begin{equation}
\begin{aligned}
\hat{V}(\mathbf{q},i\omega_m)
&=\hat{V}_{\mathbf{q}}+\hat{V}_{\mathbf{q}}\hat{\Pi}_{\mathrm{RPA}}(\mathbf{q},i\omega_m)\hat{V}_{\mathbf{q}}\\
&=\hat{V}_{\mathbf{q}}\left[\hat{\mathbb{I}}+\left[\hat{\mathbb{I}}-\hat{V}_{\mathbf{q}}\hat{\Pi}(\mathbf{q},i\omega_m)\right]^{-1}\hat{\Pi}(\mathbf{q},i\omega_m)\hat{V}_{\mathbf{q}}\right]\\
&=\hat{V}_{\mathbf{q}}\left[\left[\hat{\mathbb{I}}-\hat{V}_{\mathbf{q}}\hat{\Pi}(\mathbf{q},i\omega_m)\right]^{-1}\left[\hat{\mathbb{I}}-\hat{V}_{\mathbf{q}}\hat{\Pi}(\mathbf{q},i\omega_m)\right]+\left[\hat{\mathbb{I}}-\hat{V}_{\mathbf{q}}\hat{\Pi}(\mathbf{q},i\omega_m)\right]^{-1}\hat{\Pi}(\mathbf{q},i\omega_m)\hat{V}_{\mathbf{q}}\right]\\
&=\hat{V}_{\mathbf{q}}\left[\hat{\mathbb{I}}-\hat{V}_{\mathbf{q}}\hat{\Pi}(\mathbf{q},i\omega_m)\right]^{-1}\left[\hat{\mathbb{I}}-\hat{V}_{\mathbf{q}}\hat{\Pi}(\mathbf{q},i\omega_m)+\hat{\Pi}(\mathbf{q},i\omega_m)\hat{V}_{\mathbf{q}}\right]\\
&=\hat{V}_{\mathbf{q}}\left[\hat{\mathbb{I}}-\hat{V}_{\mathbf{q}}\hat{\Pi}(\mathbf{q},i\omega_m)\right]^{-1}.\\
\end{aligned}
\end{equation}

The ring diagram can be expressed as
\begin{equation}
\begin{aligned}
\hat{\Sigma}(\mathbf{k},i\omega_n)
&=-\frac{1}{\beta}\sum_{m}
\sum_{\mathbf{q},\mathbf{G}_1,\mathbf{G}_2}
\hat{V}(\mathbf{q},i\omega_m)
\left[
\hat{\Lambda}^*_{\mathbf{k},-\mathbf{q}-\mathbf{G}_1}
\hat{G}(\mathbf{k}-\mathbf{q},i\omega_n-i\omega_m)
\hat{\Lambda}^{\mathrm{T}}_{\mathbf{k},-\mathbf{q}-\mathbf{G}_2}
\right].\\
\end{aligned}
\end{equation}

To match the format with the Fock diagrams, we flip the sign of momentum $\mathbf{q}$ and $\mathbf{G}$,
\begin{equation}
\begin{aligned}
\hat{\Sigma}(\mathbf{k},i\omega_n)
&=-\frac{1}{\beta}\sum_{m}
\sum_{\mathbf{q},\mathbf{G}_1,\mathbf{G}_2}
\hat{V}(-\mathbf{q},i\omega_m)
\left[
\hat{\Lambda}^*_{\mathbf{k},\mathbf{q}+\mathbf{G}_1}
\hat{G}(\mathbf{k}+\mathbf{q},i\omega_n-i\omega_m)
\hat{\Lambda}^{\mathrm{T}}_{\mathbf{k},\mathbf{q}+\mathbf{G}_2}
\right],\\
\end{aligned}
\end{equation}
where
\begin{equation}
\begin{aligned}
\left[\Pi(-\mathbf{q},i\omega_m)\right]_{-\mathbf{G}_1-\mathbf{G}_2}
&=\frac{1}{\beta}\sum_{o}
\sum_{\mathbf{k}}
\mathrm{Tr}\left[
\hat{G}(\mathbf{k},i\omega_o)
\hat{\Lambda}^*_{\mathbf{k},-\mathbf{q}-\mathbf{G}_1}
\hat{G}(\mathbf{k}-\mathbf{q},i\omega_o+i\omega_m)
\hat{\Lambda}^{\mathrm{T}}_{\mathbf{k},-\mathbf{q}-\mathbf{G}_2}\right].\\
\end{aligned}
\end{equation}

\section{Conditions for Self-Consistent Solutions in the Chiral-Flat Limit}
\label{appendix:conditions}

In this section, we demonstrate that, in the chiral-flat limit, a sufficient condition for self-consistent Hartree-Fock solutions is that the order parameter $\hat Q$ commutes with the form factor $\hat\Lambda$,  i.e., $[\hat{Q}, \hat{\Lambda}] = 0$, in the spinless, valleyless model. Given the simplified model, which contains only two energy bands, we focus on the non-trivial case of half-band filling, corresponding to the only non-trivial integer filling number. At half filling, the Hartree diagrams vanish and therefore do not contribute to the competition between phases. As a result, we focus exclusively on the Fock diagrams.

According to Eq.~\eqref{eq:Fock_Q}, the Fock diagrams can be expressed as
\begin{equation}
\begin{aligned}
\hat{\Sigma}_{\mathrm{F}}(\mathbf{k}) 
&= -\frac{1}{2} \sum_{\substack{\mathbf{q},\mathbf{G}}} V_{\mathbf{q}+\mathbf{G}} 
\hat{\Lambda}^{*}_{\mathbf{k},\mathbf{q}+\mathbf{G}} \hat{Q}(\mathbf{k}+\mathbf{q}) 
\hat{\Lambda}^{\mathrm{T}}_{\mathbf{k},\mathbf{q}+\mathbf{G}} \\
&= -\frac{1}{2} \sum_{\substack{\mathbf{q},\mathbf{G}}} V_{\mathbf{q}+\mathbf{G}} \hat{Q}(\mathbf{k}+\mathbf{q}) 
\hat{\Lambda}^{*}_{\mathbf{k},\mathbf{q}+\mathbf{G}} \hat{\Lambda}^{\mathrm{T}}_{\mathbf{k},\mathbf{q}+\mathbf{G}},
\end{aligned}
\end{equation}
where we have used the commutative property and the Hermitian condition in the last line. Since we assume that the order parameters are momentum-independent, the expression simplifies to
\begin{equation}
\begin{aligned}
\hat{\Sigma}_{\mathrm{F}}(\mathbf{k}) 
&= -\frac{1}{2} \hat{Q} \sum_{\substack{\mathbf{q},\mathbf{G}}} V_{\mathbf{q}+\mathbf{G}} 
\hat{\Lambda}^{*}_{\mathbf{k},\mathbf{q}+\mathbf{G}} \hat{\Lambda}^{\mathrm{T}}_{\mathbf{k},\mathbf{q}+\mathbf{G}}.
\end{aligned}
\end{equation}

In the chiral-flat limit, the product of two form factors can be further evaluated, as per Eq.~\eqref{eq:form_factor_chiral}, yielding
\begin{equation}
\begin{aligned}
\hat{\Sigma}_{\mathrm{F}}(\mathbf{k}) 
&= -\frac{1}{2} \hat{Q} \sum_{\substack{\mathbf{q},\mathbf{G}}} V_{\mathbf{q}+\mathbf{G}} 
\left[ \left(\Lambda^{0}_{\mathbf{k},\mathbf{q}+\mathbf{G}}\right)^2 + \left(\Lambda^2_{\mathbf{k},\mathbf{q}+\mathbf{G}}\right)^2 \right] \hat{\mathbb{I}} \\
&= -F(\mathbf{k}) \hat{Q},
\end{aligned}
\end{equation}
where we define
\begin{equation}
F(\mathbf{k}) = \frac{1}{2} \sum_{\substack{\mathbf{q},\mathbf{G}}} V_{\mathbf{q}+\mathbf{G}} 
\left[ \left(\Lambda^{0}_{\mathbf{k},\mathbf{q}+\mathbf{G}}\right)^2 + \left(\Lambda^2_{\mathbf{k},\mathbf{q}+\mathbf{G}}\right)^2 \right],
\end{equation}
and $F(\mathbf{k})>0$ is a positive number.

Next, considering a general order parameter in a spinless, valleyless model, the order parameter is expressed as
\begin{equation}
\hat{Q} = Q_0\hat{\gamma}_0 + Q_1\hat{\gamma}_x + Q_2\hat{\gamma}_y + Q_3\hat{\gamma}_z.
\end{equation}
Since we focus only on the half-filling case, the constraint derived from Eq.~\eqref{eq:constrain_Q} leads to
\begin{equation}
Q_0 = 0, \quad Q_1^2 + Q_2^2 + Q_3^2 = 1.
\end{equation}
Therefore, the self-energy becomes
\begin{equation}
\hat{\Sigma}(\mathbf{k}) = -F(\mathbf{k}) \hat{Q} = -F(\mathbf{k}) \left(Q_1\hat{\gamma}_x + Q_2\hat{\gamma}_y + Q_3\hat{\gamma}_z\right).
\end{equation}
From Dyson's equation, we can write the inverse Green's function as
\begin{equation}
\begin{aligned}
\hat{G}^{-1}(\mathbf{k}, i\omega_n) 
&= \hat{G}^{-1}_0(\mathbf{k}, i\omega_n) - \hat{\Sigma}(\mathbf{k}) \\
&= i\omega_n + \mu - E_{\mathbf{k}} \hat{\gamma}_z + F_{\mathrm{F}}(\mathbf{k}) \left(Q_1\hat{\gamma}_x + Q_2\hat{\gamma}_y + Q_3\hat{\gamma}_z\right).
\end{aligned}
\end{equation}

The Green's function itself is given by
\begin{equation}
\begin{aligned}
\hat{G}(\mathbf{k}, i\omega_n) 
&= \frac{1}{\mathrm{Det}(\hat{G}^{-1})} \Big[ 
\left(i\omega_n + \mu\right)\hat{\gamma}_0 - F(\mathbf{k}) Q_1 \hat{\gamma}_x - F(\mathbf{k}) Q_2 \hat{\gamma}_y + \left(E_{\mathbf{k}} - F(\mathbf{k}) Q_3\right)\hat{\gamma}_z 
\Big],
\end{aligned}
\end{equation}
with the determinant
\begin{equation}
\mathrm{Det}(\hat{G}^{-1}) = \left[i\omega_n + \mu\right] - \left[F(\mathbf{k}) Q_1\right]^2 - \left[F(\mathbf{k}) Q_2\right]^2 - \left[E_{\mathbf{k}} - F(\mathbf{k}) Q_3\right]^2.
\end{equation}

The density matrix $\hat{\rho}(\mathbf{k})$ is computed as
\begin{equation}
\begin{aligned}
\hat{\rho}(\mathbf{k}) 
&= \frac{1}{\beta} \sum_{n} e^{-i\omega_n0^+} \hat{G}(\mathbf{k}, i\omega_n) \\
&= \frac{1}{2}\hat{\gamma}_0 + \frac{F(\mathbf{k}) Q_x \hat{\gamma}_x + F(\mathbf{k}) Q_y \hat{\gamma}_y + F(\mathbf{k}) Q_z \hat{\gamma}_z}{2 \sqrt{\left(F(\mathbf{k}) Q_1\right)^2 + \left(F(\mathbf{k}) Q_2\right)^2 + \left(F(\mathbf{k}) Q_3\right)^2}} \\
&= \frac{1}{2}\hat{\gamma}_0 + \frac{1}{2} \left(Q_x \hat{\gamma}_x + Q_y \hat{\gamma}_y + Q_z \hat{\gamma}_z\right).
\end{aligned}
\end{equation}

Consequently, the new order parameter is given by
\begin{equation}
\begin{aligned}
\hat{Q}^\prime(\mathbf{k}) 
&= 2\left(\hat{\rho}(\mathbf{k}) - \frac{1}{2} \hat{\mathbb{I}}\right) \\
&= Q_x \hat{\gamma}_x + Q_y \hat{\gamma}_y + Q_z \hat{\gamma}_z \\
&= \hat{Q},
\end{aligned}
\end{equation}
which remains unchanged in the self-consistent loop. It can also be shown that both $\hat{\gamma}_0$ and $\hat{\gamma}_y$ satisfy the commutative condition.

\section{Evaluating Diagrams}
\label{appendix:diagrams}

In this section, we will formulate detailed procedures about how to obtain different symmetry-breaking phases in Table.~(\ref{tb:ground_state_chiralflat}). The key idea is to start from symmetry-breaking ansatz and then evaluate the self-energy. By comparing both sides of Dyson's equation, a self-consistent condition can be found and calculated analytically. We will mainly focus on two cases: (i) Spin-polarized state at $\nu=0$, and (ii) Inter-valley coherence state at $\nu=-2$. The rest of the symmetry-breaking phases follow similar procedures therefore we do not show them here. 

\subsection{Spin polarized state at $\nu=0$}

We start with the spin-polarized state at $\nu=0$, in which the self-energy ansatz is assumed to be
\begin{equation}\label{eq:ansatz_SP}
\hat{\Sigma}(\mathbf{k})=\Sigma_\mathrm{H}(\mathbf{k})+\Sigma_\mathrm{F}(\mathbf{k})\hat{\sigma}_z,
\end{equation}
where $\Sigma_{\mathrm{H}}(\mathbf{k})$ and $\Sigma_{\mathrm{F}}(\mathbf{k})$ are in general complex functions depend only on momentum need to be solved self-consistently. From Dyson's equation, we have
\begin{equation}
\begin{aligned}
\hat{G}^{-1}(\mathbf{k},i\omega_n)
&=\hat{G}^{-1}_0(\mathbf{k},i\omega_n)-\hat{\Sigma}(\mathbf{k})\\
&=i\omega_n+\mu-\Sigma_{\mathrm{H}}(\mathbf{k})-E_{\mathbf{k}}\hat{\gamma}_{z}-\Sigma_\mathrm{F}(\mathbf{k})\hat{\sigma}_z.
\end{aligned}
\end{equation}
The inversion of $\hat{G}^{-1}(\mathbf{k},i\omega_n)$ is then straightforward,
\begin{equation}
\begin{aligned}
\hat{G}(\mathbf{k},i\omega_n)
&=\frac{\frac{1}{2}(\hat{\sigma}_0+\hat{\sigma}_z)\left[E_{\mathbf{k}}\hat{\gamma}_z+i\omega_n+\mu-\Sigma_\mathrm{F}(\mathbf{k})-\Sigma_{\mathrm{H}}(\mathbf{k})\right]}{\left[i\omega_n+\mu-\Sigma_\mathrm{F}(\mathbf{k})-\Sigma_{\mathrm{H}}(\mathbf{k})\right]-E_{\mathbf{k}}^2}+\frac{\frac{1}{2}(\hat{\sigma}_0-\hat{\sigma}_z)\left[E_{\mathbf{k}}\hat{\gamma}_z+i\omega_n+\mu+\Sigma_\mathrm{F}(\mathbf{k})-\Sigma_{\mathrm{H}}(\mathbf{k})\right]}{\left[i\omega_n+\mu+\Sigma_\mathrm{F}(\mathbf{k})-\Sigma_{\mathrm{H}}(\mathbf{k})\right]^2-E_{\mathbf{k}}^2}.\\
\end{aligned}
\end{equation}
The quasiparticle dispersion can be found by the poles of the Green's function (2 fold denigrate)
\begin{equation}
\begin{cases}
i\omega_n=-\mu+\Sigma_{\mathrm{H}}(\mathbf{k})+\Sigma_{\mathrm{F}}(\mathbf{k})\pm E_{\mathbf{k}}\\
i\omega_n=-\mu+\Sigma_{\mathrm{H}}(\mathbf{k})-\Sigma_{\mathrm{F}}(\mathbf{k})\pm E_{\mathbf{k}}\\
\end{cases}.
\end{equation}
In the chiral-flat limit, $E_{\mathbf{k}}\rightarrow0$, we arrive at two bands with 4 fold denigrate each
\begin{equation}\label{eq:hf_dispersion}
E^{\mathrm{HF}}_{\mathbf{k},\pm}=-\mu+\Sigma_{\mathrm{H}}(\mathbf{k})\pm\Sigma_{\mathrm{F}}(\mathbf{k}).
\end{equation}
Evaluating the Hartree diagram gives
\begin{equation}\label{eq:hartree_SP}
\begin{aligned}
\hat{\Sigma}_{\mathrm{H}}(\mathbf{k})
&=\sum_{\mathbf{G}}
V_{\mathbf{G}}
\hat{\Lambda}^{\mathrm{T}}_{\mathbf{k},\mathbf{G}}
\sum_{\mathbf{k}^{\prime}}
\frac{1}{\beta}\sum_{m}
\mathrm{Tr}\left[\hat{\Lambda}^{*}_{\mathbf{k}^{\prime},\mathbf{G}}\left(\hat{G}(\mathbf{k}^{\prime},i\omega_m)-\frac{1}{i\omega_n}\right)\right]\\
&=2\sum_{\mathbf{G}}
V_{\mathbf{G}}
\hat{\Lambda}^{\mathrm{T}}_{\mathbf{k},\mathbf{G}}
\sum_{\mathbf{k}^{\prime}}
\Lambda^{0}_{\mathbf{k}^{\prime},\mathbf{G}}\left[
f\left(-E_{\mathbf{k}^{\prime}}-\mu-\Sigma_{\mathrm{F}}(\mathbf{k}^{\prime})+\Sigma_{\mathrm{H}}(\mathbf{k}^{\prime})\right)
+f\left(E_{\mathbf{k}^{\prime}}-\mu-\Sigma_{\mathrm{F}}(\mathbf{k}^{\prime})+\Sigma_{\mathrm{H}}(\mathbf{k}^{\prime})\right)\right.\\
&\quad\quad\quad\quad\quad\quad\quad\left.+f\left(-E_{\mathbf{k}^{\prime}}-\mu+\Sigma_{\mathrm{F}}(\mathbf{k}^{\prime})+\Sigma_{\mathrm{H}}(\mathbf{k}^{\prime})\right)
+f\left(E_{\mathbf{k}^{\prime}}-\mu+\Sigma_{\mathrm{F}}(\mathbf{k}^{\prime})+\Sigma_{\mathrm{H}}(\mathbf{k}^{\prime})\right)-2
\right]\\
&=4\sum_{\mathbf{G}}
V_{\mathbf{G}}
\hat{\Lambda}^{\mathrm{T}}_{\mathbf{k},\mathbf{G}}
\sum_{\mathbf{k}^{\prime}}
\Lambda^{0}_{\mathbf{k}^{\prime},\mathbf{G}}\left[
f\left(E^{\mathrm{HF}}_{\mathbf{k}^{\prime},-}\right)
+f\left(E^{\mathrm{HF}}_{\mathbf{k}^{\prime},+}\right)-1
\right]\\
&=0,
\end{aligned}
\end{equation}
where in the last line we have we have used $f\left(E^{\mathrm{HF}}_{\mathbf{k},\pm}\right)=(1\mp1)/2$, which replaced the filled bands with 1 and empty bands with 0. The fact that Hartree vanish at charge neutral can be also understood by symmetry argument \cite{Ezzi_arxiv2024_Hartree}. Similarly, evaluate the Fock diagram gives
\begin{equation}\label{eq:fock_SP}
\begin{aligned}
\hat{\Sigma}_{\mathrm{F}}(\mathbf{k},i\omega_n)
&=-\frac{1}{\beta}\sum_{m}
\sum_{\mathbf{q},\mathbf{G}}
V_{\mathbf{q}+\mathbf{G}}
\hat{\Lambda}^{*}_{\mathbf{k},\mathbf{q}+\mathbf{G}}
\left[\hat{G}(\mathbf{k}+\mathbf{q},i\omega_n-i\omega_m)-\frac{1}{i\omega_n}\right]
\hat{\Lambda}^{\mathrm{T}}_{\mathbf{k},\mathbf{q}+\mathbf{G}}\\
&=-\frac{1}{2}
\sum_{\mathbf{q},\mathbf{G}}
V_{\mathbf{q}+\mathbf{G}}
\left[
\left(\Lambda^0_{\mathbf{k}, \mathbf{q}+\mathbf{G}}\right)^2
+\left(\Lambda^2_{\mathbf{k}, \mathbf{q}+\mathbf{G}}\right)^2
\right]
\left[\frac{1}{2}(\hat{\sigma}_0+\hat{\sigma}_z)\left[2f\left(E^{\mathrm{HF}}_{\mathbf{k}+\mathbf{q},+}\right)-1\right]+\frac{1}{2}(\hat{\sigma}_0-\hat{\sigma}_z)\left[2f\left(E^{\mathrm{HF}}_{\mathbf{k}+\mathbf{q},-}\right)-1\right]\right]\\
&=\frac{1}{2}
\sum_{\mathbf{q},\mathbf{G}}
V_{\mathbf{q}+\mathbf{G}}
\left[
\left(\Lambda^0_{\mathbf{k}, \mathbf{q}+\mathbf{G}}\right)^2
+\left(\Lambda^2_{\mathbf{k}, \mathbf{q}+\mathbf{G}}\right)^2
\right]
\left[\frac{1}{2}(\hat{\sigma}_0+\hat{\sigma}_z)-\frac{1}{2}(\hat{\sigma}_0-\hat{\sigma}_z)\right]\\
&=\frac{1}{2}
\sum_{\mathbf{q},\mathbf{G}}
V_{\mathbf{q}+\mathbf{G}}
\left[
\left(\Lambda^0_{\mathbf{k}, \mathbf{q}+\mathbf{G}}\right)^2
+\left(\Lambda^2_{\mathbf{k}, \mathbf{q}+\mathbf{G}}\right)^2
\right]
\hat{\sigma}_z.\\
\end{aligned}
\end{equation}
By comparing Eq.~(\ref{eq:ansatz_SP}) with Eq.~(\ref{eq:hartree_SP}) and Eq.~(\ref{eq:fock_SP}), we finally obtain the solution
\begin{equation}
\begin{aligned}
\Sigma_{\mathrm{H}}(\mathbf{k})&=0, \\
\Sigma_{\mathrm{F}}(\mathbf{k})&=\frac{1}{2}\sum_{\mathbf{q},\mathbf{G}}
V_{\mathbf{q}+\mathbf{G}}\left[
\left(\Lambda^0_{\mathbf{k}, \mathbf{q}+\mathbf{G}}\right)^2
+\left(\Lambda^2_{\mathbf{k}, \mathbf{q}+\mathbf{G}}\right)^2
\right].\\
\end{aligned}
\end{equation}

The density matrix in a one-particle basis can be computed via Eq.~(\ref{eq:density_matrix}),
\begin{equation}
\hat{\rho}(\mathbf{k})=\frac{1}{\beta}\sum_{n}e^{-i\omega_n0^-}\hat{G}(\mathbf{k},i\omega_n)=\frac{1}{2}(1-\hat{\sigma}_z),
\end{equation}
with the order parameter defined by Eq.~\eqref{eq:order_parameter},
\begin{equation}
\hat{Q}(\mathbf{k}) = 2\left(\hat{\rho}(\mathbf{k})-\frac{1}{2}\hat{\mathbb{I}}\right)=-\hat{\sigma}_z,
\end{equation}
and the ground energy can be found by Eq.~(\ref{eq:total_energy}),
\begin{equation}
\begin{aligned}
E_{\mathrm{tot}}
&=\sum_{\mathbf{k}}\mathrm{Tr}\left[\left(\hat{\mathcal{H}}_0(\mathbf{k})
+\frac{1}{2}\hat{\Sigma}(\mathbf{k})\right)\hat{\rho}(\mathbf{k})\right]\\
&=\frac{1}{2}
\sum_{\mathbf{k}}\mathrm{Tr}\left[\hat{\Sigma}(\mathbf{k})\hat{\rho}(\mathbf{k})\right]\\
&=\frac{1}{2}
\sum_{\mathbf{k}}\mathrm{Tr}\left[\Sigma_\mathrm{F}(\mathbf{k})\hat{\sigma}_z\frac{1}{2}(1-\hat{\sigma}_z)\right]\\
&=-2\sum_{\mathbf{k}}\Sigma_\mathrm{F}(\mathbf{k}).\\
\end{aligned}
\end{equation}

\subsection{Inter-valley coherence state at $\nu=-2$}

Next, we consider a more complicated case where two of eight bands are filled. According to Table.~(\ref{tb:ground_state_chiralflat}), we assume that self-energy has the form of
\begin{equation}\label{eq:ansatz_IVC}
\begin{aligned}
\hat{\Sigma}(\mathbf{k})
&=-2\Sigma_{\mathrm{H}}(\mathbf{k})+\Sigma_\mathrm{F}(\mathbf{k})\left[\frac{1}{2}(1+\hat{\sigma}_z)+\frac{1}{2}(1-\hat{\sigma}_z)\left(\hat{\tau}_x\cos{\theta_{\mathrm{IVC}}}+\hat{\tau}_y\sin{\theta_{\mathrm{IVC}}}\right)\hat{\gamma}_y\right]\\
&=-2\Sigma_{\mathrm{H}}(\mathbf{k})+\Sigma_\mathrm{F}(\mathbf{k})\left[\frac{1}{2}(1+\hat{\sigma}_z)+\frac{1}{2}(1-\hat{\sigma}_z)\hat{\tau}_{\mathrm{IVC}}\hat{\gamma}_y\right],\\
\end{aligned}
\end{equation}
where we have introduced $\hat{\tau}_{\mathrm{IVC}}=\hat{\tau}_x\cos{\theta_{\mathrm{IVC}}}+\hat{\tau}_y\sin{\theta_{\mathrm{IVC}}}$ with the phase of inter-valley coherence state $\theta_{\mathrm{IVC}}$. $\Sigma_{\mathrm{H}}(\mathbf{k})$ and $\Sigma_{\mathrm{F}}(\mathbf{k})$ are in general complex functions depend only on momentum need to be solved self-consistently. From Dyson's equation, we have
\begin{equation}
\begin{aligned}
\hat{G}^{-1}(\mathbf{k},i\omega_n)
&=\hat{G}^{-1}_0(\mathbf{k},i\omega_n)-\hat{\Sigma}(\mathbf{k})\\
&=i\omega_n+\mu+2\Sigma_{\mathrm{H}}(\mathbf{k})-E_{\mathbf{k}}\hat{\gamma}_{z}-\Sigma_\mathrm{F}(\mathbf{k})\left[\frac{1}{2}(1+\hat{\sigma}_z)+\frac{1}{2}(1-\hat{\sigma}_z)\hat{\tau}_{\mathrm{IVC}}\hat{\gamma}_y\right].
\end{aligned}
\end{equation}
The inversion of $\hat{G}^{-1}(\mathbf{k},i\omega_n)$ is then straightforward,
\begin{equation}
\begin{aligned}
\hat{G}(\mathbf{k},i\omega_n)
&=\frac{\frac{1}{2}(1+\hat{\sigma}_z)\left[E_{\mathbf{k}}\hat{\gamma}_z+i\omega_n+\mu+2\Sigma_{\mathrm{H}}(\mathbf{k})-\Sigma_\mathrm{F}(\mathbf{k})\right]}{\left[i\omega_n+\mu+2\Sigma_{\mathrm{H}}(\mathbf{k})-\Sigma_\mathrm{F}(\mathbf{k})\right]^2-E_{\mathbf{k}}^2}
+\frac{\frac{1}{2}(1-\hat{\sigma}_z)\left[E_{\mathbf{k}}\hat{\gamma}_z+i\omega_n+\mu+2\Sigma_{\mathrm{H}}(\mathbf{k})+\Sigma_\mathrm{F}(\mathbf{k})\hat{\tau}_{\mathrm{IVC}}\hat{\gamma}_y\right]}{\left[i\omega_n+\mu+2\Sigma_{\mathrm{H}}(\mathbf{k})\right]^2-\Sigma^2_\mathrm{F}(\mathbf{k})-E_{\mathbf{k}}^2}.\\
\end{aligned}
\end{equation}

Apparently, we have four poles which correspond to the eight quasiparticle dispersion (2 fold denigrate)
\begin{equation}
\begin{aligned}
i\omega_n&=-\mu-2\Sigma_{\mathrm{H}}(\mathbf{k})+\Sigma_{\mathrm{F}}(\mathbf{k})\pm E_{\mathbf{k}},\\
i\omega_n&=-\mu-2\Sigma_{\mathrm{H}}(\mathbf{k})\pm\sqrt{E^2_{\mathbf{k}}+\Sigma^2_{\mathrm{F}}(\mathbf{k})}.\\
\end{aligned}
\end{equation}

In the chiral-flat limit, $E_{\mathbf{k}}\rightarrow0$, we arrive at
\begin{equation}
E^{\mathrm{HF}}_{\mathbf{k},\pm}=-\mu-2\Sigma_{\mathrm{H}}(\mathbf{k})\pm\Sigma_{\mathrm{F}}(\mathbf{k}).
\end{equation}
with six bands degenerate in $E^{\mathrm{HF}}_{\mathbf{k},+}$ and two bands degenerate in $E^{\mathrm{HF}}_{\mathbf{k},-}$. Evaluating the Hartree diagram gives
\begin{equation}\label{eq:hartree_IVC}
\begin{aligned}
\hat{\Sigma}_{\mathrm{H}}(\mathbf{k})
&=\sum_{\mathbf{G}}
V_{\mathbf{G}}
\hat{\Lambda}^{\mathrm{T}}_{\mathbf{k},\mathbf{G}}
\sum_{\mathbf{k}^{\prime}}
\frac{1}{\beta}\sum_{m}
\mathrm{Tr}\left[\hat{\Lambda}^{*}_{\mathbf{k}^{\prime},\mathbf{G}}\left(\hat{G}(\mathbf{k}^{\prime},i\omega_m)-\frac{1}{i\omega_n}\right)\right]\\
&=2\sum_{\mathbf{G}}
V_{\mathbf{G}}
\hat{\Lambda}^{\mathrm{T}}_{\mathbf{k},\mathbf{G}}
\sum_{\mathbf{k}^{\prime}}
\Lambda^{0}_{\mathbf{k}^{\prime},\mathbf{G}}\left[
f\left(-\mu-E_{\mathbf{k}^{\prime}}-2\Sigma_{\mathrm{H}}(\mathbf{k}^{\prime})+\Sigma_{\mathrm{F}}(\mathbf{k}^{\prime})\right)
+f\left(-\mu+E_{\mathbf{k}^{\prime}}-2\Sigma_{\mathrm{H}}(\mathbf{k}^{\prime})+\Sigma_{\mathrm{F}}(\mathbf{k}^{\prime})\right)\right.\\
&\quad\quad\quad\quad\quad\quad\quad\left.+f\left(-\mu-2\Sigma_{\mathrm{H}}(\mathbf{k}^{\prime})-\sqrt{E^2_{\mathbf{k}^{\prime}}+\Sigma^2_{\mathrm{F}}(\mathbf{k}^{\prime})}\right)
+f\left(-\mu-2\Sigma_{\mathrm{H}}(\mathbf{k}^{\prime})+\sqrt{E^2_{\mathbf{k}^{\prime}}+\Sigma^2_{\mathrm{F}}(\mathbf{k}^{\prime})}\right)-2
\right]\\
&=2\sum_{\mathbf{G}}
V_{\mathbf{G}}
\hat{\Lambda}^{\mathrm{T}}_{\mathbf{k},\mathbf{G}}
\sum_{\mathbf{k}^{\prime}}
\Lambda^{0}_{\mathbf{k}^{\prime},\mathbf{G}}\left[
f\left(3E^{\mathrm{HF}}_{\mathbf{k}^{\prime},+}\right)
+f\left(E^{\mathrm{HF}}_{\mathbf{k}^{\prime},-}\right)-2
\right]\\
&=-2\sum_{\mathbf{G}}
V_{\mathbf{G}}
\hat{\Lambda}^{\mathrm{T}}_{\mathbf{k},\mathbf{G}}
\sum_{\mathbf{k}^{\prime}}
\Lambda^{0}_{\mathbf{k}^{\prime},\mathbf{G}}\\
&=-2\sum_{\mathbf{k}^{\prime},\mathbf{G}}
V_{\mathbf{G}}
\Lambda^{0}_{\mathbf{k}^{\prime},\mathbf{G}}
\left(\Lambda^0_{\mathbf{k}, \mathbf{G}} \hat{\gamma}_0 - \Lambda^2_{\mathbf{k}, \mathbf{G}} i\hat{\gamma}_y\right)\\
&=-2\sum_{\mathbf{k}^{\prime},\mathbf{G}}
V_{\mathbf{G}}
\Lambda^{0}_{\mathbf{k}^{\prime},\mathbf{G}}
\Lambda^0_{\mathbf{k}, \mathbf{G}} \hat{\gamma}_0,\\
\end{aligned}
\end{equation}
where we have used the property $\Lambda^{2}_{\mathbf{k}, \mathbf{G}}=-\Lambda^{2}_{\mathbf{k}, -\mathbf{G}}$ (See Eq.~(C24) in Ref.~\cite{Bernevig_PRB2021_HartreeFock}), which will cancel the value from opposite $\mathbf{G}$ components. Similarly, evaluating the Fock diagram gives
\begin{equation}\label{eq:fock_IVC}
\begin{aligned}
\hat{\Sigma}_{\mathrm{F}}(\mathbf{k},i\omega_n)
&=-\frac{1}{\beta}\sum_{m}
\sum_{\mathbf{q},\mathbf{G}}
V_{\mathbf{q}+\mathbf{G}}
\hat{\Lambda}^{*}_{\mathbf{k},\mathbf{q}+\mathbf{G}}
\left[\hat{G}(\mathbf{k}+\mathbf{q},i\omega_n-i\omega_m)-\frac{1}{i\omega_n}\right]
\hat{\Lambda}^{\mathrm{T}}_{\mathbf{k},\mathbf{q}+\mathbf{G}}\\
&=-\frac{1}{2}
\sum_{\mathbf{q},\mathbf{G}}
V_{\mathbf{q}+\mathbf{G}}
\left[
\left(\Lambda^0_{\mathbf{k}, \mathbf{q}+\mathbf{G}}\right)^2
+\left(\Lambda^2_{\mathbf{k}, \mathbf{q}+\mathbf{G}}\right)^2
\right]
\left[\frac{1}{2}(1+\hat{\sigma}_z)\left[2f\left(E^{\mathrm{HF}}_{\mathbf{k}+\mathbf{q},+}\right)-1\right]\right.\\
&\left.+\frac{1}{2}(1-\hat{\sigma}_z)\left[f\left(E^{\mathrm{HF}}_{\mathbf{k}+\mathbf{q},+}\right)+f\left(E^{\mathrm{HF}}_{\mathbf{k}+\mathbf{q},-}\right)-1+\hat{\tau}_{\mathrm{IVC}}\hat{\gamma}_y\biggl\{f\left(E^{\mathrm{HF}}_{\mathbf{k}+\mathbf{q},+}\right)-f\left(E^{\mathrm{HF}}_{\mathbf{k}+\mathbf{q},-}\right)\biggr\}\right]\right]\\
&=\frac{1}{2}
\sum_{\mathbf{q},\mathbf{G}}
V_{\mathbf{q}+\mathbf{G}}
\left[
\left(\Lambda^0_{\mathbf{k}, \mathbf{q}+\mathbf{G}}\right)^2
+\left(\Lambda^2_{\mathbf{k}, \mathbf{q}+\mathbf{G}}\right)^2
\right]
\left[\frac{1}{2}(1+\hat{\sigma}_z)+\frac{1}{2}(1-\hat{\sigma}_z)\hat{\tau}_{\mathrm{IVC}}\hat{\gamma}_y\right].\\
\end{aligned}
\end{equation}

\begin{table*}[!t]
    \renewcommand{\arraystretch}{1.25} 
    \centering
    \begin{tabular}{| c | c | c | c |} 
    \hline
    $\nu$ & Phase & Order parameter [Eq.~(\ref{eq:order_parameter})] & Self-energy [Eq.~(\ref{eq:self_energy_HF})] \\ [0.5ex] 
    \hline
    \multirow{5}{*}{0} & Polarized & $\pm\hat{\sigma}_z, \pm\hat{\tau}_z, \pm\hat{\tau}_z\hat{\sigma}_z$ & $\mp\hat{\Sigma}^{\mathrm{\RNum{1}}}_{\mathrm{F}}\hat{\sigma}_z, \mp\hat{\Sigma}^{\mathrm{\RNum{1}}}_{\mathrm{F}}\hat{\tau}_z, \mp\hat{\Sigma}^{\mathrm{\RNum{1}}}_{\mathrm{F}}\hat{\tau}_z\hat{\sigma}_z$ \\ 
    \cline{2-4}
     & \multirow{2}{*}{Hall} & $\pm\hat{\gamma}_y, \pm\hat{\tau}_z\hat{\gamma}_y$ & $\mp\Sigma^{\mathrm{\RNum{2}}}_{\mathrm{F}}\hat{\gamma}_y, \mp\Sigma^{\mathrm{\RNum{2}}}_{\mathrm{F}}\hat{\tau}_z\hat{\gamma}_y$ \\ 
    \cline{3-4}
    & & $\pm\hat{\sigma}_z\hat{\gamma}_y, \pm\hat{\tau}_z\hat{\sigma}_z\hat{\gamma}_y$ & $\mp\Sigma^{\mathrm{\RNum{2}}}_{\mathrm{F}}\hat{\sigma}_z\hat{\gamma}_y, \mp\Sigma^{\mathrm{\RNum{2}}}_{\mathrm{F}}\hat{\tau}_z\hat{\sigma}_z\hat{\gamma}_y$ \\ 
    \cline{2-4}
    & K-IVC & $\pm\hat{\gamma}_{y}\hat{\tau}_{\mathrm{IVC}}$ & $\mp\hat{\Sigma}^{\mathrm{\RNum{1}}}_{\mathrm{F}}\hat{\gamma}_{y}\hat{\tau}_{\mathrm{IVC}}$ \\ 
    \cline{2-4}
    & $\mathcal{T}$-IVC & $\pm\hat{\gamma}_{0}\hat{\tau}_{\mathrm{IVC}}$ & $\mp\Sigma^{\mathrm{\RNum{2}}}_{\mathrm{F}}\hat{\gamma}_{0}\hat{\tau}_{\mathrm{IVC}}$ \\ 
    \cline{1-4}
    \multirow{3}{*}{-1} & Hall & ${\color{green}\pm}\left(1-\hat{\tau}_{{\color{blue}\pm}}\hat{\sigma}_{{\color{red}\pm}}\right)\hat{\gamma}_y-\hat{\tau}_{{\color{blue}\pm}}\hat{\sigma}_{{\color{red}\pm}}$ & ${\color{green}\mp}\Sigma^{\mathrm{\RNum{2}}}_{\mathrm{F}}\left(1-\hat{\tau}_{{\color{blue}\pm}}\hat{\sigma}_{{\color{red}\pm}}\right)\hat{\gamma}_y+\hat{\Sigma}^{\mathrm{\RNum{1}}}_{\mathrm{F}}\hat{\tau}_{{\color{blue}\pm}}\hat{\sigma}_{{\color{red}\pm}}-\Sigma_\mathrm{H}$ \\ 
    \cline{2-4}
    & K-IVC+Hall & ${\color{yellow}\pm}\hat{\tau}_{\mathrm{IVC}} \hat{\sigma}_{{\color{red}\pm}} \hat{\gamma}_{y} {\color{green}\pm} \hat{\tau}_{{\color{blue}\pm}}\hat{\sigma}_{{\color{red}\mp}}\hat{\gamma}_y - \hat{\tau}_{{\color{blue}\mp}}\hat{\sigma}_{{\color{red}\mp}}$ & ${\color{yellow}\mp}\hat{\Sigma}^{\mathrm{\RNum{1}}}_{\mathrm{F}}\hat{\tau}_{\mathrm{IVC}}\hat{\sigma}_{{\color{red}\pm}}\hat{\gamma}_{y}{\color{green}\mp}\Sigma^{\mathrm{\RNum{2}}}_{\mathrm{F}}\hat{\tau}_{{\color{blue}\pm}}\hat{\sigma}_{{\color{red}\mp}}\hat{\gamma}_y+\hat{\Sigma}^{\mathrm{\RNum{1}}}_{\mathrm{F}}\hat{\tau}_{{\color{blue}\mp}}\hat{\sigma}_{{\color{red}\mp}}-\Sigma_\mathrm{H}$ \\ 
    \cline{2-4}
    & $\mathcal{T}$-IVC+Hall & ${\color{yellow}\pm}\hat{\tau}_{\mathrm{IVC}} \hat{\sigma}_{{\color{red}\pm}} \hat{\gamma}_{0} {\color{green}\pm} \hat{\tau}_{{\color{blue}\pm}}\hat{\sigma}_{{\color{red}\mp}}\hat{\gamma}_y - \hat{\tau}_{{\color{blue}\mp}}\hat{\sigma}_{{\color{red}\mp}}$ & ${\color{yellow}\mp}\Sigma^{\mathrm{\RNum{2}}}_{\mathrm{F}}\hat{\tau}_{\mathrm{IVC}}\hat{\sigma}_{{\color{red}\pm}}\hat{\gamma}_{0}{\color{green}\mp}\Sigma^{\mathrm{\RNum{2}}}_{\mathrm{F}}\hat{\tau}_{{\color{blue}\pm}}\hat{\sigma}_{{\color{red}\mp}}\hat{\gamma}_y+\hat{\Sigma}^{\mathrm{\RNum{1}}}_{\mathrm{F}}\hat{\tau}_{{\color{blue}\mp}}\hat{\sigma}_{{\color{red}\mp}}-\Sigma_\mathrm{H}$ \\ 
    \cline{1-4}
    \multirow{4}{*}{-2} & Polarized & ${\color{red}\pm}\hat{\tau}_{{\color{blue}\pm}}\hat{\sigma}_z-\hat{\tau}_{{\color{blue}\mp}}$ & $\hat{\Sigma}^{\mathrm{\RNum{1}}}_{\mathrm{F}}({\color{red}\mp}\hat{\tau}_{{\color{blue}\pm}}\hat{\sigma}_z+\hat{\tau}_{{\color{blue}\mp}})-2\Sigma_\mathrm{H}$ \\ 
    \cline{2-4}
    & Hall & ${\color{red}\pm}\hat{\tau}_{{\color{blue}\pm}}\hat{\sigma}_z\hat{\gamma}_y-\hat{\tau}_{{\color{blue}\mp}}$ & ${\color{red}\mp}\Sigma^{\mathrm{\RNum{2}}}_{\mathrm{F}}\hat{\tau}_{{\color{blue}\pm}}\hat{\sigma}_z\hat{\gamma}_y+\hat{\Sigma}^{\mathrm{\RNum{1}}}_{\mathrm{F}}\hat{\tau}_{{\color{blue}\mp}}-2\Sigma_\mathrm{H}$ \\ 
    \cline{2-4}
    & K-IVC & ${\color{yellow}\pm}\hat{\sigma}_{{\color{red}\pm}}\hat{\gamma}_{y} \hat{\tau}_{\mathrm{IVC}}-\hat{\sigma}_{{\color{red}\mp}}$ & ${\color{yellow}\mp}\hat{\Sigma}^{\mathrm{\RNum{1}}}_{\mathrm{F}}\hat{\tau}_{\mathrm{IVC}}\hat{\sigma}_{{\color{red}\pm}}\hat{\gamma}_{y}+\hat{\Sigma}^{\mathrm{\RNum{1}}}_{\mathrm{F}}\hat{\sigma}_{{\color{red}\mp}}-2\Sigma_\mathrm{H}$ \\ 
    \cline{2-4}
    & $\mathcal{T}$-IVC & ${\color{yellow}\pm}\hat{\sigma}_{{\color{red}\pm}}\hat{\gamma}_{0} \hat{\tau}_{\mathrm{IVC}}-\hat{\sigma}_{{\color{red}\mp}}$ & ${\color{yellow}\mp}\Sigma^{\mathrm{\RNum{2}}}_{\mathrm{F}}\hat{\tau}_{\mathrm{IVC}}\hat{\sigma}_{{\color{red}\pm}}\hat{\gamma}_{0}+\hat{\Sigma}^{\mathrm{\RNum{1}}}_{\mathrm{F}}\hat{\sigma}_{{\color{red}\mp}}-2\Sigma_\mathrm{H}$ \\ 
    \cline{1-4}
    -3 & Hall & ${\color{green}\pm}\hat{\tau}_{{\color{blue}\pm}}\hat{\sigma}_{{\color{red}\pm}}\hat{\gamma}_y-\hat{\tau}_{{\color{blue}\mp}}-\hat{\tau}_{{\color{blue}\pm}}\hat{\sigma}_{{\color{red}\mp}}$ & ${\color{green}\mp}\Sigma^{\mathrm{\RNum{2}}}_{\mathrm{F}}\hat{\tau}_{{\color{blue}\pm}}\hat{\sigma}_{{\color{red}\pm}}\hat{\gamma}_y+\hat{\Sigma}^{\mathrm{\RNum{1}}}_{\mathrm{F}}(\hat{\tau}_{{\color{blue}\mp}}+\hat{\tau}_{{\color{blue}\pm}}\hat{\sigma}_{{\color{red}\mp}})-3\Sigma_\mathrm{H}$ \\ 
    \hline
\end{tabular}
\caption{Analytical expressions for the order parameter and corresponding self-energy for different symmetry-breaking states at each integer filling factor $\nu$. Here, $\hat{\tau}$, $\hat{\sigma}$, and $\hat{\gamma}$ represent the Pauli matrices in the valley, spin, and band bases, respectively. We also introduce $\hat{\tau}_{\pm} = \frac{1}{2}(1 \pm \hat{\tau}_z)$ and $\hat{\sigma}_{\pm} = \frac{1}{2}(1 \pm \hat{\sigma}_z)$ for notational convenience. The color-coded terms are used to distinguish between different components of the order parameters and the corresponding self-energies for clarity. The Fock energy is modified, with the matrix $\hat{\gamma}_0$ and $\hat{\gamma}_{y}\hat{\tau}_{\mathrm{IVC}}$ being replaced by the first group solution, as described in Eq.~\eqref{eq:fock_group1}, and the matrix $\hat{\gamma}_y$ and $\hat{\gamma}_{0}\hat{\tau}_{\mathrm{IVC}}$ being replaced by the first group solution, as outlined in Eq.~\eqref{eq:fock_group2}.}
\label{tb:ground_state_nonchiralflat}
\end{table*}

By comparing Eq.~(\ref{eq:ansatz_IVC}) with Eq.~(\ref{eq:hartree_IVC}) and Eq.~(\ref{eq:fock_IVC}), we finally obtain the solution
\begin{equation}
\begin{aligned}
\Sigma_{\mathrm{H}}(\mathbf{k})&=\sum_{\mathbf{k}^{\prime},\mathbf{G}}
V_{\mathbf{G}}
\Lambda^{0}_{\mathbf{k}^{\prime},\mathbf{G}} \Lambda^0_{\mathbf{k}, \mathbf{G}}\\
\Sigma_{\mathrm{F}}(\mathbf{k})&=\frac{1}{2}\sum_{\mathbf{q},\mathbf{G}}
V_{\mathbf{q}+\mathbf{G}}
\left[
\left(\Lambda^0_{\mathbf{k}, \mathbf{q}+\mathbf{G}}\right)^2
+\left(\Lambda^2_{\mathbf{k}, \mathbf{q}+\mathbf{G}}\right)^2
\right].\\
\end{aligned}
\end{equation}

The density matrix in a one-particle basis can be computed via Eq.~(\ref{eq:density_matrix}),
\begin{equation}
\hat{\rho}(\mathbf{k})=\frac{1}{4}(1-\hat{\sigma}_z)\left( 1 - \hat{\tau}_{\mathrm{IVC}}\hat{\gamma}_y \right),
\end{equation}
with the order parameter defined by Eq.~\eqref{eq:order_parameter},
\begin{equation}
\hat{Q}(\mathbf{k}) = 2\left(\hat{\rho}(\mathbf{k})-\frac{1}{2}\hat{\mathbb{I}}\right)=-\frac{1}{2}(1-\hat{\sigma}_z)\hat{\tau}_{\mathrm{IVC}}\hat{\gamma}_y-\frac{1}{2}(1+\hat{\sigma}_z),
\end{equation}
and the ground energy can be found by Eq.~(\ref{eq:total_energy}),
\begin{equation}
\begin{aligned}
E_{\mathrm{tot}}
&=\frac{1}{2}
\sum_{\mathbf{k}}\mathrm{Tr}\left[\hat{\Sigma}(\mathbf{k})\hat{\rho}(\mathbf{k})\right]\\
&=\frac{1}{2}
\sum_{\mathbf{k}}\mathrm{Tr}\left[\biggl\{-2\Sigma_{\mathrm{H}}(\mathbf{k})+\Sigma_\mathrm{F}(\mathbf{k})\left[\frac{1}{2}(1+\hat{\sigma}_z)+\frac{1}{2}(1-\hat{\sigma}_z)\hat{\tau}_{\mathrm{IVC}}\hat{\gamma}_y\right]\biggr\}\frac{1}{4}(1-\hat{\sigma}_z)\left( 1 - \hat{\tau}_{\mathrm{IVC}}\hat{\gamma}_y \right)\right]\\
&=-2 \sum_{\mathbf{k}}\Sigma_{\mathrm{H}}(\mathbf{k})-\sum_{\mathbf{k}}\Sigma_{\mathrm{F}}(\mathbf{k}).\\
\end{aligned}
\end{equation}

\section{Analytical solutions for non-chiral flat limit}
\label{appendix:solutions_non-chiralflat}
In the main text, we analyzed the symmetry-breaking phases at charge neutrality in the non-chiral flat limit and our results were summarized in Table~\ref{tb:ground_state_chiralflat}.  We can also extend our results away from charge neutrality.  We note that Eq.~\eqref{eq:diagram_hartree} and Eq.~\eqref{eq:diagram_fock} are completely general and apply at any filling.  Solving for non-zero filling we find that all the order parameters remain the same as the chiral-flat band limit.  However, the solution for the Fock-induced self-energy is modified.  For for non-zero integer filling the ground state could be mixture of phases classified in the chiral flat limit.  We find that for all terms with the matrix $\hat{\gamma}_0$ and $\hat{\gamma}_{y}\hat{\tau}_{\mathrm{IVC}}$ (Group 1) we need to replace 
$\Sigma_\mathrm{F}$ with $
\hat{\Sigma}^{\mathrm{\RNum{1}}}_{\mathrm{F}}(\mathbf{k})=
\Sigma^1_{\mathrm{F}}(\mathbf{k})
+\hat{\gamma}_x\hat{\tau}_z
\Sigma^2_{\mathrm{F}}(\mathbf{k})
+\hat{\gamma}_z\hat{\tau}_z\Sigma^3_{\mathrm{F}}(\mathbf{k})$, where $\Sigma^i_{\mathrm{F}}(\mathbf{k})$ was defined below Eq.~\eqref{eq:fock_group1_componants}.  Similarly for all terms with the matrix  $\hat{\gamma}_y$ and $\hat{\gamma}_{0}\hat{\tau}_{\mathrm{IVC}}$ (Group 2) need to replace $\Sigma_\mathrm{F}$  with the terms in Eq.~\eqref{eq:fock_group2}.  With this modification, we can then obtain analytical results for non-chiral flat limit at any integer filling.  Our results are summarized in Table~\ref{tb:ground_state_nonchiralflat}.

\begin{figure*}
    \centering
    \includegraphics[width=1\columnwidth]{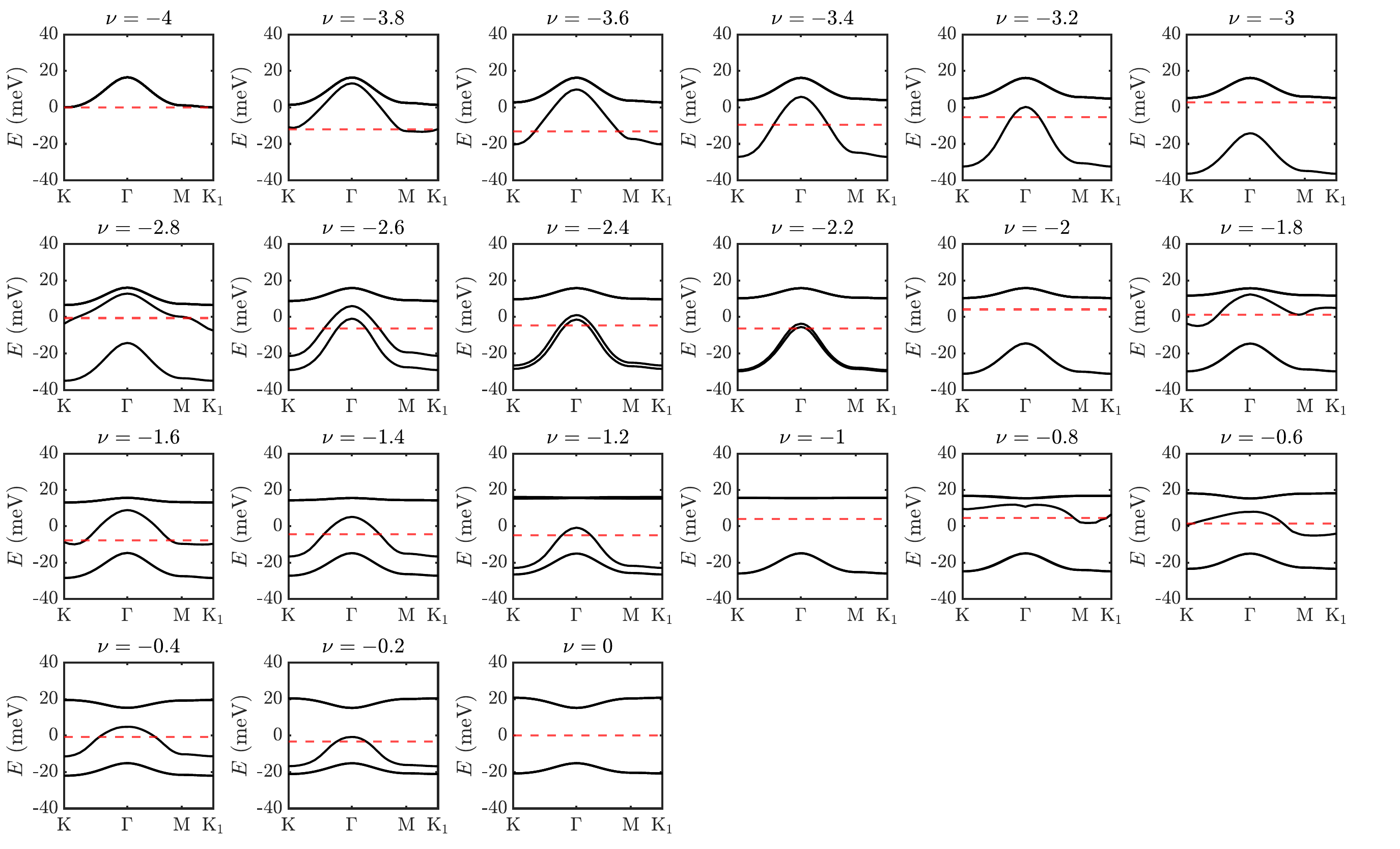}
    \caption{Numerical results for Hartree-Fock band structures (black solid line) and the corresponding chemical potential (red dashed line) in the chiral flat limit are presented as a function of the filling factor $\nu$.}
    \label{fig:hf_bands}
\end{figure*}

\section{Numerical results for for Hartree-Fock band structures at chiral flat limit}

To understand the origin of the cascade feature, we present numerical results for the Hartree-Fock band structures (black solid line) and the corresponding chemical potential (red dashed line) in the chiral flat limit as a function of the filling factor $\nu$, as shown in Fig.~\eqref{fig:hf_bands}.  For example, starting at $\nu=-3$ and increasing $\nu$ by doping with electrons, the chemical potential, interpreted as the energy of the highest occupied level decreases until the band is almost completely occupied (see $\nu = -2.2$).  This trend arises from the exchange energy, which becomes increasingly negative as the electron density grows, significantly outweighing the small upward shift of the Fermi level within the band. However, at integer filling factors (see e.g. $\nu=-2$), a new band becomes occupied resulting in a sharp positive jump in $\nu$. This corresponds to the band gap.  In the GW calculation, the RPA ring diagrams provide additional screening to the Fock contribution, thereby reducing the magnitude of the oscillations while preserving the overall increase of the chemical potential with filling.

\section{Gap equation for metal-insulator transition}
\label{appendix:MIT}

Interestingly, we find that the metal–insulator transition is highly similar to the superconducting-metal transition. This similarly implies a similar mechanism for both two phenomena. At charge neutral, the Dyson's equation can be written down as, from Eq.~(\ref{eq:hartree_SP}) and Eq.~(\ref{eq:fock_SP}),
\begin{equation}\label{eq:dyson_full}
\begin{aligned}
\Sigma_{\mathrm{H}}(\mathbf{k})&=\frac{1}{2}\sum_{\mathbf{q},\mathbf{G}}
V_{\mathbf{q}+\mathbf{G}}\left[\left(\Lambda^{0}_{\mathbf{k},\mathbf{q}+\mathbf{G}}\right)^2+\left(\Lambda^{2}_{\mathbf{k},\mathbf{q}+\mathbf{G}}\right)^2\right]\{f\left[\Sigma_{\mathrm{H}}(\mathbf{q})-\Sigma_{\mathrm{F}}(\mathbf{q})\right]+f\left[\Sigma_{\mathrm{H}}(\mathbf{q})+\Sigma_{\mathrm{F}}(\mathbf{q})\right]-1\},\\
\Sigma_{\mathrm{F}}(\mathbf{k})&=\frac{1}{2}\sum_{\mathbf{q},\mathbf{G}}
V_{\mathbf{q}+\mathbf{G}}\left[\left(\Lambda^{0}_{\mathbf{k},\mathbf{q}+\mathbf{G}}\right)^2+\left(\Lambda^{2}_{\mathbf{k},\mathbf{q}+\mathbf{G}}\right)^2\right]\{f\left[\Sigma_{\mathrm{H}}(\mathbf{q})-\Sigma_{\mathrm{F}}(\mathbf{q})\right]+f\left[\Sigma_{\mathrm{H}}(\mathbf{q})+\Sigma_{\mathrm{F}}(\mathbf{q})\right]\}.\\
\end{aligned}
\end{equation}

This is a self-consistent equation set which is hard to solve at first glance. However, we can simplify the equations above by noticing that the Hartree self-energy is zero at the charge neutral point, i.e., $\Sigma_{\mathrm{H}}(\mathbf{k})=0$. Therefore, Eq.~(\ref{eq:dyson_full}) become
\begin{equation}\label{eq:dyson_simplified}
\begin{aligned}
\Sigma_{\mathrm{F}}(\mathbf{k})
&=\frac{1}{2}\sum_{\mathbf{q},\mathbf{G}}
V_{\mathbf{q}+\mathbf{G}}
\left[\left(\Lambda^{0}_{\mathbf{k},\mathbf{q}+\mathbf{G}}\right)^2+\left(\Lambda^{2}_{\mathbf{k},\mathbf{q}+\mathbf{G}}\right)^2\right]
\tanh{\frac{\Sigma_{\mathrm{F}}(\mathbf{q})}{2T}.}\\
\end{aligned}
\end{equation}

A strong similarity between Eq.~(\ref{eq:dyson_simplified}) with the ``BCS" gap equation inspired us to define the ``BCS"-like gap by assuming the gap is a constant within BZ, i.e., $\Delta=\Sigma_{\mathrm{F}}(\mathbf{k})$, therefore, we have
\begin{equation}
1=\frac{U_{\mathrm{F}}}{2}\frac{\tanh{(\Delta/2T)}}{\Delta},
\end{equation}
where we have introduced the pairing strength
\begin{equation}
U_{\mathrm{F}}=\sum_{\mathbf{k},\mathbf{q},\mathbf{G}}
V_{\mathbf{q}+\mathbf{G}}
\left[\left(\Lambda^{0}_{\mathbf{k},\mathbf{q}+\mathbf{G}}\right)^2+\left(\Lambda^{2}_{\mathbf{k},\mathbf{q}+\mathbf{G}}\right)^2\right],
\end{equation}
and take the limit $\Delta\rightarrow0$, we finally get
\begin{equation}
\begin{aligned}
1&=\lim_{\Delta\rightarrow0}\frac{U_{\mathrm{F}}}{2}\frac{\tanh{(\Delta/2T)}}{\Delta}\\
&=\lim_{\Delta\rightarrow0}\frac{U_{\mathrm{F}}}{2}\frac{1}{2T}\left(1-\tanh^2{\frac{\Delta}{2T}}\right) =\frac{U_{\mathrm{F}}}{4T},
\end{aligned}
\end{equation}
which gives the solution of transition temperature $T=\frac{U_{\mathrm{F}}}{4}$.

\end{document}